\documentclass[fleqn,usenatbib]{mnras}

\usepackage{newtxtext,newtxmath}

\usepackage[T1]{fontenc}

\DeclareRobustCommand{\VAN}[3]{#2}
\let\VANthebibliography\thebibliography
\def\thebibliography{\DeclareRobustCommand{\VAN}[3]{##3}\VANthebibliography}

\usepackage{preamble}
\usepackage{subfiles}

\title[Search for VMS and determination of stellar and nebular properties $Z\lesssim0.1\,Z_\odot$]{Stars and gas in the most metal-poor galaxies -- II. Search for VMS and determination of stellar population and nebular properties at $Z\lesssim0.1\,Z_\odot$}
\author[A. Wofford et al.]{\parbox{\textwidth}{Aida Wofford\orcidlink{0000-0001-8289-3428}$^{1}$,\thanks{E-mail: awofford@astro.unam.mx}
Oskar Arangur\'e\orcidlink{0009-0004-7070-1782}$^{1}$,
Christophe Morisset\orcidlink{0000-0001-5801-6724}$^{1,2}$,
Calum Hawcroft\orcidlink{0000-0003-0145-8964}$^{3}$,
Claus Leitherer\orcidlink{0000-0003-2685-4488}$^{3}$,
Svea Hernández\orcidlink{0000-0003-4857-8699}$^{3}$,
Gustavo Bruzual\orcidlink{0000-0003-3458-2275}$^{4}$ and
Stephane Charlot\orcidlink{0000-0003-3458-2275}$^{5}$}
\\
$^{1}$Instituto de Astronom\'ia, Universidad Nacional Aut\'onoma de M\'exico, Unidad Acad\'emica en Ensenada, Km 103 Carr. Tijuana$-$Ensenada, Ensenada, B.C.,\\C.P. 22860, M\'exico\\
$^{2}$Instituto de Ciencias Físicas, Universidad Nacional Autónoma de México, Av. Universidad s/n, 62210 Cuernavaca, Mor., México\\
$^{3}$Space Telescope Science Institute, 3700 San Martin Drive, Baltimore, MD 21218, USA\\
$^{4}$Instituto de Radioastronomía y Astrofísica, Universidad Nacional Autónoma de México, Campus Morelia, 58089 Morelia, Michoacán, México\\
$^{5}$Sorbonne Universit\'e, CNRS, UMR 7095, Institut d'Astrophysique de Paris, 98 bis bd Arago, 75014 Paris, France
}

\author[A. Wofford et al.]{\parbox{\textwidth}{Aida Wofford\orcidlink{0000-0001-8289-3428}$^{1}$,\thanks{E-mail: awofford@astro.unam.mx}
Oskar Arangur\'e\orcidlink{0009-0004-7070-1782}$^{1}$,
Christophe Morisset\orcidlink{0000-0001-5801-6724}$^{1,2}$,
Calum Hawcroft\orcidlink{0000-0003-0145-8964}$^{3}$,
Claus Leitherer\orcidlink{0000-0003-2685-4488}$^{3}$,
Svea Hernández\orcidlink{0000-0003-4857-8699}$^{3}$,
Gustavo Bruzual\orcidlink{0000-0003-3458-2275}$^{4}$ and
Stephane Charlot\orcidlink{0000-0003-3458-2275}$^{5}$}
\\
$^{1}$Instituto de Astronom\'ia, Universidad Nacional Aut\'onoma de M\'exico, Unidad Acad\'emica en Ensenada, Km 103 Carr. Tijuana$-$Ensenada, Ensenada, B.C.,\\C.P. 22860, M\'exico\\
$^{2}$Instituto de Ciencias Físicas, Universidad Nacional Autónoma de México, Av. Universidad s/n, 62210 Cuernavaca, Mor., México\\
$^{3}$Space Telescope Science Institute, 3700 San Martin Drive, Baltimore, MD 21218, USA\\
$^{4}$Instituto de Radioastronomía y Astrofísica, Universidad Nacional Autónoma de México, Campus Morelia, 58089 Morelia, Michoacán, México\\
$^{5}$Sorbonne Universit\'e, CNRS, UMR 7095, Institut d'Astrophysique de Paris, 98 bis bd Arago, 75014 Paris, France
}

\date{Accepted XXX. Received YYY; in original form ZZZ}

\pubyear{2015}

\begin{document}
\label{firstpage}
\pagerange{\pageref{firstpage}--\pageref{lastpage}}
\maketitle


\begin{abstract}

Constraining the properties of OB-star populations and ionized gas at low metallicity is essential for understanding a broad range of astrophysical objects, from stellar black holes to the first galaxies. Rest-frame UV spectra of nearby metal-poor galaxies provide crucial constraints on theoretical predictions of low-metallicity stars and gas. However, such spectra are scarce at $Z\lesssim0.1\,Z_\odot$, and candidate Very Massive Stars (VMS) remain elusive.

We analyse optical and new UV spectra of nearby star-forming galaxies with $Z\lesssim0.1\,Z_\odot$, selected for their nebular He\2 $\lambda4686$ emission and lack of Wolf-Rayet blue and red bumps, making them promising targets for VMS searches. Some galaxies exhibit stellar wind features (group 1), while others do not (group 2). For group 1, we compare STARBURST99, GALAXEV, and BPASS predictions with observed stellar lines to constrain OB-population ages and masses. The inferred ages are too old for VMS to be present. We also use pySTARBURST99 and GALAXEV photoionization models to investigate the ionized gas in a subsample, finding no evidence for VMS-dominated ionization.

The absence of detected VMS may reflect their rarity, our limited sample size, or shortcomings in current models. Moreover, reproducing the extreme optical and UV line ratios simultaneously requires unrealistic N/O, ionization parameters, or Lyman continuum escape fractions, suggesting that multiple ionizing-source populations may be needed.

\end{abstract} 

\begin{keywords}
    galaxies: dwarf,
    galaxies: stellar content,
    (ISM:) H II regions,
    ultraviolet: galaxies,
    ultraviolet: ISM
\end{keywords}


\section{Introduction}\label{sec:introduction}

Constraining the properties of OB-type star populations and ionized
gas at low metallicity is of large relevance for a variety of fields in astrophysics, since it relates to topical questions such as how early galaxies formed, chemically-enriched the universe, and transformed the ionization structure of the IGM; and how the most  metal-poor massive stars evolve and give rise to black holes. We expect low-metallicity massive stars to have weaker winds, higher luminosities, and altered Wolf-Rayet (WR) phases, which shape their energy/momentum output, ionizing spectrum, chemical yields and evolution endpoints \citep{Pauli2026arXiv260825003P}. Crucial constraints for theoretical predictions of low-metallicity massive stars and the ionized gas they imply, can be obtained from ultraviolet spectra of nearby metal-poor star-forming (SF) galaxies. This is because the UV is where massive-star peak in emission and where some of their main diagnostics are located \citep{Roman-Duval2025ApJ...985..109R}. Unfortunately, the nearest most metal-poor SF galaxies are either too far or not star-forming enough to obtain representative samples of individual star spectra  \citep{Garcia2021ExA....51..887G, Senchyna2026ASPC..542....3S}. Thus, for such galaxies, one has to resort to spectra of the integrated light of  stars, gas and dust in galaxies, which poses the challenge of disentangling their contributions to the signal. 

The most metal-poor SF galaxies known are arbitrarily defined to have ionized-gas oxygen abundance ratios of ${\rm O/H\lesssim0.1\times\,(O/H)_\odot}$ \citep{Kunth2000A&ARv..10....1K}. They are commonly referred to as extremely metal-poor galaxies (XMPGs). For reference, if we adopt the solar reference set of \cite{Asplund2009}, XMPGs have $\rm{12+log(O/H)\lesssim7.69}$; and if we use oxygen as a gauge of metallicity, Z, here defined as the mass ratio of metals to all elements, then XMPGs have Z$\lesssim0.0014$.  Two of the most well studied XMPGs are I Zw 18, which is considered to be the Rosetta Stone of XMPGs\citep{Kunth1986ApJ...300..496K, Izotov1990Natur.343..238I, Izotov1997ApJ...487L..37I, Garnett1997ApJ...481..174G, Brown2002ApJ...579L..75B, Lecavelier2004A&A...413..131L, Lebouteiller2013, James2014, Kehrig2016MNRAS.459.2992K, Dorottya2025A&A...703A.131S}; and SBS 0335-052E, which is an extreme He\2 emitter \citep{Izotov1990Natur.343..238I, James2014, Kehrig2018MNRAS.480.1081K,  Wofford2021MNRAS.500.2908W, Herenz2023A&A...670A.121H}. Some of the nearest most-metal poor XMPGs have been found by the Subaru
EMPRESS survey \citep{Nakajima2022ApJS..262....3N} or the Dark Energy Spectroscopic Instrument \citep{Levi2013arXiv1308.0847L, DESI2024AJ....167...62D, Scholte2026MNRAS.551g1381S}. Unfortunately, some of these galaxies are too faint for UV spectroscopy. 

Since known nearby-XMPGs suitable for UV spectroscopy are rare, and obtaining rest-frame UV observations for suitable systems requires space-based observations, which are hard to obtain and expensive, UV spectra of nearby XMPGs are scarce. Some past works that analyse UV spectra of nearby samples include  \cite{Senchyna2019b}, who combined archival data with observations of six local XMPGs. These authors found that high-ionization nebular C\4 $\lambda\lambda1548$, 1551 emission is ubiquitous in XMPGs with high specific star-formation rates (sSFRs) and highlight the importance of this emission as a signpost of massive stars in the re-ionization era. However, \cite{Mingozzi2022ApJ...939..110M} studied the CLASSY \citep{Berg2022ApJS..261...31B} sample, which includes some XMPGs. These authors warn against classifying different sources of ionization by using C\4. This is because this doublet is affected by resonant scattering and stellar wind absorption which make its profile hard to interpret. On the other hand,  \cite{Dorottya2025A&A...703A.131S} found that chemically homogeneously evolving stars produce C\4~emission consistent with that of I Zw 18.

In addition to the C\4  doublet, some UV XMPG spectra may show the He\2 $\lambda$1640 emission line, which has an optical counterpart at 4686 \AA. These He\2 lines are among the highest-ionization that young stellar populations can plausibly produce, which makes them interesting for studying the content of such populations. The lines form via recombination and since He$^{+}$ ionization requires photons of $\ge54.4$ eV, the presence of He\2 lines flag hard-ionization sources capable of producing extreme-UV/soft-X-ray photons. The lines can be produced by massive stars \citep{Gutkin2016MNRAS.462.1757G, Plat2019MNRAS.490..978P, Wofford2021MNRAS.500.2908W}; the accretion disks of Active Galactic Nuclei \citep[AGN;][]{Feltre2016MNRAS.456.3354F} or stellar black holes in high-mass X-ray binaries \citep[HMXB,][]{Senchyna2020, Komarova2024ApJ...967..117K}; or shocked regions \cite[e.g.][]{Allen2008ApJS..178...20A, Alarie2019RMxAA..55..377A}. Based on a study of galaxies observed as part of the Sloan Digital Sky Survey (SDSS) Data Release (DR) 7, \cite{Shirazi2012MNRAS.421.1043S} found that the presence of strong nebular (narrow) He\2$\,\lambda4686$ emission in SF and starburst galaxies is more frequent at low Z. Harder stellar ionizing continuum at low metallicity could explain the higher frequency of narrow-line He\2 emitters at low Z. \cite{Shirazi2012MNRAS.421.1043S} also found that at low metallicity, models of photoionization by single and binary stars able to reproduce strong spectral lines, were not able to predict some of the observed ratios of He\2$\,\lambda4686$/H$\beta$. This is known as the He\2~problem. At cosmic noon and beyond ($z\approx2-4$), \cite{Saxena2020A&A...636A..47S} studied a sample of SF galaxies with $Z\approx0.1 Z_\odot$. They found that narrow ($FWHM<1000$ km/s) He\2 emitters and objects with no He\2 emission share the same stellar masses, SFRs, and rest-frame UV magnitudes; and concluded that the emitters may be experiencing a recent burst of star-formation, or be powered by additional sources of He\2 ionizing photons. 

There are several stellar sources of He\2$^+$ ionizing photons. Classical (hydrogen-free) Wolf-Rayet (cWR) stars is one of them. cWRs descend from O-type stars with initial masses of $\gtrsim25$ M$_\odot$ (depending somewhat on metallicity). They are divided into those with strong emission lines of helium and nitrogen (WN subtype) and those with strong helium, carbon, and oxygen (WC and WO subtypes). 
The hot dense winds of these stars can produce broad ($FWHM\geq1000$ km/s) stellar He\2 emission in galaxy spectra. In principle, at very low metallicity, weaker line-driven winds suppress the formation of cWR stars, although they could form via the binary channel. Interestingly, \cite{Garnett1991ApJ...373..458G} observed two He\2 nebulae centred on WR stars, including one in galaxy IC 1613, which has an ionized-gas oxygen abundance of ${\rm O/H=0.1\times\,(O/H)_\odot}$ \citep{Garcia2014ApJ...788...64G}, which puts IC 1613 in the XMPG regime. Models by \cite{Schaerer1998ApJ...497..618S} show that nebular He\2$\,\lambda4686$ emission from a low-metallicity starburst should be attributed to the presence of WC/WO stars and/or hot WN stars evolving to become WC/WO stars (see also \citealt{Berg2024ApJ...971...87B}). The studies of \cite{Garnett1991ApJ...373..458G} and \cite{Schaerer1998ApJ...497..618S} imply that some of the He\2 ionizing photons of cWR stars can escape the winds of these stars and produce nebular He\2 emission. 

High-mass X-ray binaries, which are stellar black holes accreting from a companion star, are another stellar source of H\2 ionizing photons. However, photoionization models by \cite{Senchyna2020MNRAS.494..941S} with input spectral energy distributions containing the joint contribution of stellar atmospheres and a multicolour disc model for HMXBs, show that HMXBs are not the dominant source of He$^{+}$ ionization in metal-poor star-forming galaxies. 

Stripped stars, which are defined by \citet{Gotberg2023ApJ...959..125G} as massive stars ($8-25$ M$_\odot$) stripped of their hydrogen-rich envelopes via binary interactions are also stellar sources of He\2~ionizing photons \citep{Beryl2025arXiv251116752H}. However, population-synthesis tests generally find that the photon budget from these sources is insufficient at low metallicity to solve the He\2 problem on its own \citep{Saxena2020A&A...636A..47S, Gonzalez2025A&A...702A...3G}.

A fourth stellar source of He\2~ionization is Very Massive Stars (VMS), defined by \cite{Vink2015HiA....16...51V} to be $>100\,$M$_\odot$. These stars could form via stellar mergers on the Main Sequence (MS). VMS are of particular interest because they efficiently spread nucleosynthesis products through their boosted stellar winds, they are predicted to explode as pair-instability supernovae or to form heavy black-holes from direct collapse, and they dominate certain spectral lines in the integrated light of starbursts, including lines in the UV \citep{Crowther2010MNRAS.408..731C, Vink2022ARA&A..60..203V}. Although VMS are rare due to their short lifetimes ($<3$ Myr; see example fig.~10 of \citealt{Wofford2023MNRAS.523.3949W}) and the shape of the IMF, direct or indirect evidence of the presence of VMS has been found all the way to cosmic dawn, where they may have played a key role in the formation of the first galaxies \citep{Martins2025A&A...698A.262M}. Only about twenty of them are known in the Galaxy and the Large Magellanic Cloud \citep{Martins2025A&A...698A.262M}. According to \cite{Senchyna2021MNRAS.503.6112S}, and excess of VMS may be required to mitigate  the discrepancy between model fits to the strongest wind features and the optical nebular lines of nearby SF galaxies with cWR features. At moderately low metallicities, the UV range alone is not sufficient to distinguish between VMS- and cWR-dominated sources because their spectra share similar features \citep{Wofford2023MNRAS.523.3949W}. However,  \citet{Martins2023A&A...678A.159M} show that  the presence of N\3 4640 + C\3 4650 ``blue-bump'' emission of similar strength to the He\2 4686 emission, plus broad C\4 5808-12 ``red-bump'' emission, is a key indication of the presence of cWR. In contrast, a cluster with VMS will have He\2 4686  emission that is stronger than the blue-bump  and no broad C\4 5808-12 emission. In XMPGs, candidate VMS are yet to be found. Interestingly, \cite{Grafener2015A&A...578L...2G} predict that nebular He\2 emission can originate in the weak winds of VMS, but such models are not currently implemented in predictions of integrated spectra. He\2 emission from VMS would help  mitigate the He\2 problem \citep{Wofford2021MNRAS.500.2908W}. 

If one is able to establish that the nebular He\2 emission of an XMPG is primarily due to the presence of massive stars or their evolved descendants, one can use the He\2 emission to constrain the properties of these sources. This may be possible by comparing spectral synthesis models (which combine synthetic populations with matched stellar-atmosphere models) to observations of direct stellar lines; or by comparing models of photoionization by massive stars to observations of the nebular lines that they imply. 

In the XMP regime, spectral synthesis and photoionization models are poorly constrained and VMS are yet to be found. In order to mitigate this problem, we obtained UV spectra of nearby XMPGs that are good targets for VMS searches. Our observations cover from P\5~1118 to C\3]~1909 and contain a rich set of stellar and nebular diagnostics. The sample is split into two groups depending on whether the observed C\4~lines are dominated by stellar or nebular components. We test different spectral synthesis models against the  stellar lines of the first group, and models of photoionization by AGN, shocks, and massive stars against the observed nebular lines of both groups. In this paper we address two main questions: i) What are the properties of the OB-star populations and the ionized gas in our sample of XMPGs, according to independent models; and ii) is He\2~emission a good tracer of XMP VMS?

In \S~\ref{sec:observations} we describe the sample selection, optical and UV observations, and UV data reduction. In \S~\ref{sec:sp_properties}, we divide the galaxies into two groups one with P-Cygni like and another one with nebular like C\4 1550 profiles. We also discuss the apparently broad He\2 1640 profiles of the three galaxies with P-Cygni like profiles. In \S~\ref{sec:models}, given that spectral synthesis models are poorly constrained in the XMP regime, we discuss differences in predicted age and metallicity sensitive spectral profiles predicted by the independent models used to later derive OB-star population properties. Details about the models are given in Appendix~\ref{app:model_details}. In \S~\ref{sec:stars} we determine the properties of the OB-star populations using direct stellar lines and different models. In \S~\ref{sec:gas}, we analyse the location of our galaxies in various optical and UV diagnostic diagrams, relative to  models of photoionization by AGN, shocks, and massive stars. Independent models of the ionizing output of stellar populations are used as input in the photoionization models. Finally, in \S~\ref{sec:conclusion} we provide a summary and conclusions.

Hereafter, for simplicity, we drop the $\lambda$ that precedes the wavelength in \AA~from the  notation of spectral lines, e.g., we use He\2~1640 instead of He\2~$\lambda$1640. 

\section{Sample, observations, and data reduction}\label{sec:observations}

\subsection{Sample selection and optical spectra}

We searched the literature for a sample of galaxies meeting favourable conditions for XMP VMS searches, and sufficiently bright for UV observations. The galaxies have narrow He\2~4686 emission and no blue/red WR bumps. The latter is for making it  easier to distinguish between VMS and cWR stars in the UV, since in this range, they can show some similar features. This assumes that in the XMP regime, VMS produce  nebular He\2 emission when present at ages < 3 Myr. We settled on a sample of seven XMPGs and one galaxy with sub-SMC metallicity. Three galaxies in our sample leverage Chandra observations. From the Chandra observations, \cite{Prestwich2013} conclude that two of our galaxies (J1119 and J1132) host Ultra Luminous X-ray (ULX) sources, while J0944 does not. 

\begin{figure*}
    \centering
    \includegraphics[width=1\textwidth]{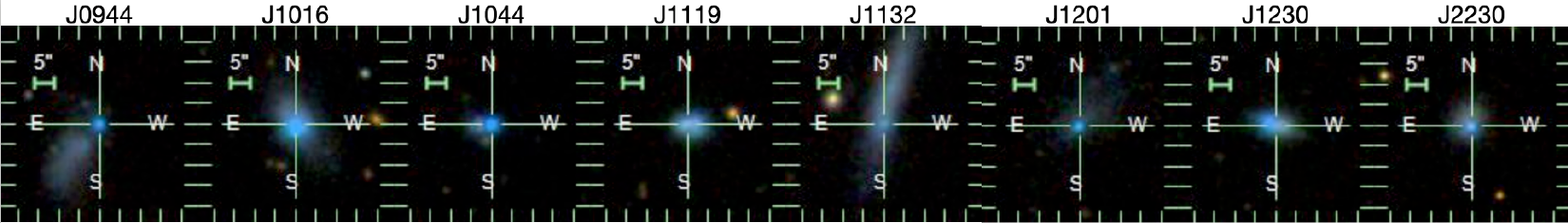}
    \caption{50"x50" SDSS DR15 g, r, i composite images of the galaxies in our sample centred on the COS coordinates. The COS PSA is 2.5" in diameter.}
    \label{fig:sdss_postage_stamps}
\end{figure*}

\begin{figure*}
    \centering
    \includegraphics[width=0.15\linewidth]{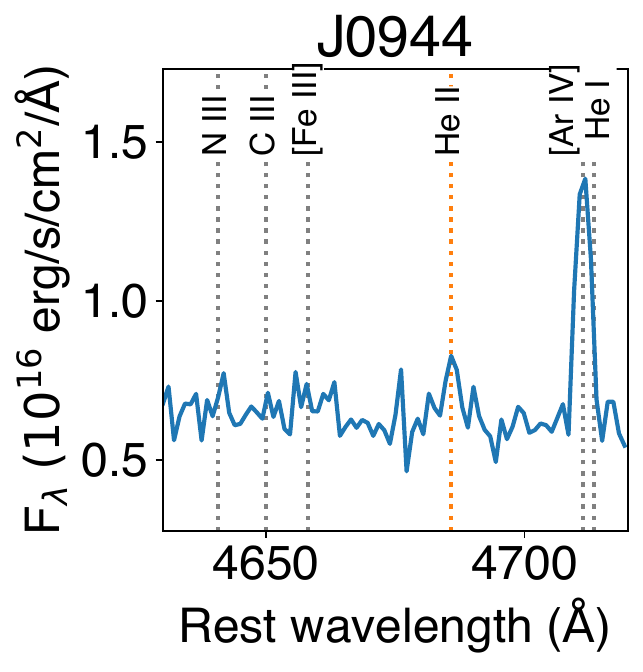}%
    \includegraphics[width=0.15\linewidth]{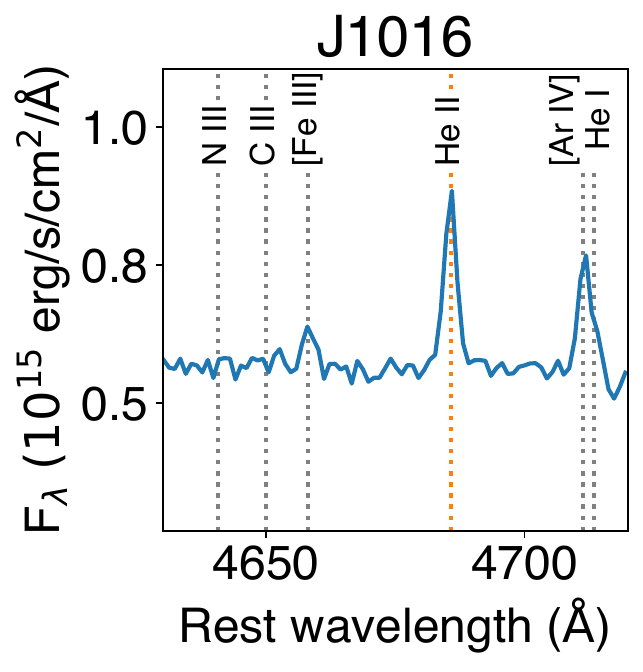}%
    \includegraphics[width=0.15\linewidth]{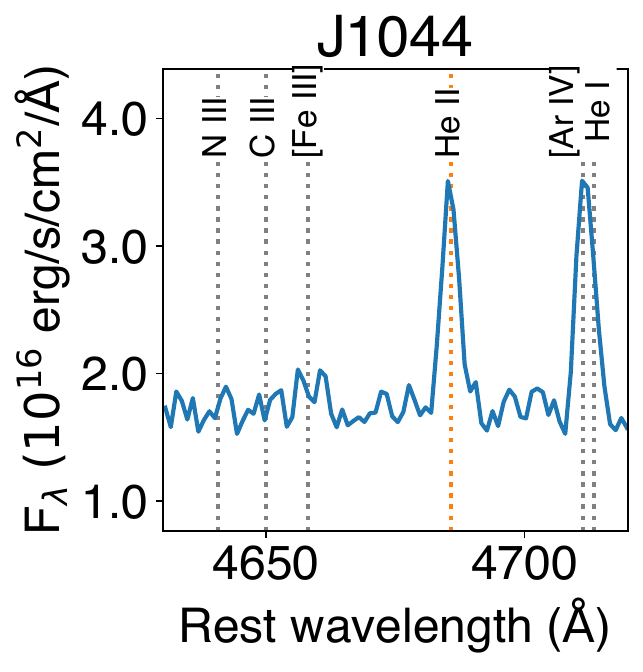}%
    \includegraphics[width=0.15\linewidth]{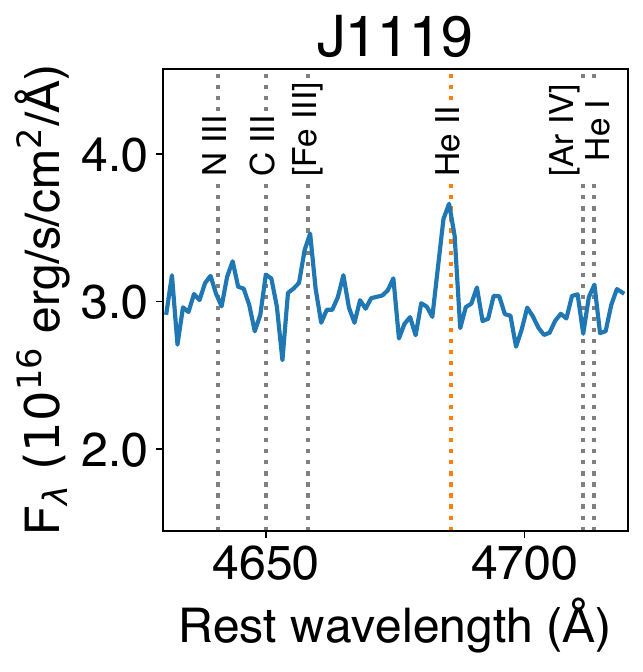}\\
    \includegraphics[width=0.15\linewidth]{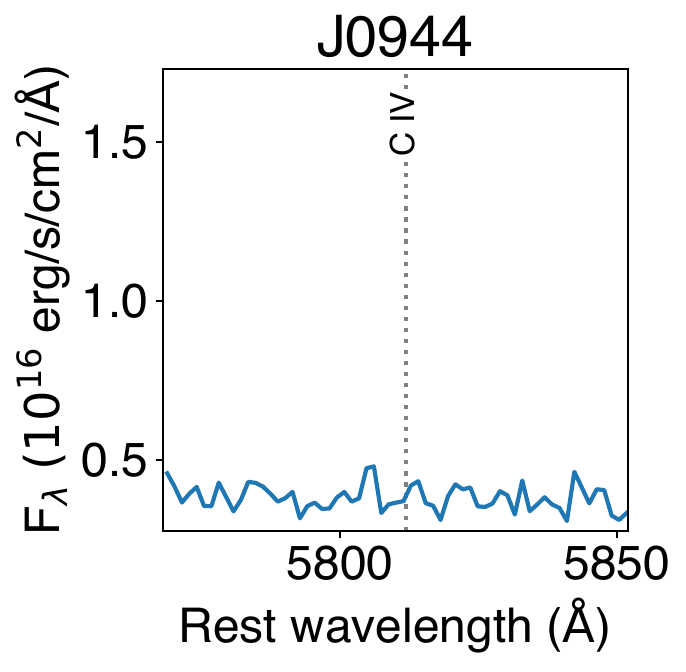}%
    \includegraphics[width=0.15\linewidth]{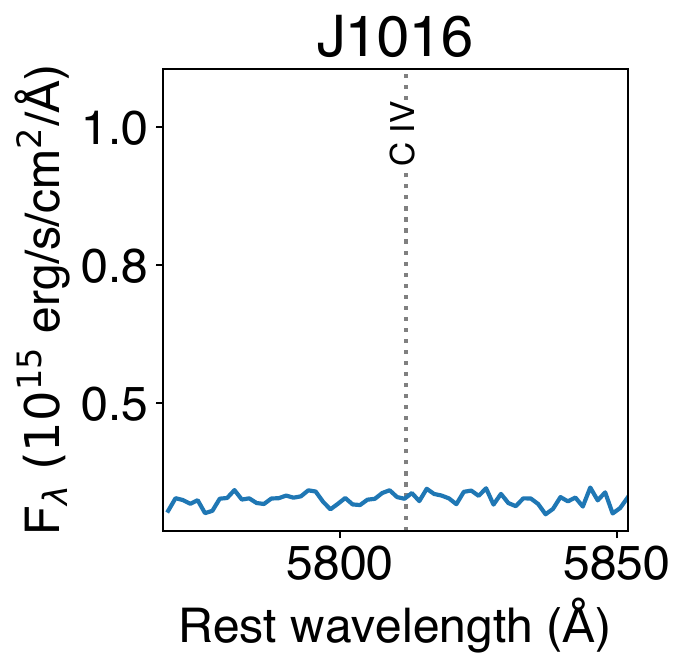}%
    \includegraphics[width=0.15\linewidth]{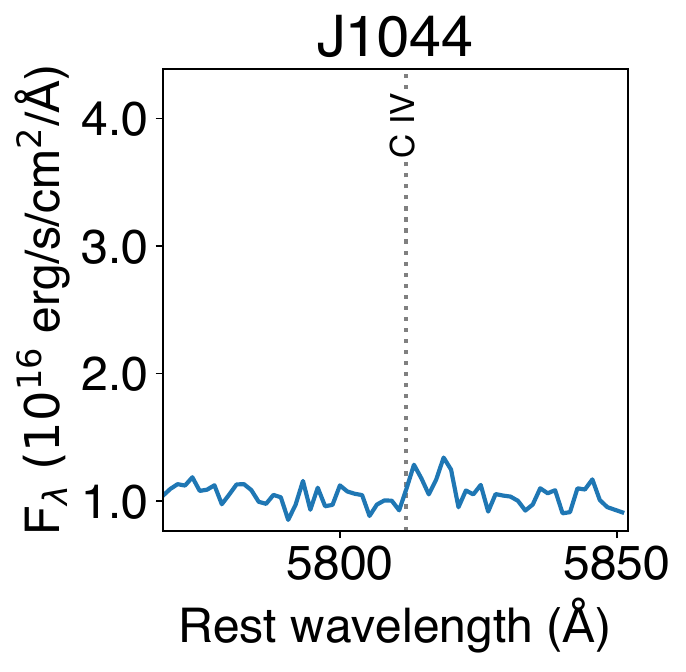}%
    \includegraphics[width=0.15\linewidth]{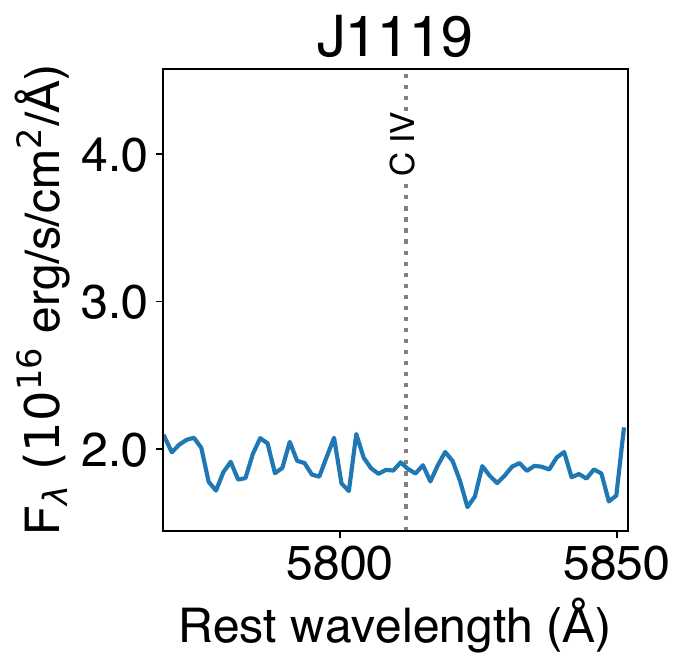}\\
    \includegraphics[width=0.15\linewidth]{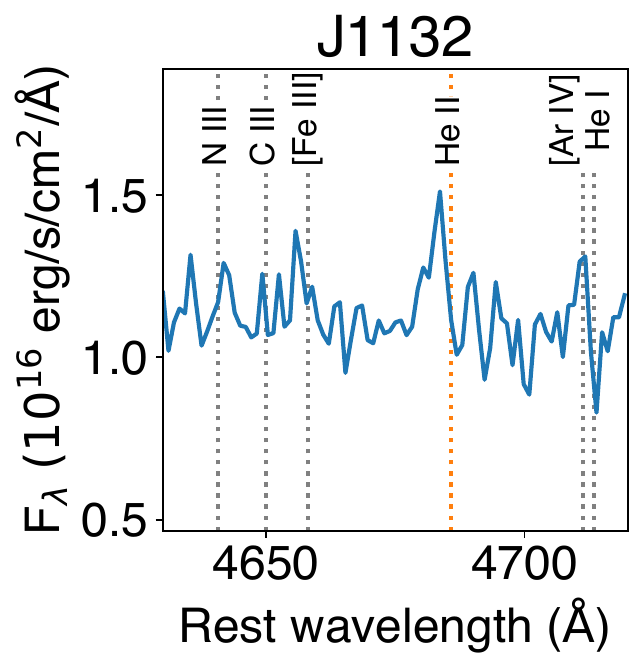}%
    \includegraphics[width=0.15\linewidth]{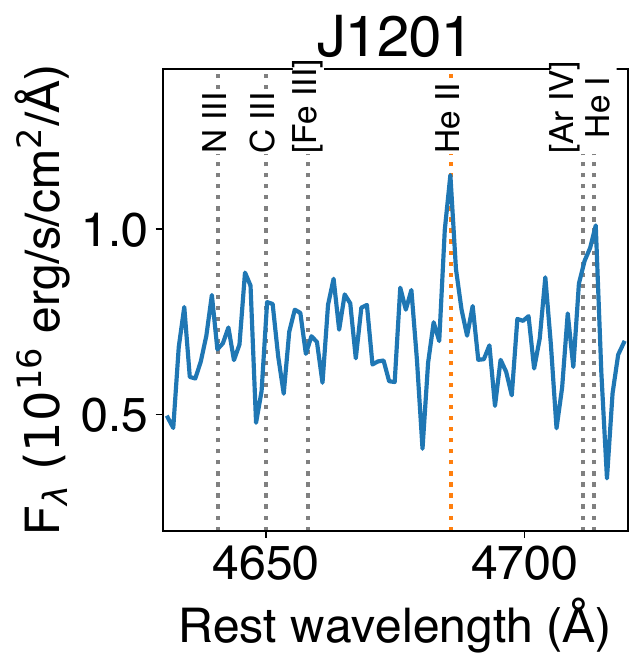}%
    \includegraphics[width=0.15\linewidth]{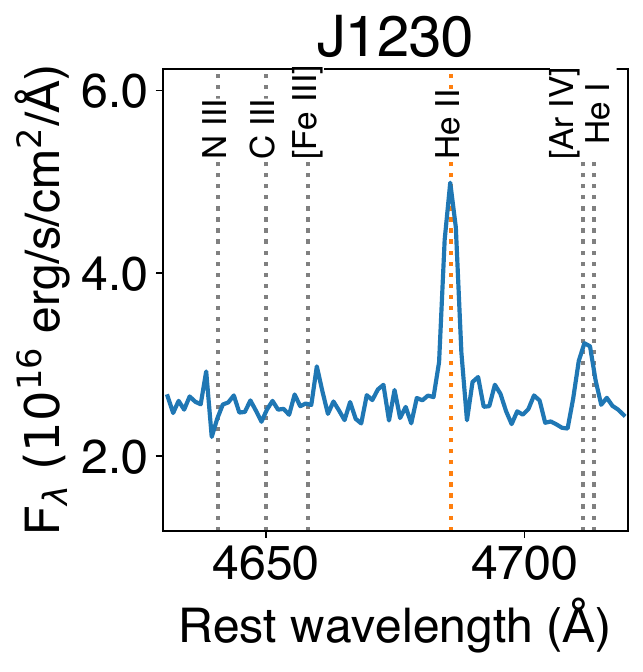}%
    \includegraphics[width=0.15\linewidth]{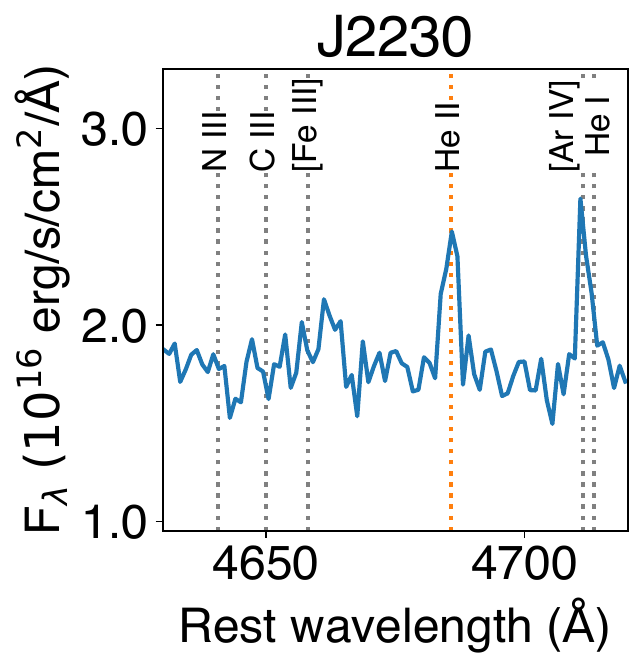}\\
    \includegraphics[width=0.15\linewidth]{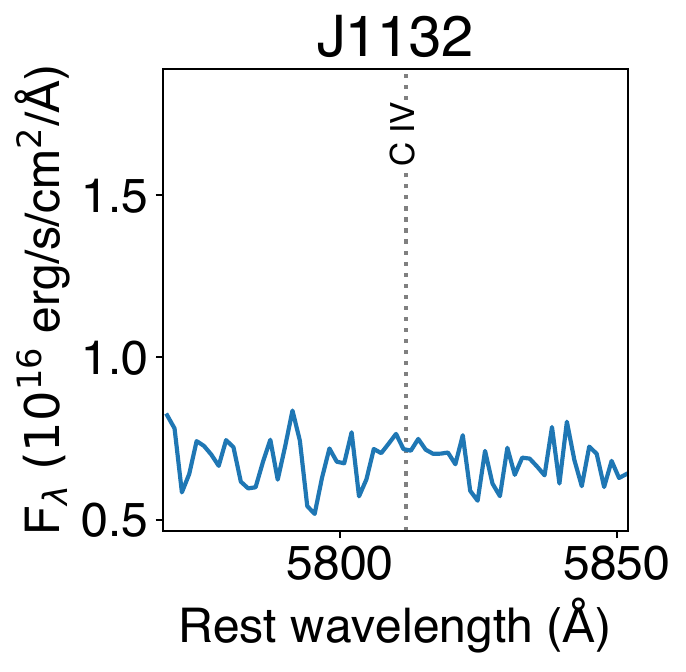}%
    \includegraphics[width=0.15\linewidth]{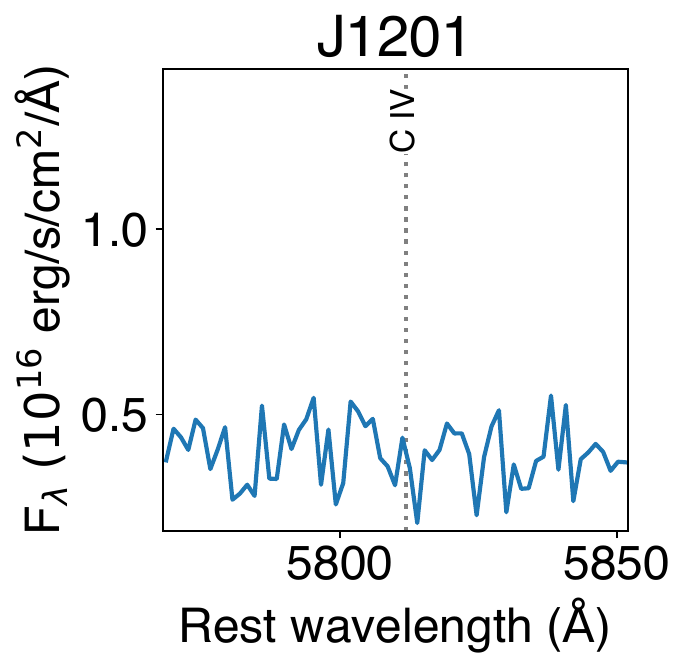}%
    \includegraphics[width=0.15\linewidth]{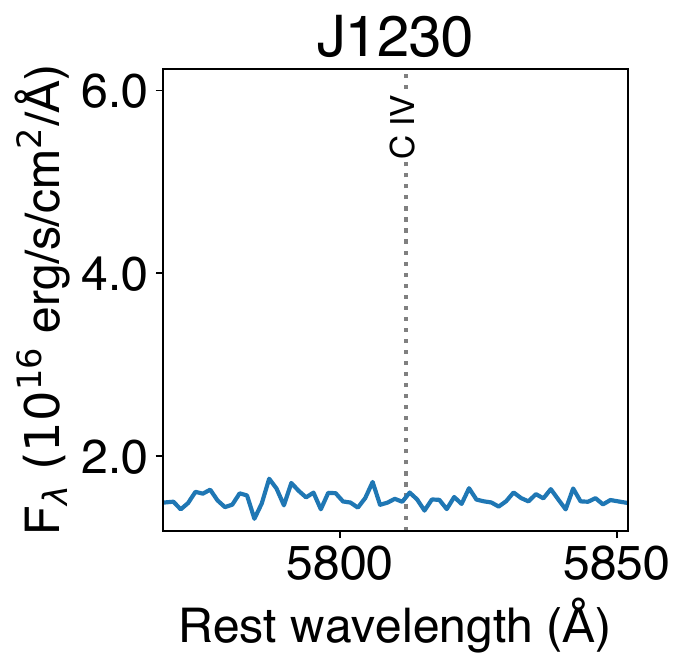}%
    \includegraphics[width=0.15\linewidth]{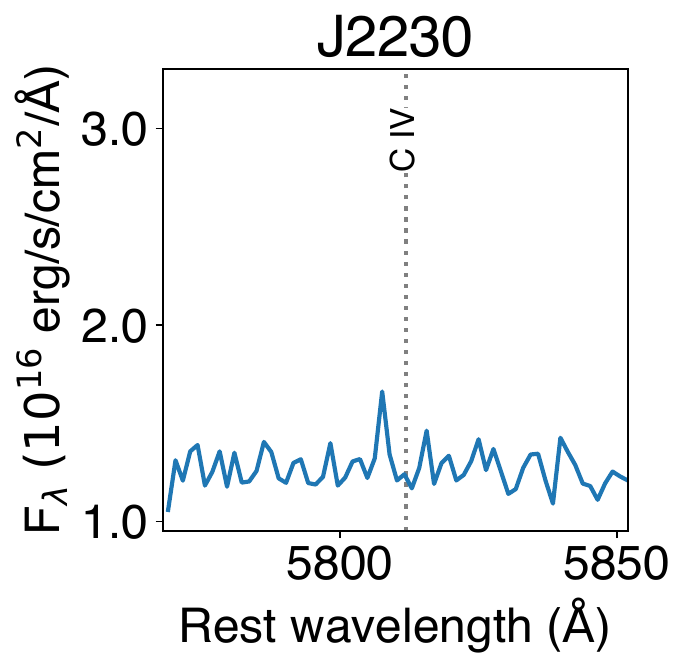}
    \caption{For the galaxies in our sample, SDSS observations of the ``blue-bump'' region (N\3 4640+C\3 4650) and emission lines: [Fe\3] 4659, He\2 4686, [Ar IV] 4711, and He\1 4713 (first and third rows); and the ``red-bump'' region (C\4 5808-12; second and fourth rows).}
    \label{fig:sdss_4686}
\end{figure*}

Known properties of the galaxies are summarized in Tab.~\ref{tab:known_galaxy_properties}. Figure~\ref{fig:sdss_postage_stamps} shows composite SDSS g, r, i postage stamps of the targets. The galaxies look blue and in two cases have tidal tails. As shown in Fig.~\ref{fig:sdss_4686}, the galaxies show narrow He\2~4686 emission ($FWHM\approx165-235$ km/s) in observations of the 4th edition of the Sloan Digital Sky Survey (SDSS IV / DR14, \citealt{Abolfathi2018ApJS..235...42A}) or for galaxy J1132, the 7th edition (SDSS I / II DR7; \citealt{Abazajian2009ApJS..182..543A}), obtained with the BOSS (Baryon Oscillation Spectroscopic Survey) spectrograph. Figure~\ref{fig:sdss_4686} also shows that the galaxies do not have ``blue-bumps'' stronger than the He\2 4686 emission or broad ``red bumps'', although J1044 shows some narrow C\4~5012 emission. According to \cite{Martins2023A&A...678A.159M}, these optical properties are favourable for finding candidate VMS, with the caveat that the galaxies in the latter work are not XMPGs. 

In addition to using optical SDSS spectra for our sample selection, we used it to determine the dominant sources of ionization in the galaxies and to search for evidence of candidate VMS.  

\begin{table*}
    \centering
    \begin{tabular}{llllllll}
    \hline
        SDSS ID & RA$^{\rm(a)}$ & Dec$^{\rm(a)}$  & z$^{\rm(a)}$ & D$^{\rm(b)}$ &  $^{\rm(c)}$log & $^{\rm(c)}$log &$^{\rm(d)}$12+\\
          \hfill & hh:mm:ss & dd:mm:ss & \hfill & Mpc  & (M$_*$ [M$_\odot$]) & (SFR [M$_\odot$/yr]) & log(O/H)\\
        \hline
J094416.59+541134.3	&	09:44:16.6104	&	+54:11:34.260	&	0.0055	&	$29.77\pm2.10$	&		 N/A 	&	-1.76$^{+0.03}_{-0.03}$	&	$7.43\pm 0.01^{(1)}$\\
J101624.51+375445.9	&	10:16:24.516	&	+37:54:45.972	&	0.0039	&	$23.05\pm1.65$	&	6.72$^{+0.27}_{-0.22}$	&	-1.17$^{+0.18}_{-0.18}$	&	$7.56\pm0.01$$^{(2)}$ \\
J104457.79+035313.1	&	10:44:57.7968	&	+03:53:13.128	&	0.0129	&	$63.61\pm4.48$	&		6.80$^{+0.41}_{-0.26}$	&	-0.59$^{+0.11}_{-0.14}$	&	$7.45\pm0.02$$^{(3)}$\\
J1119+5130 & 11:19:34.356 & +51:30:12.096  & 0.0045 & $26.78\pm1.90$ & 6.77$^{+0.15}_{-0.28}$ & -1.58$^{+0.21}_{-0.12}$ & $7.57\pm0.04$$^{(2)}$ \\
J113202.11+572251.7 & 11:32:02.64 & +57:22:36.39 & 0.0053 & $29.97\pm2.12$ & 7.31$^{+0.23*}_{-0.26}$ & -1.07$^{-0.27}_{+0.31}$ & $7.58\pm0.08$$^{(2)}$\\
J120122.32+021108.5	&	12:01:22.308	&	+02:11:08.304	&	0.0033	&	$9.02\pm0.68$	&		6.09	&	-1.97	&	$7.45\pm0.03^{(3)}$\\
J123048.61+120242.8	&	12:30:48.5208	&	+12:02:42.072	&	0.0042	&	$14.98\pm1.07$	&	 $5.6^{+0.20}_{-0.20}$	&	$-2.20^{+0.10}_{-0.10}$	&	$7.81\pm0.08^{(4)}$\\
J2230-0006	&	22:30:36.7992	&	-00:06:36.972	&	0.0053	&	$25.34\pm1.80$	&	 N/A 	&	 N/A 	&	7.62$^{(5)}$\\
\hline
    \end{tabular}
    \caption{Galaxy properties. $^{\rm(a)}$SDSS J2000 right ascension (RA) and declination (Dec), and redshift (z). $^{\rm(b)}$Hubble flow corrected (Virgo + Great Attractor + Shapley) distances (D) from NASA/IPAC Extragalactic Database. $^{\rm(c)}$Mass in stars (M$_*$) and star formation rate (SFR) from \citet[][FUV, J0944]{Prestwich2013}; \citet[][J1201]{Berg2016}; \citet[][SDSS, J1230]{Senchyna2017}; and \citet[][SDSS, rest of galaxies]{Brinchmann2004MNRAS.351.1151B}. N/A=not available. $^{\rm(d)}$Refs. for direct-method oxygen abundances: $^{(1)}$\citet{Izotov1999ApJ...511..639I}, $^{(2)}$\citet{Berg2022ApJS..261...31B}; $^{(3)}$\citet{Berg2016}; $^{(4)}$\citet{Senchyna2017}; $^{(5)}$\citet[][tab. 4, ID: A228-00]{Kinman1981ApJ...243..127K}.}
    \label{tab:known_galaxy_properties}
\end{table*}

\subsection{New UV observations}

\begin{figure*}
    \centering
    \includegraphics[width=0.22\linewidth]{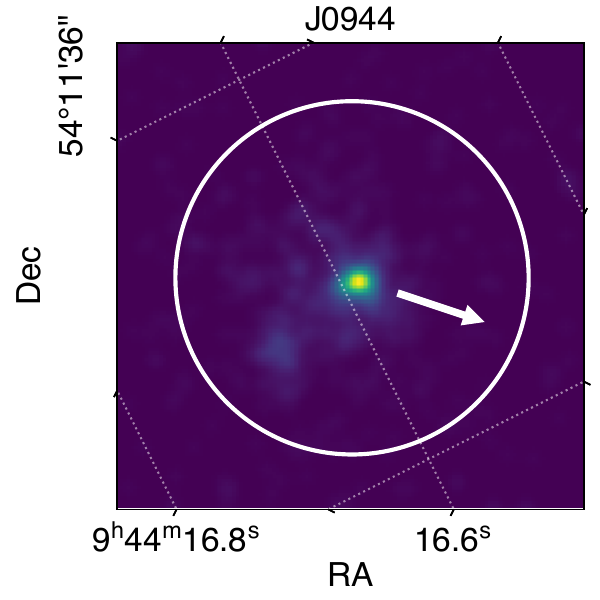}%
    \includegraphics[width=0.22\linewidth]{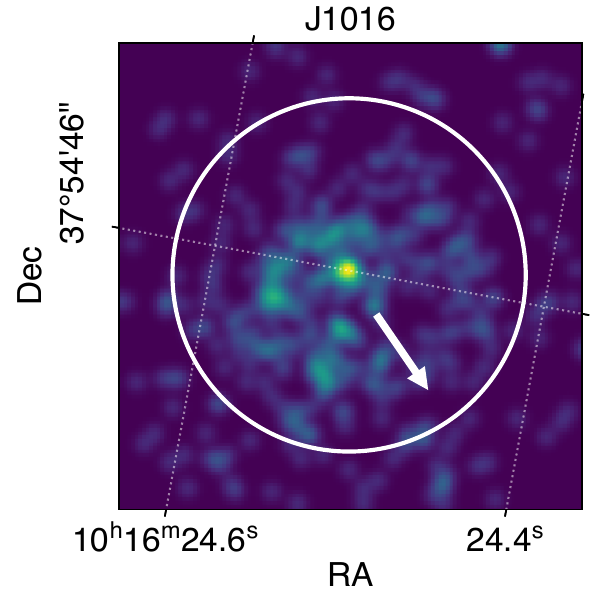}%
    \includegraphics[width=0.22\linewidth]{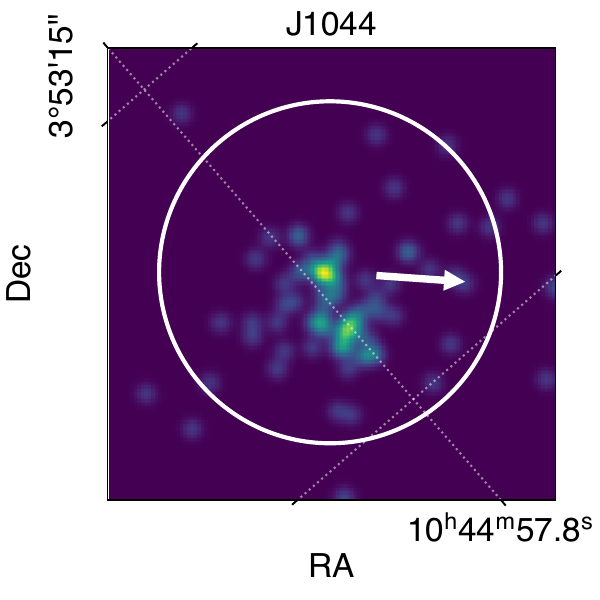}%
    \includegraphics[width=0.22\linewidth]{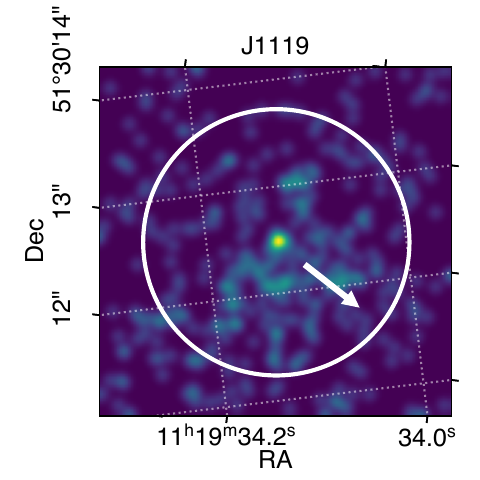}\\
    \includegraphics[width=0.22\linewidth]{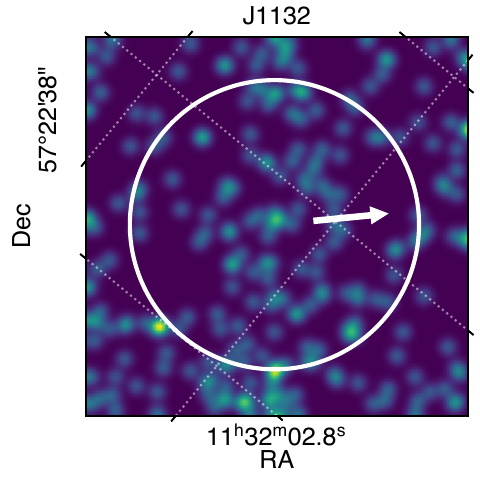}%
    \includegraphics[width=0.22\linewidth]{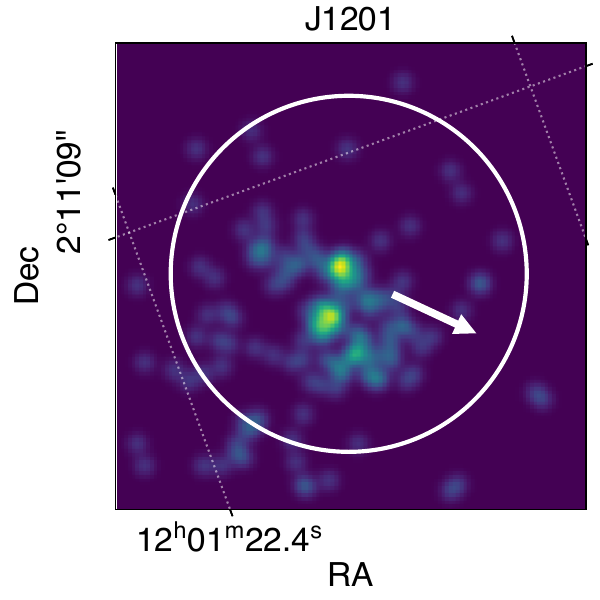}%
    \includegraphics[width=0.22\linewidth]{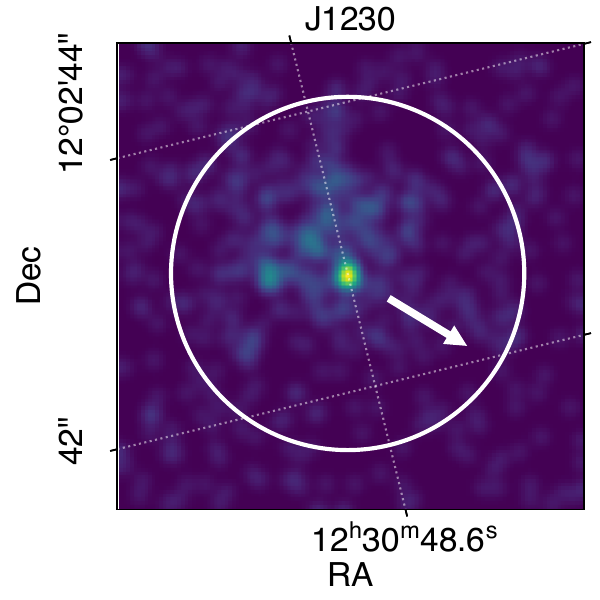}%
    \includegraphics[width=0.22\linewidth]{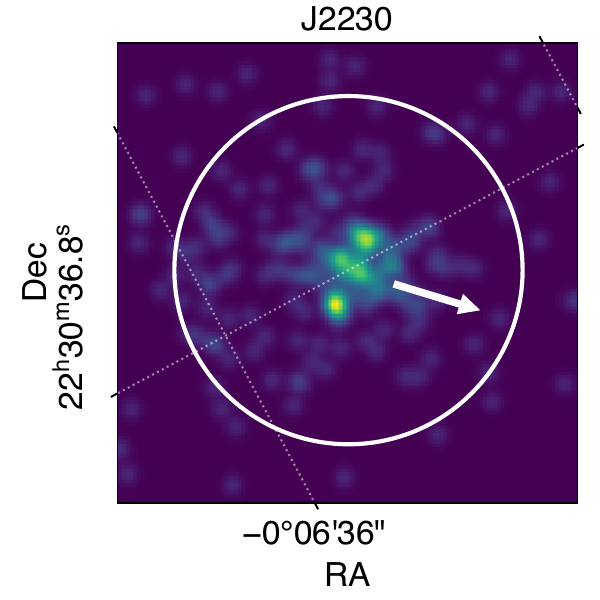}%
    \caption{COS NUV TA images of the galaxies. The horizontal and vertical axes give the J2000 RA and Dec, respectively, except for J2230, where the software inverted the axes due to the negative RA. The images are shown in linear scale, Gaussian-smoothed with a $\sigma$ of 2 pixels, and with enhanced contrast. Roughly speaking, North is up and East to the left except for J2230, where East is up and North is to the left. The white circle represents the PSA footprint. The white arrow shows the dispersion direction. No ghost TA images with a 0.5" offset are observed in the dispersion direction. The observed structure is real.}
    \label{fig:ta_images}    
\end{figure*}

We used COS G140L ($\lambda/FWHM\approx2000-5000$) at central wavelength 1105 \AA~and life time position 4, and the circular Primary Science Aperture (PSA), which has a diameter of 2.5", to obtain UV spectra with continuous observed-frame coverage from $1121$ to $2148$~\AA. The observing sequence was as follows. 1) Perform target-acquisition (TA) procedures in order to ensure that the target is well centered in the PSA. This avoids throughput losses, produces reliable wavelength zero points, and outputs an image of the target over an NUV band of wavelength range, $1700-3200$ \AA. The plate scale of the NUV detector is 0.075 arcsec.  Depending on the target's brightness, the TA image was obtained with MIRRORB (brightest targets) or MIRRORA. MIRRORB has an attenuation factor of $\sim25$ relative to MIRRORA.  2) Obtain spectra in TIME-TAG mode using the FP-POS=ALL setting and FLASH=YES so that the spectra can be corrected for flat-field anomalies and OSM drifts. 3) Obtain additional spectra during subsequent orbits in order to achieve the desired signal-to-noise ratio (SNR). We display the TA images in Figure~\ref{fig:ta_images}, which each show a compact source in the middle of the aperture except for J1132, which appears diffuse in nature, with no bright clumps.

\subsection{Reduction of UV spectra}

We retrieved the individual spectral data sets from the Mikulski Archive for Space Telescopes and initially processed them on-the-fly with version 3.3.4 (2018-01-08) of the CalCOS pipeline. The observation log is presented in Table~\ref{tab:obs_log}. CalCOS processes the data for detector noise, thermal drifts, geometric distortions, orbital Doppler shifts, count-rate nonlinearity, and pixel-to-pixel variations in sensitivity. A standard wavelength scale is applied using the on board wavelength calibration. With central wavelength 1105 \AA, the G140L data fall on one detector: segment A. Since multiple exposures are contained within a visit, these are combined into a single summed spectrum. The final products are one-dimensional, flux calibrated, heliocentric-velocity corrected, background subtracted, and combined spectra which are stored in x1dsum files. We worked directly with the x1dsum files. 

\begin{table}
	\centering
	\begin{tabular}{llllr} 
		\hline
		Short & Dataset & Mirror & Exp. & CENWAVE\\
		ID & \hfill & Grating  & Time (s) & PID G160M \\        
		\hline
        J0944 & LDKM04ZBQ   & MIRRA &  9.0 & N/A\\
        \hfill 			 & LDKM04010  & G140L &  2404 & N/A\\
		J1016 & LDKM07JXQ  & MIRRB &  8.0 & 1589\\         
        \hfill 			 & LDKM07010  & G140L &  2267 & 14120\\
		J1044 & LDKM01F5Q  & MIRRA &  0.6 & 1533\\
		\hfill 			 & LDKM01010  & G140L &  2221 & 15646\\    
		J1119 & LDKM05UGQ  & MIRRB  & 13.0 & 1623 \\        
        \hfill 			 & LDKM05010  & G140L  & 2364 & 14679\\        
        J1132 & LDKM03WPQ  &  MIRRB  & 15 & 1623 \\
        \hfill 		     & LDKM03010  & G140L  & 2415 & 14679\\
        J1201 & LDKM06HXQ & MIRRA  & 2.5 & 1533 \\        
        \hfill 			 & LDKM06010  & G140L  & 5057 & 16677 \\
        J1230 & LDKM02UMQ  & MIRRB  & 20 & 1623 \\
        \hfill		     & LDKM02010  & G140L  & 2191 & 14168\\
		J2230 & LDKM08BHQ & MIRRA  & 2.6 & N/A\\ 
        \hfill 			 & LDKM08010  & G140L  & 2224 & N/A\\ 
		\hline
	\end{tabular}  
 \caption{Observing log of COS program 15220 (PI Wofford). The last column gives the CENWAVE and PID of the ancillary G160M  data used in this work.}
 \label{tab:obs_log}
\end{table}

For a point-source well-centred on the PSA, the instrumental $FWHM$ value of the G140L grating is 0.48 \AA~in the FUV and 0.24 \AA~in the NUV, respectively, as derived by multiplying the dispersion of the grating (0.08 \AA~pix$^{-1}$) by the size of the spectral resolution element of the associated detector (6 pixels for FUV XDL and 3 pixels for NUV MAMA). Since CalCOS outputs oversampled data, we resample the data to the nominal dispersion of the grating. Wavelength zero point shifts can occur due to imperfect centering and the flux profiles of the galaxies in the PSA. For the data taken with G140L, we looked for wavelength offsets by determining the centroids of Galactic UV absorption lines towards our targets. Offsets of $\le-50$ km s$^{-1}$ are likely due to intermediate or high-velocity clouds (IVC or HVC, respectively; e.g., \citealt{Wofford2013ApJ...765..118W}). We did not correct the spectra of any offsets.

\subsection{Correction for Lyman-alpha absorption}\label{sec:lya_corr}

The N\5~1239,\,43 doublet of low-redshift galaxies suffers from strong contamination with the H\1~1216 \lya~absorption in the Galactic plane (e.g., \citealt{Wofford2013ApJ...765..118W}). In addition, the \lya~profile of the galaxy itself (intrinsic \lya, which can be an absorption and/or emission feature) may also contaminate the N\5\,1240 doublet. As shown in Tab.~\ref{tab:known_galaxy_properties}, our eight galaxies have high galactic latitudes. Thus we expect weaker \lya~from the MW. They also have low redshifts, so we expect the intrinsic \lya~to contaminate the N\5\,1240 doublet. This is a problem because as shown in \S~\ref{sec:stars}, the 
N\5\,1240 doublet can be a key diagnostic of a stellar population's age. In particular, we detected N\5 1240 P-Cyngi like profiles in galaxies J1016, J1230, and J2230.

In order to correct N\5\,1240 for the Galactic + intrinsic \lya~ absorption, we follow the procedure that is outlined in \S~2.4 of \citet{Wofford2023MNRAS.523.3949W}. For each galaxy, we fix the velocity and H\1~column density of the MW, log N(H\1)$_{\rm MW}$, in the direction of the target to the values that are provided in columns 3 and 4 of Tab.~\ref{tab:lya_properties}. The N(H\1)$_{\rm MW}$ values are based on the H\1 map by the \cite{HI4PI2016A&A...594A.116H}.  The intrinsic log N(H\1) values and their errors are provided in columns 5 and 6 of the table. For J1044, the only way to fit the H\1~ absorption was by assuming a covering fraction $<1$. For the rest of the galaxies, based on the morphology of the absorption (the fluxes were ~0 on the trough), we assumed a covering fraction of 1, and obtained reasonable fits. For galaxy J1016 we were unable to constrain the intrinsic H\1~column density.

\begin{figure*}
    \centering
    \includegraphics[width=0.23\linewidth]{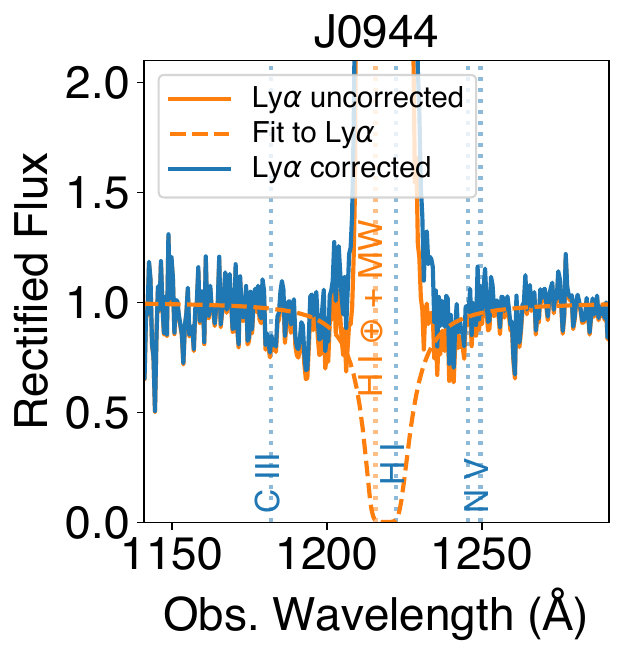}%
    \includegraphics[width=0.23\linewidth]{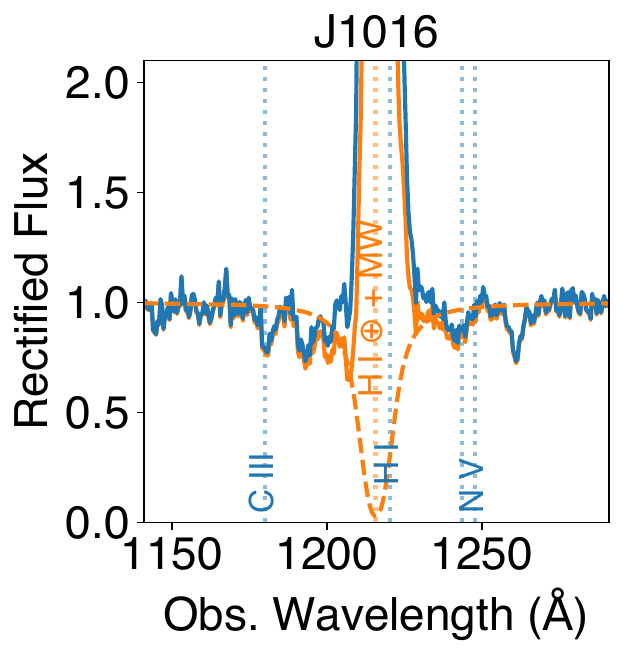}%
    \includegraphics[width=0.23\linewidth]{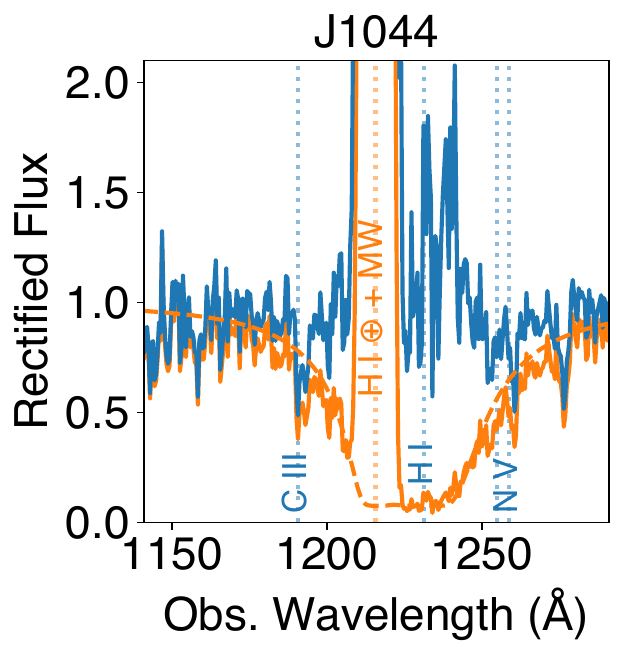}%
    \includegraphics[width=0.23\linewidth]{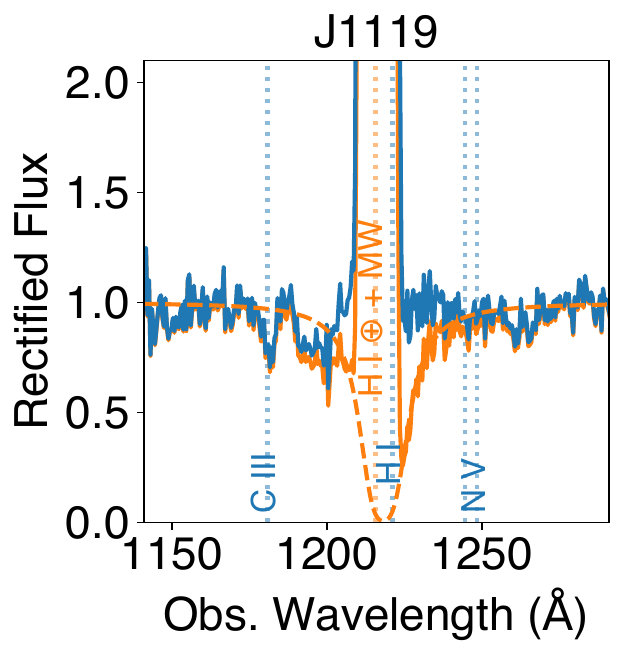}\\
    \includegraphics[width=0.23\linewidth]{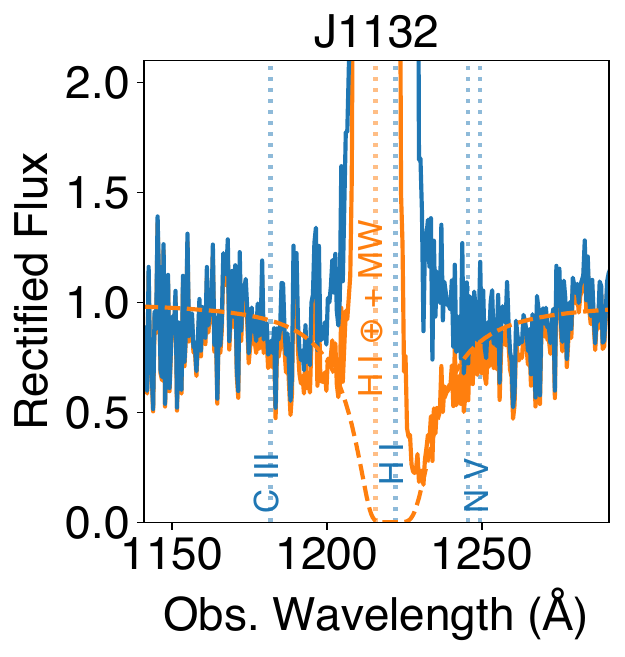}%
    \includegraphics[width=0.23\linewidth]{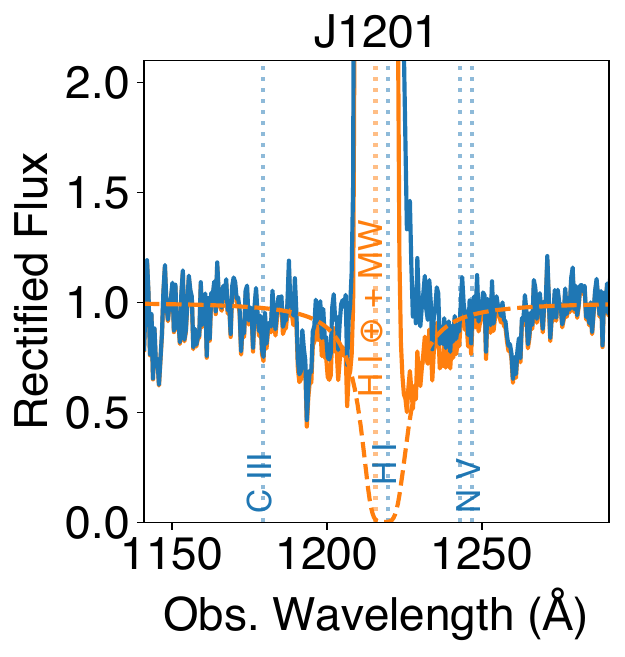}%
    \includegraphics[width=0.23\linewidth]{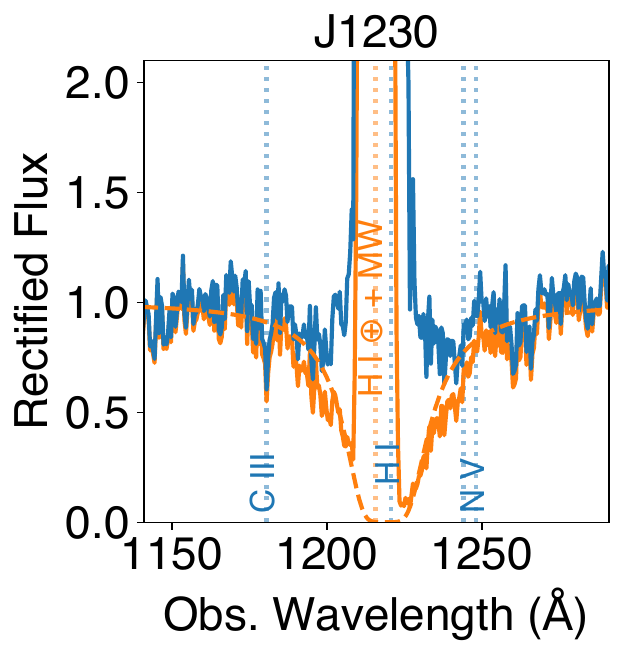}%
    \includegraphics[width=0.23\linewidth]{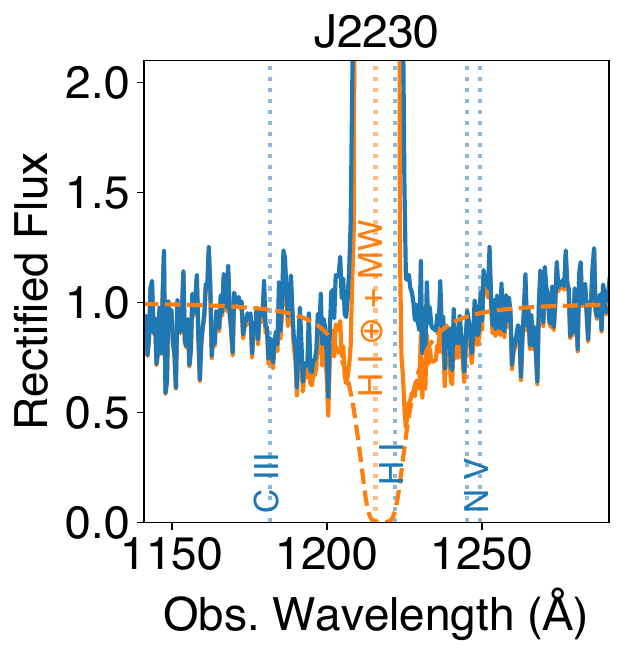}
    \caption{COS G140L spectra around the H\1~\lya~blend, which includes foreground (geocoronal + MW absorption) + intrinsic (absorption or emission) components. We show the spectra before and after  the \lya~correction (solid orange and blue curves, respectively). The dashed-orange curves are the fits to the \lya~absorptions. For galaxies whose flux does not reach zero on the trough, we adopt a covering factor of <1. We mark the positions of the intrinsic C\3 1175, \lya, and N\5~1239, 1243 lines; and of the foreground geocoronal + MW \lya~lines.}
    \label{fig:after_lya_corr}
\end{figure*}

For each galaxy, Fig.~\ref{fig:after_lya_corr} shows the observed spectra around the H\1\,$\lambda1216$ \lya~blend (geocoronal emission + MW absorption + intrinsic line), and Voigt-profile fits to the MW + intrinsic \lya~absorption blend. Owing to the positions of the targets far from the MW's disk, the log N(H\1) values from the MW absorptions have low log N(H\1) values between 20.0 and 20.7. Only J1044 shows a large intrinsic log N(H\1) value of 22.0 $\pm$ 0.0240. The latter galaxy has Ly$\alpha$ emission reported in \citet{Hu2023ApJ...956...39H}, which we confirm in this work (see Fig.~\ref{fig:after_lya_corr}), although with a different morphology due to the lower spectral resolution of our observations. 

\begin{table}
    \centering
    \begin{tabular}{lccccc}
    \hline
Short	& Gal. Lat. &vel	&	N$_{\rm H\,I,\,MW}$ &	Log  &	Log 	\\
ID & deg & km/s & cm$^{-2}$ & N$_{\rm H\,I,\,MW}$ & N$_{\rm H\,I,\,gxy}$   \\
\hline
J0944 & 46.57741$^*$  &		1794	&	1.29E+20	&	20.1	&	20.9	$\pm$	0.0667	 \\
J1016 & 55.86592 &		1196	&	1.40E+20	&	20.1	&	17.2	$\pm$	14.4209	 \\
J1044  & 51.90994 &		3887$^*$	&	3.78E+20	&	20.6	&	22.0 $\pm$ 0.0240$^*$	 \\
J1119 & 60.10620&		1495	&	1.14E+20	&	20.1	&	21.0	$\pm$ 0.0346	 \\
J1132 & 56.64759 & 1495	&	9.83E+19$^*$ &	20.0$^*$	&	21.5	$\pm$	0.0325	 \\
J1230 & 74.15947$^*$ &		897	&	1.77E+20	&	20.2	&	21.4	$\pm$	0.0350	 \\
J2230  & -46.72487 &		1495	&	4.63E+20$^*$ &	20.7$^*$	&	20.8	$\pm$	0.0837$^*$	 \\
\hline
    \end{tabular}
    \caption{Properties of the MW and intrinsic absorption \lya~lines. $\rm{^{(a)}}$Covering factor: one if the flux is zero at the trough and N/A if unable to constrain the column density of the intrinsic \lya~profile. For each column, the minimum and maximum absolute values are marked with asterisks.}
    \label{tab:lya_properties}
\end{table}

\subsection{Effective spectral resolution}

In this work we compare spectral synthesis models to direct stellar lines from three galaxies in our sample (J1016, J1230, and J2230). For the comparison, the models have to match the effective spectral resolution of the observations. Here we derive its value for above three galaxies.

\begin{figure}
    \centering
    \includegraphics[width=1\linewidth]{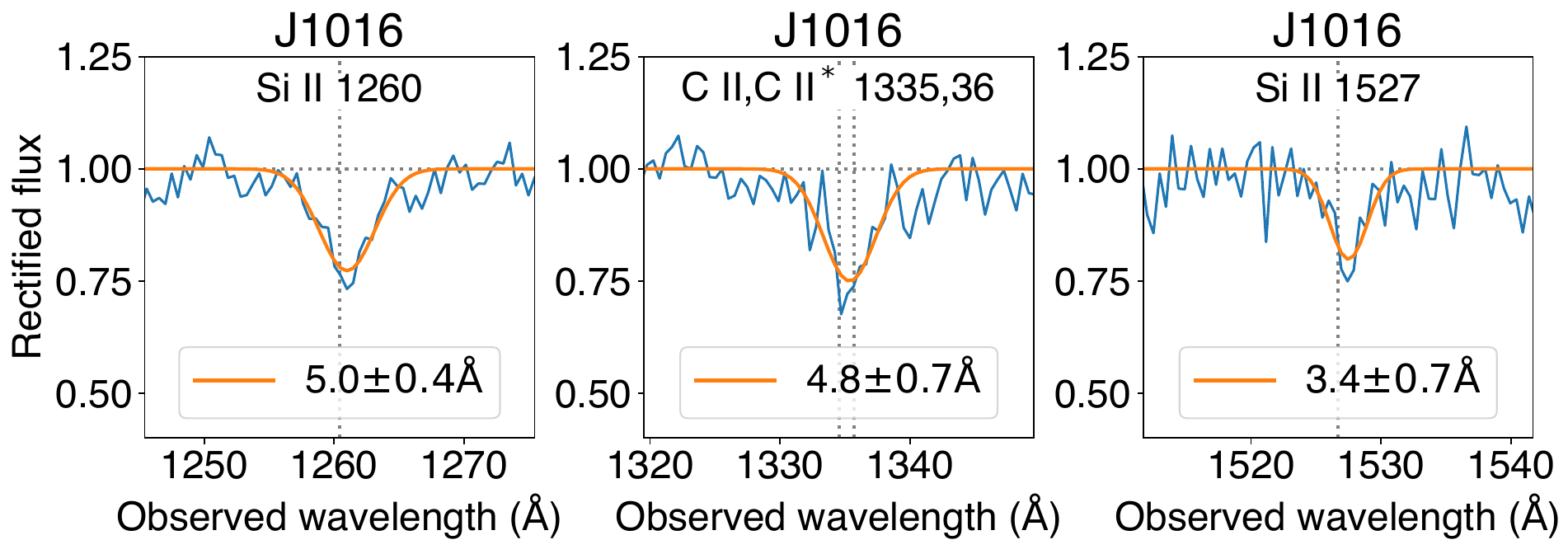}
    \includegraphics[width=1\linewidth]{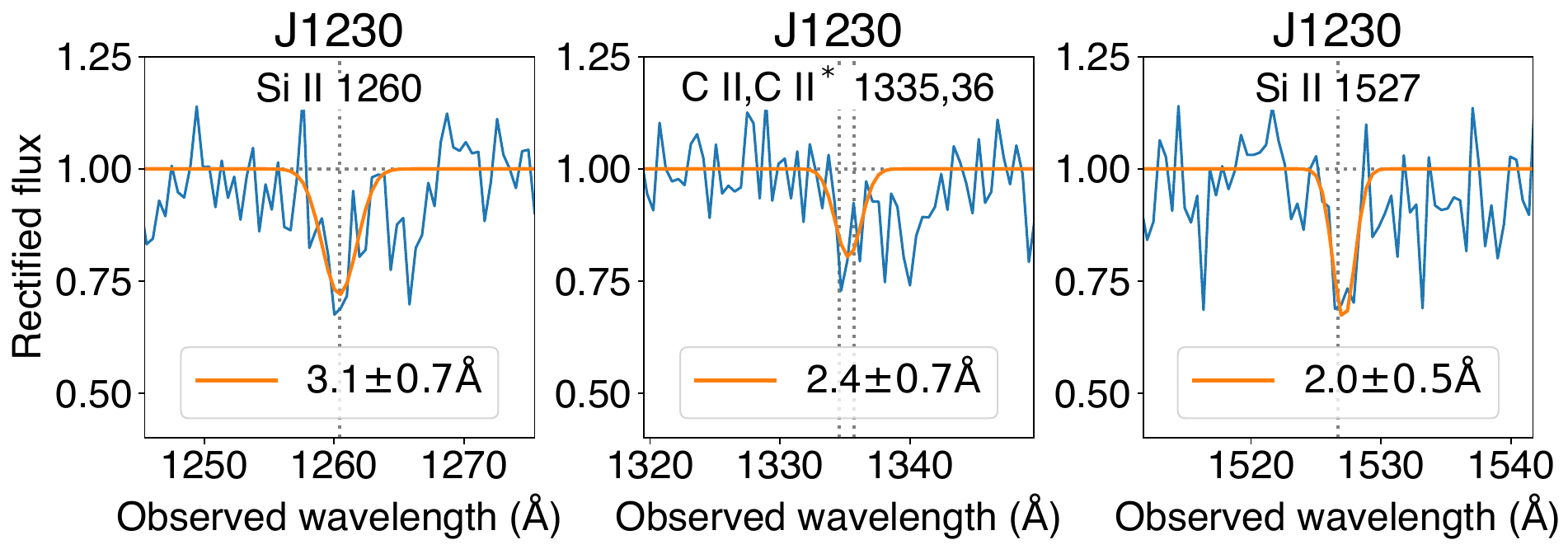}
    \includegraphics[width=1\linewidth]{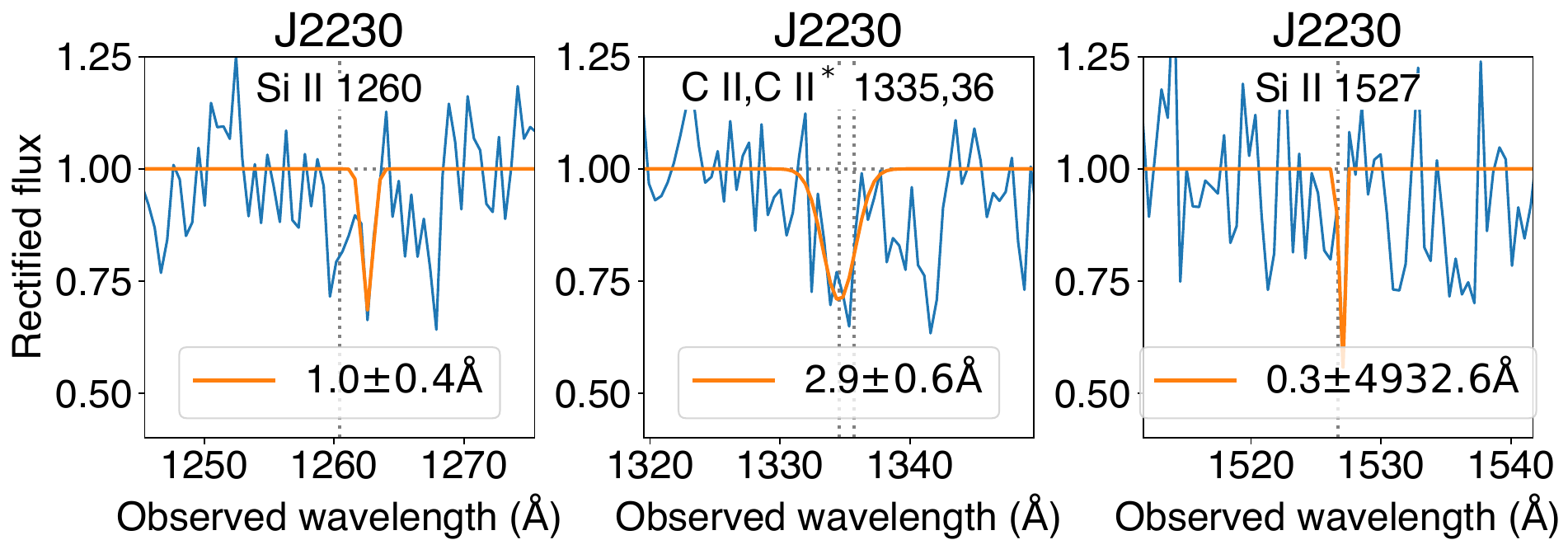}
    \caption{For the galaxies dominated by P-Cygni like C\4 1550 profiles, fits to MW absorption lines used to determine the effective spectral resolution. We use vertical lines to indicate the rest-frame positions of the lines. The legend gives the FWHM values.}
    \label{fig:mw_fwhm}
\end{figure}

The effective spectral resolution of COS is determined by a combination of instrumental and observational effects. For point sources, the resolution is limited by the line spread function, which is set by the instrument and its configuration (grating choice and CENWAVE). FUV detector effects such as non-uniform pixel response and fixed-pattern noise do not change resolution directly but they affect measured line widths at low SNR. Extended sources and targets that are mis-centred in the aperture broadens lines. For the galaxies with P-Cygni stellar wind profiles (J1016, J1230, and J2230), we adopt the procedure described in \citealt{Wofford2013ApJ...765..118W} to estimate the effective spectral resolution of the COS observations. For this purpose, we fit inverse Gaussians to three strong MW absorption lines: Si\2~1260, Si\2~1527, and C\2~1335; and computing the mean of their FWHM values. Note that C\2~1335 is blended with the weaker nearby C\2$^*$~1336 absorption. C\2~and Si\2~require energies of 11.26 and 8.15 eV to form, originate in neutral gas, and are expected to have similar values of the FWHM. Figure~\ref{fig:mw_fwhm} shows the fits to the above MW absorption lines. We only include the line in the average if its FWHM value is in the range$\gtrsim2-5$~\AA~and its velocity shift is less than $\pm$200 km/s, as other values are due  to poor fits to the data. Thus, we exclude  the two Si\2~lines of J2230. For galaxies J1016, J1230, and J2230, the average FWHM values are: 4.4, 2.6, and 3 \AA, or equivalently, $FHWM = 805$, 476, and 549 km/s, respectively. 

\section{COS G140L spectra and He II 1640 emission}\label{sec:sp_properties}

\begin{figure*}
    \centering
    \includegraphics[width=1\textwidth]{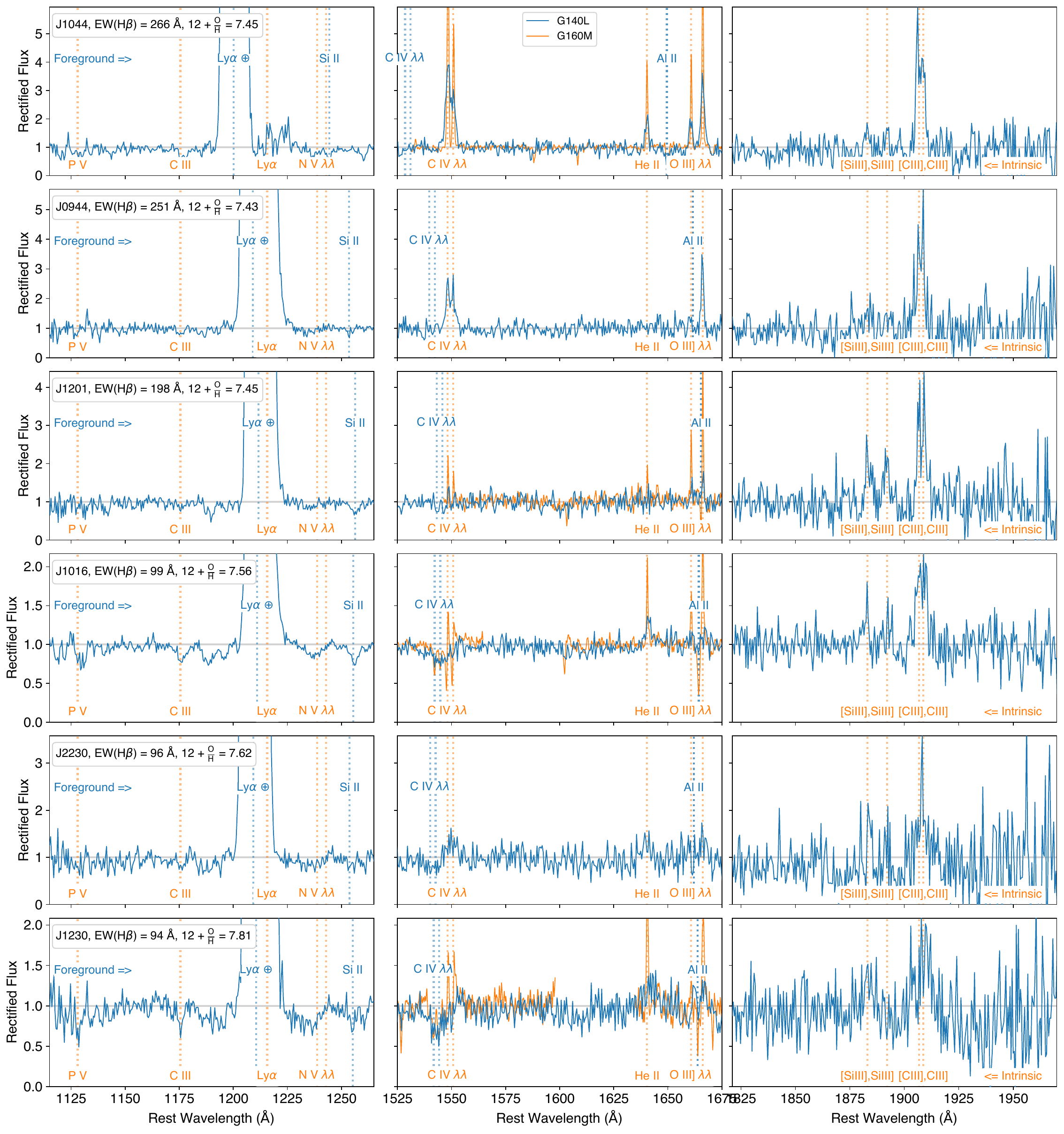}
    \caption{COS G140L observations of galaxies (blue curves) in order of decreasing EW(H$\beta$) from top to bottom, as given by the legend. For galaxies with ancillary G160M data, we show these data in orange. We mark the positions of selected foreground and intrinsic lines.}
    \label{fig:g140l-sp}
\end{figure*}

In Fig.~\ref{fig:g140l-sp}, we show portions of the COS G140L spectra of six of the galaxies in our sample (blue curves). The spectra of the galaxies are arranged in order of decreasing value of the equivalent width, EW(H$\beta$), from top to bottom, as given by the legends. The spectra of the other two galaxies (J1132 and J1119) in our sample pose some challenges and are discussed in Appendix~\ref{app:bad_galaxies}. Ancillary higher resolution G160M data available for four of our galaxies are also shown in Figs.~\ref{fig:g140l-sp} and~\ref{fig:bad_uv_sp} for comparison (orange curves, see program IDs of the G160M data in Tab.~\ref{tab:obs_log}).

Figure~\ref{fig:g140l-sp} shows that there are two groups of galaxies. The top three constitute the first group and includes the most metal-poor galaxies in our sample, which also have the highest values of EW(H$\beta$). These galaxies show weak C\4~1548,\,1551 (C\4~1550) absorptions and no P-Cygni like N\5~1239,\,1243 (N\5~1240) profiles. At low metallicity and ages of up to 10 Myr, BPASS spectral synthesis models with and without binaries predict similarly large values of EW(H$\beta$) (see fig. 38 of \citealt{Eldridge2017PASA...34...58E}). The bottom three galaxies constitute the second group and includes galaxies with P-Cygni like N\5 and C\4 profiles, which we attribute to stellar winds. 

Regarding the first group of galaxies, J1044 has a strong narrow He\2 1640 emission line in both the G140L and G160M spectra. However, galaxies J0940 and J1201 do not have He\2 1640 emission, even though they have weak He\2~4686 emission in their SDSS spectra (see Fig.~\ref{fig:sdss_4686}).

\begin{figure*}
    \centering
    \includegraphics[width=0.3\linewidth]{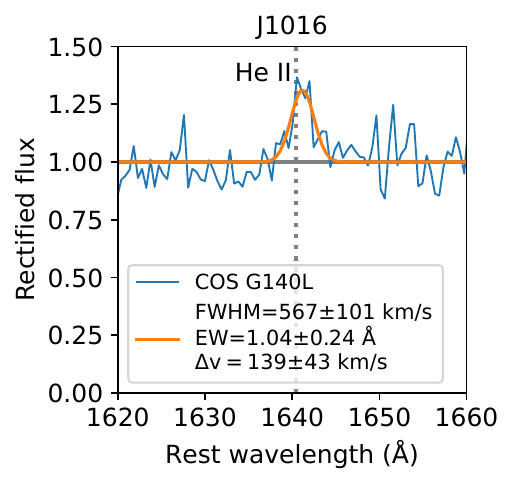}
    \includegraphics[width=0.3\linewidth]{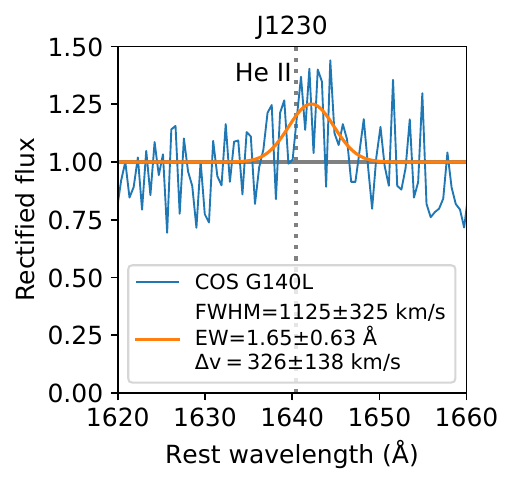}
    \includegraphics[width=0.3\linewidth]{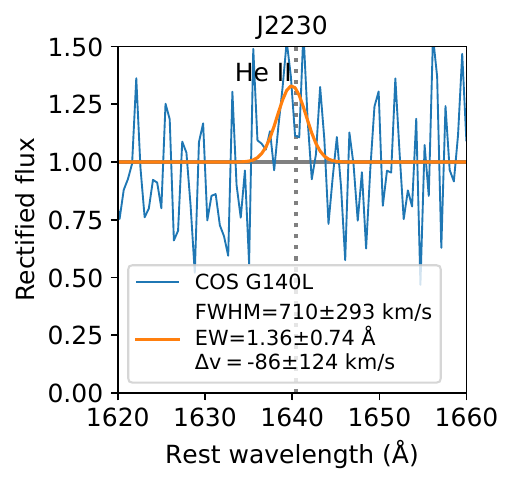}
    \caption{Properties of He\2~1640 emission lines for galaxies with P-Cygni like profiles.}
    \label{fig:broadHe2}
\end{figure*}

Regarding the second group of galaxies, their G140L He\2 1640 profiles of look broad even though they have narrow optical He\2 emission (see Fig.~\ref{fig:sdss_4686}). For J1016 and J1230, which have G160M data, they look broader relative to their G160M counterparts. We note that the aperture position angles of the G140L and G160M observations are somewhat different (-10.9 vs. 10.5 for J1016 and 15.4 vs -13.8 deg for J1230). Thus, it is possible that the G140L spectra may be picking up broad He\2~ emission while the G160M is not. We fitted Gaussians + a linear continuum to the He\2 1640 lines of the galaxies in this group. The fits are shown in Fig.~\ref{fig:broadHe2}. Taking into consideration the effective spectral resolution and SNR of the data, the He\2 1640 lines are consistent with being narrow within the uncertainties. In \S~\ref{sec:stars}, we use the stellar UV lines of these three galaxies to derive properties of their OB-star populations. 

Figure~\ref{fig:g140l-sp} also shows that the six galaxies have weak stellar P\5~1128; and C\3~1175 absorptions and detections of O\3~1661, 1666; [Si\3], Si\3] 1883, 1892; and/or [C\3],C\3] 1907, 1909 high-ionization nebular emission lines. Finally, the Fig. shows that J1044 has intrinsic \lya~emission, in agreement with the finding  of \cite{Hu2023ApJ...956...39H}. 

\section{Spectral Synthesis Models}\label{sec:models}

Spectral synthesis models combine the output from population synthesis models with spectral libraries, and predict the spectral evolution of stellar populations. We use spectral synthesis models to derive the ages of the OB-star populations in our three galaxies with P-Cygni profiles.  Several past works have compared predictions from this type of models in the non-ionizing UV, to verify their consistency. However, most of these works have focused on the super-XMP regime or do not compare the models used in this work. For single stellar populations (SSPs\footnote{In an SSP stars are coeval and born instantaneously}) of solar metallicity ($Z_\odot$), \cite{Eldridge2017PASA...34...58E} compare the UV outcomes of their Binary Population and Spectral Synthesis (BPASS v2.1) code, with GALAXEV BC03 \citep{Bruzual2003MNRAS.344.1000B}, STARBURST99 (S99) v00 \citep{Leitherer1999ApJS..123....3L, Leitherer2014ApJS..212...14L}, and \citealt{Maraston2005MNRAS.362..799M} models. They conclude that BPASS binary models do not predict a radically different non-ionizing UV spectrum from the other codes but that rather, the diagnostic power of BPASS lies chiefly in their enhanced, longer-lived ionizing UV emission and the nebular consequences of that harder radiation field. In addition, \cite{Eldridge2017PASA...34...58E} find that differences that remain in the non-ionizing UV are primarily due to details of stellar-atmosphere libraries, stellar-wind and photospheric features, metallicity, IMF choice, and the assumed age/star-formation history. However, these authors do not explore how these differences affect derived stellar population properties such as age, reddening and mass in stars. In a follow-up paper, \cite{Stanway2018MNRAS.479...75S} compare BC03 predictions with their updated BPASS v2.2, at 0.4$Z_\odot$ and $Z_\odot$. They conclude that meaningful differences emerge at particular ages and in the infrared, UV, and ionizing continuum. \cite{Jones2023ApJ...959...20J} present a full spectrum fitting tool for Simultaneous Estimates of Star-cluster Age, Metallicity, Mass, and Extinction (SESAMME) and test the consistency of their tool, when used with S99 and BPASS models, applied to observed UV spectroscopy, against published physical properties of star clusters in two spiral galaxies. They find good agreement in the properties within the uncertainties. Finally, \cite{Berg2024ApJ...971...87B} compare the physical properties obtained with S99 and BPASS when analysing COS UV spectra of star-forming regions in spiral galaxy M101. They find that neither S99 nor BPASS have WR prescriptions that can produce sufficiently broad He\2 emission in these regions, and that some differences in the properties inferred from the different codes exist. 

Since in the XMP regime, the consistency of STARBURST99, GALAXEV and BPASS models has not been extensively studied, we star by comparing  SSP spectra predicted by these codes. We do not test more complicated star formation histories because in the XMP regime models are poorly constrained. The models used in this work are described in Appendix~\ref{app:model_details}. Table~\ref{tab:model_summary} provides a summary of the input parameters to the models. 

\begin{table*}
    \centering

    \begin{tabular}{llllccc}
        \toprule
        Code & Model ID & Stellar physics & IMF & $M_\mathrm{up}$ ($M_\odot$) & $Z$ & Age range (Myr), time step\\
        \midrule
        \multirow{2}{*}{STARBURST99}
            & v00-l & Single, Non-rotating ($v/v_\mathrm{crit} = 0$) & Kroupa & 100 & 0.001, 0.002 & 1-9, $\Delta t$ (Myr)=0.1 \\
            & v40-l & Single, Rotating ($v/v_\mathrm{crit} = 0.4$)   & Kroupa & 100 & 0.001, 0.002 & 1-9, $\Delta t$ (Myr)=0.1 \\
        \addlinespace
        \multirow{2}{*}{GALAXEV}
            & CB19-100 & Single, Non-rotating & Chabrier & 100 & 0.001, 0.004 & 1-9, $\Delta t$ (Myr)=0.1 \\
            & CB19-300 & Single, Non-rotating & Chabrier & 300 & 0.001, 0.004 & 1-9, $\Delta t$ (Myr)=0.1 \\
        \addlinespace
        \multirow{4}{*}{BPASS v2.2.1}
            & sin-100 & Single, Non-rotating      & Chabrier & 100 & 0.001, 0.004 & 1-9, log(t/yr)=0.1 \\
            & sin-300 & Single, Non-rotating      & Chabrier & 300 & 0.001, 0.004 & 1-9, log(t/yr)=0.1 \\
            & bin-100 & Mass-exchanging binary & Chabrier & 100 & 0.001, 0.004 & 1-9, log(t/yr)=0.1 \\
            & bin-300 & Mass-exchanging binary & Chabrier & 300 & 0.001, 0.004 & 1-9, log(t/yr)=0.1 \\
        \bottomrule
    \end{tabular}
    \caption{Parameters of the spectral synthesis models that are described in Appendix~\ref{app:model_details}, compared against each-other in Fig.~\ref{fig:model_comparison}, and compared against UV observations of galaxies with P-Cygni profiles in Figs.~\ref{fig:j1016}-\ref{fig:j2230}.}
    \label{tab:model_summary}
\end{table*}

\subsection{Evolution of selected synthetic line profiles}\label{subsec:syn_lines}

Weak P-Cygni like profiles of the N\5 1240 and C\4 1550 resonance doublets originating in stellar winds are observed in UV spectra of individual O-type stars of the SMC (see fig. 1 of \citealt{Lennon2010ASSP...15...19L}), which has a metallicity of $Z\approx0.2\times Z_\odot$ \citep{Roman-Duval2019ApJ...871..151R}. In synthetic spectra of young ($\lesssim10$ Myr) stellar populations of MW, LMC ($Z\approx0.5\times Z_\odot$; \citealt{Roman-Duval2019ApJ...871..151R}) or SMC metallicity, the N\5 and C\4 doublets decrease in strength as age increases and the O stars disappear from the population; and decrease as metallicity decreases. On the other hand, at these metallicities, broad He\2 emission is present when either VMS or cWR stars are present  \citep{Crowther2006ASPC..353..157C, Crowther2007ARA&A..45..177C, Crowther2016MNRAS.458..624C, Martins2025A&A...698A.262M}. Due to distance, no representative samples of individual O-star spectra exist at sub-SMC metallicity. However, according to some XMP continuous star-formation  models, a weak N\5~1240 P-Cygni like profile is present at 50 Myr. In this regime, the predicted emission component of the N\5 doublet is stronger than that of the C\4 profile (compare figs. 29 and 30 of \citealt{Leitherer2010ApJS..189..309L}). Therefore, in the XMP regime, N\5 is expected to be a more sensitive indicator of the presence of O stars and thus of young age. An absence of N\5\,1240 or C\4\,1550 P-Cygni like profiles could indicate that either the O stars have disappeared or were never there to begin with. On the other hand, since the strength of stellar winds depends on metallicity \citep{Crowther2006ASPC..353..157C}, in the XMP regime, broad stellar-wind He\2 emission, is expected to be very weak or absent, specially if single-star evolution is considered, although cWR stars could form via the binary channel. On the other hand, \cite{Martins2025A&A...698A.262M} predict significant He\2 emission in a spectral synthesis model with VMS of $Z=0.1\times Z_\odot$, that they compute without scaling of the stellar mass loss rate with metallicity.

\begin{figure*}
    \centering
    \includegraphics[width=1\linewidth]{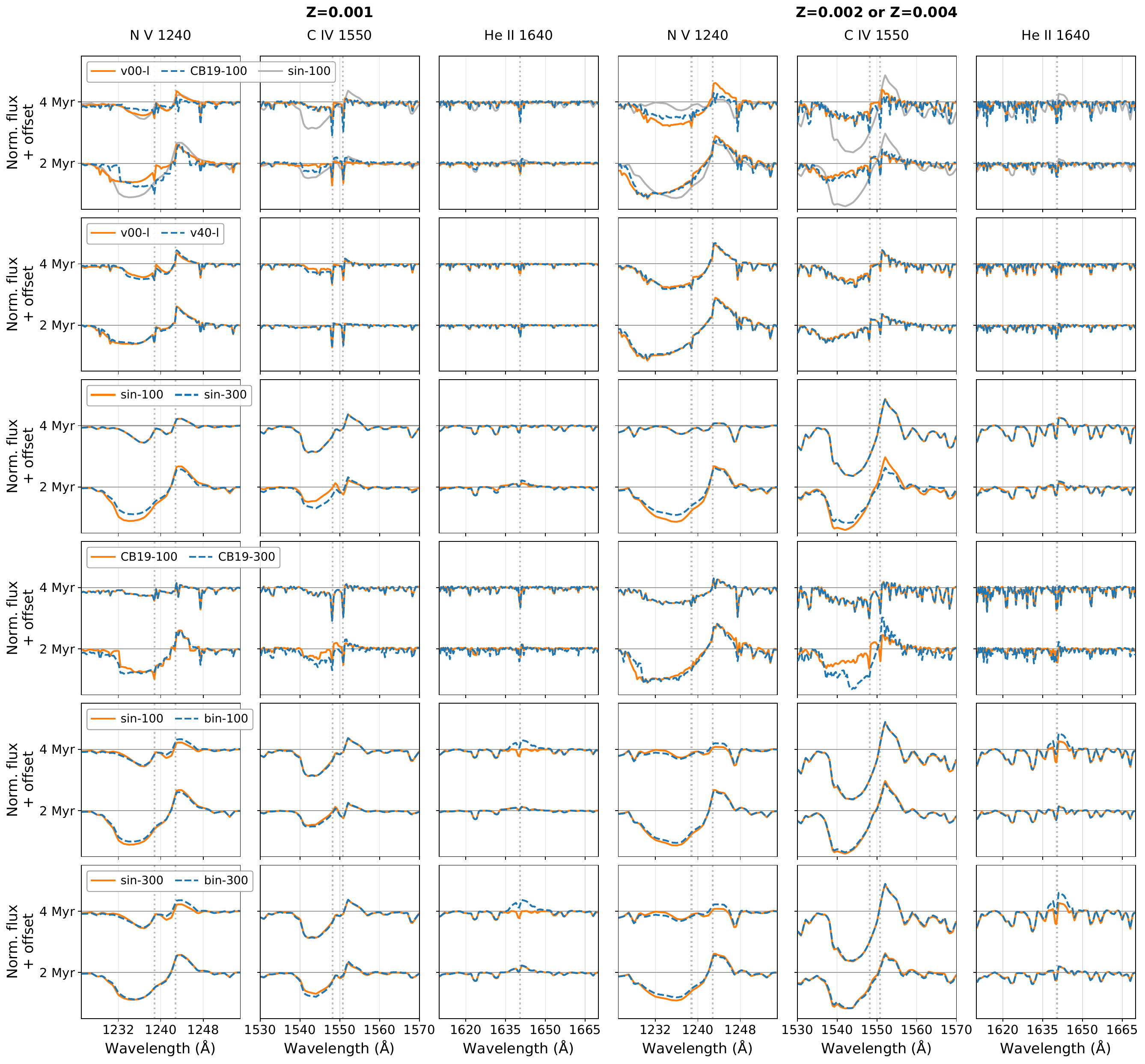}
    \caption{Comparison of N\5 1240, C\4 1550 and He\2 1640 line profiles from models given by the legends, with properties summarized in Tab.~\ref{tab:model_summary}. We do this at metallicities of $Z = 0.001$ and $Z = 0.002$ (v00-l and v40-l) or $Z=0.004$ (rest of models); and ages of 2 Myr (when VMS might be present) and 4~Myr (when they are absent). The spectra have been resampled to a common wavelength array and normalized. At 2 Myr and Z=0.001, the N\5 profile is stronger than at 4 Myr.}
    \label{fig:model_comparison}
\end{figure*}

In Fig.~\ref{fig:model_comparison}, we compare predictions from the models listed in Tab.~\ref{tab:model_summary} for the N\5 1240, C\4 1550, and He\2 1640 line profiles. We do this at ages of 2 and 4 Myr, and two metallicities\footnote{when the v00-l and v40-l have Z=0.002, the other models have Z=0.004. This is because models with Z=0.004 are not available for V00-l and v40-l.}. In models with VMS, these stars are still present at 2 Myr while they are not at 4 Myr. The first row from the top of Fig.~\ref{fig:model_comparison} shows that at the two metallicities there is a stronger C\4 profile in the sin-100 relative to the v00-l and CB19-100 models. This is attributed to differences in the chemical composition of the BPASS models (private communication). On the other hand, the v00-l and CB19-100 show more consistent predictions. The second row shows that  the N\5 and C\4 profiles look practically identical in the v00-l and v40-l models. This is also the case at 4 Myr for the sin-100 and sin-300 models (third row), and for the CB19-100 and CB19-300 models (fourth row), since at that age, the VMS are gone.

The fifth row shows that at the highest metallicity and 2 Myr, the C\4 absorption of the CB19-300 model is stronger relative to that of the CB19-100 model. Interestingly, the sixth row shows that this behaviour is reversed in the sin-100 and sin-300 BPASS models. The fifth and sixth rows also show that weak He\2 emission is present in the BPASS models with binaries (bin-100 and bin-300), which is strongest at 4 Myr.  

At 2 Myr and $Z=0.001-0.004$, the classical He\2~$\lambda$1640 emission line characteristic of VMS is absent from the CB19-300 spectrum, and it is very weak in the sin-300 and bin-300 predictions. A notable difference in the CB19-300 model is the strength of C\4~$\lambda$1550, but this line is also sensitive to age. Although the CB19-300 lines shown in Fig.~\ref{fig:model_comparison} are not useful for directly identifying candidate VMS via stellar features, it is important to check if the the ionizing fluxes predicted by this model can be used to search for VMS, indirectly through the high-ionization nebular emission lines they produce, as we do in \cite{Wofford2021MNRAS.500.2908W} and \S~\ref{sec:gas}. Using these fluxes as input to photoionization models for the above purpose only makes sense if the stellar populations are sufficiently young to host VMS, i.e., are $< 3$ Myr. 

\subsection{Evolution of beta slope}\label{subsec:models}

We now look at the evolution of the slope ($\beta$) of the massive-star Raleigh-Jeans portion of the SSP continuum predicted by the different models. This is relevant to the correction for reddening due to dust of the UV observations. The $\beta$ slope is obtained by fitting a power-law to this region, avoiding strong-line features. This $\beta$ slope is sensitive to reddening due to dust \citep{Calzetti1994ApJ...429..582C} and has been used by various authors to reddening-correct observations of local starburst galaxies obtained with different space telescopes, assuming the foreground dust screen model. This is done by comparing the slope of the observations to slopes of the dust-free models. The reddening correction is a necessary step prior to determining the mass in stars of a stellar population. Different authors have used different wavelength ranges to determine the value of $\beta$, depending on the coverage of their observations. \cite{Calzetti1994ApJ...429..582C} observed 39 starburst galaxies with the International Ultraviolet Explorer (IUE) and used the $1250-1950$ or $1250-2600$ \AA~ranges, depending on the configuration of the spectrograph. Note that rest-frame space UV observations of nearby galaxies are often contaminated with geocoronal H I Ly$\alpha$ emission and/or MW Ly$\alpha$ absorption, and/or intrinsic Ly$\alpha$ absorption or emission lines. These may affect the range between slightly bluer than Ly$\alpha$ and slightly redder than the N\5~1240 doublet. This is why some authors chose to fit the continuum starting at 1250 \AA. On the other hand, older stellar populations contribute to wavelengths in the 0.27-0.28 $\mu m$ range, in particular F stars. This sets the red limit used by \cite{Calzetti1994ApJ...429..582C} to 2600 \AA. \cite{Pellerin2007MNRAS.381..228P} observed 24 starburst galaxies with the Far Ultraviolet Spectroscopic Explorer (FUSE) and used the bluer $\approx1155-1165$ \AA~range (their tab.~2) to fit the continuum. As a third example, \cite{Chandar2004ApJ...604..153C} observed 18 starburst galaxies with \hst's Space Telescope Imaging Spectrograph and used the 1240-1600 \AA~range (their tab. 2). This was in spite of the observations providing continuous coverage from 1175 to 3000 \AA, perhaps because it is difficult to achieve a good power-law fit over the wide wavelength range from 1175 to 2600 \AA. After all, a sum of massive stars modelled as black bodies, is not a black body, so a polynomial fit may be better than a power law over such wide wavelength range.
According to fig.~31 of \citet{Leitherer1995ApJS...96....9L} and fig.~15, of \citealt{Hawcroft2025ApJS..280....5H}), when considering models instead of observations, the value of $\beta$ is somewhere in the range from -3 to -2, when measured from 1200-1900 \AA~and 1250-1750 \AA, respectively. In the above works, this holds for the first 10 Myr of a young stellar population with a Kroupa IMF having an upper mass limit of up to 120 M$_\odot$ This result considers single stellar evolution with and without rotation, MW-to-SMC metallicities, and the limiting star-forming cases of a single burst or continuous star formation.  

\begin{figure*}
    \centering
    \includegraphics[width=0.24\textwidth]{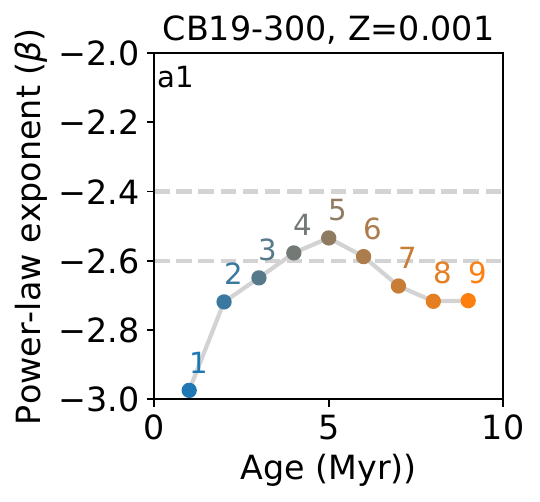}
    \includegraphics[width=0.24\textwidth]{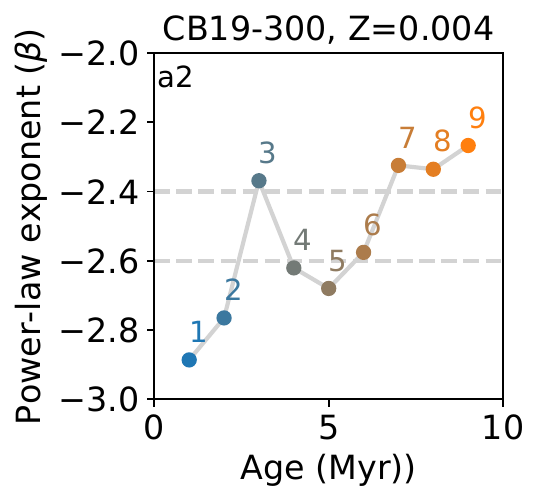}
    \includegraphics[width=0.24\textwidth]{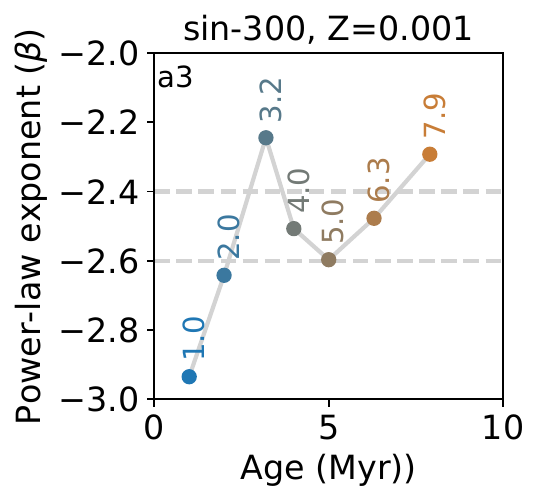}
    \includegraphics[width=0.24\textwidth]{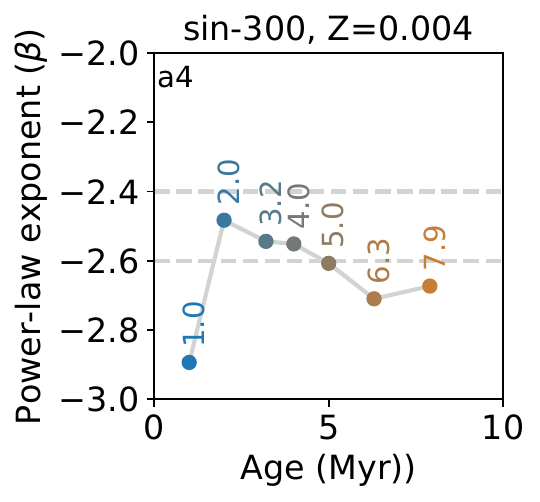}\\

    \includegraphics[width=0.24\textwidth]{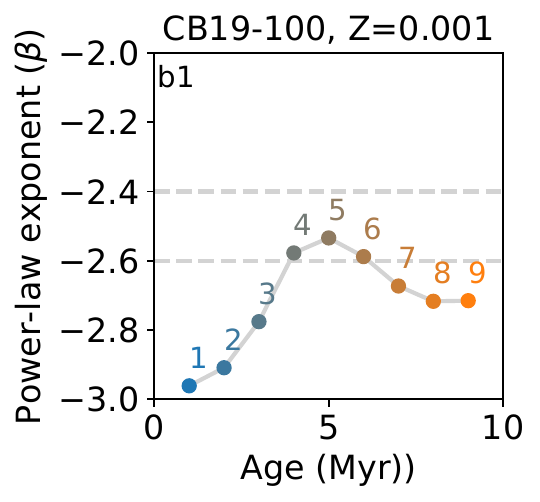}
    \includegraphics[width=0.24\textwidth]{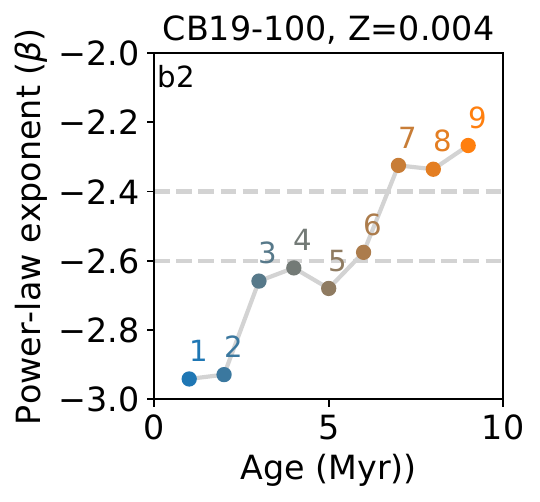}
    \includegraphics[width=0.24\textwidth]{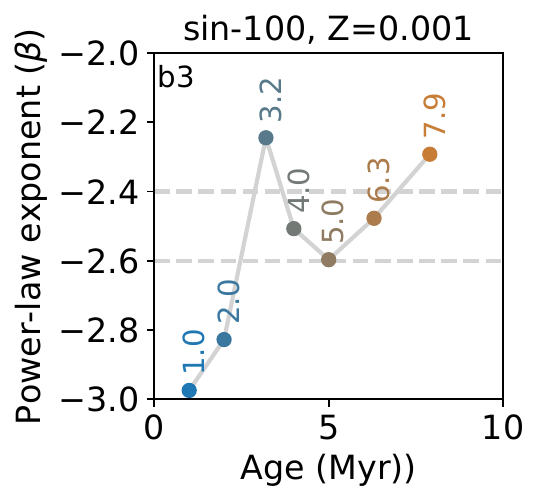}
    \includegraphics[width=0.24\textwidth]{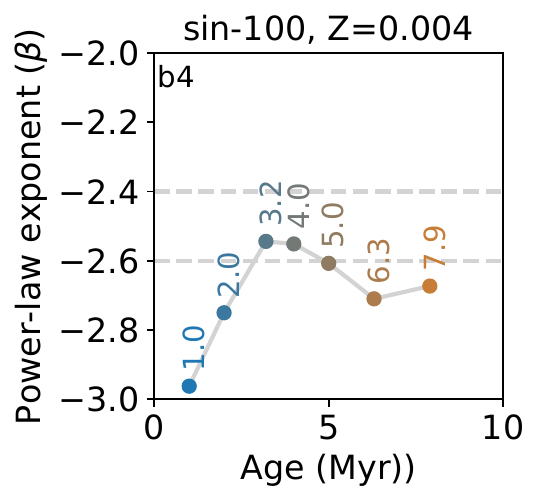}\\

    \includegraphics[width=0.24\textwidth]{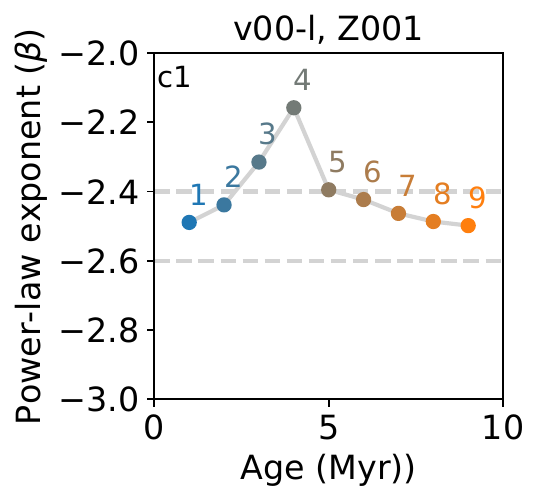}
    \includegraphics[width=0.24\textwidth]{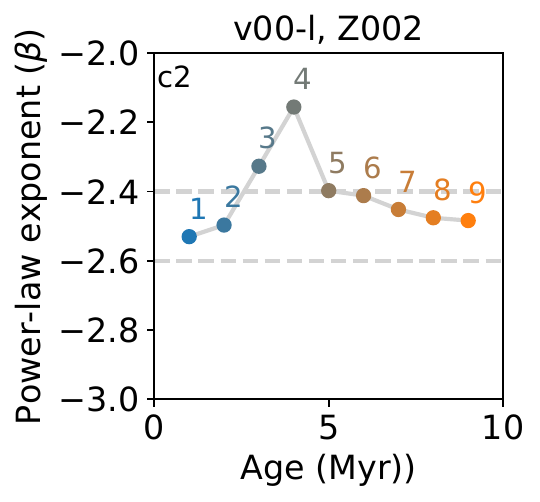}
    \includegraphics[width=0.24\textwidth]{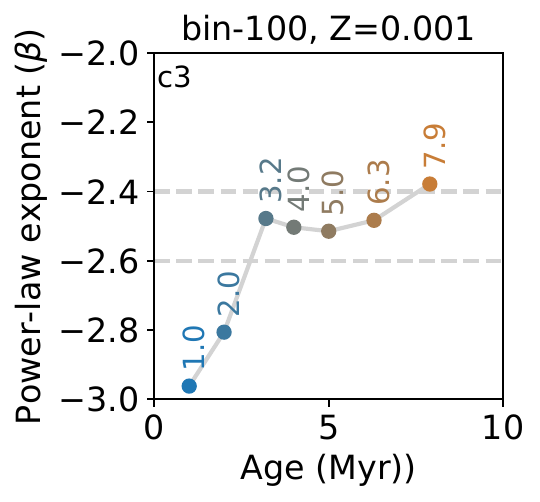}
    \includegraphics[width=0.24\textwidth]{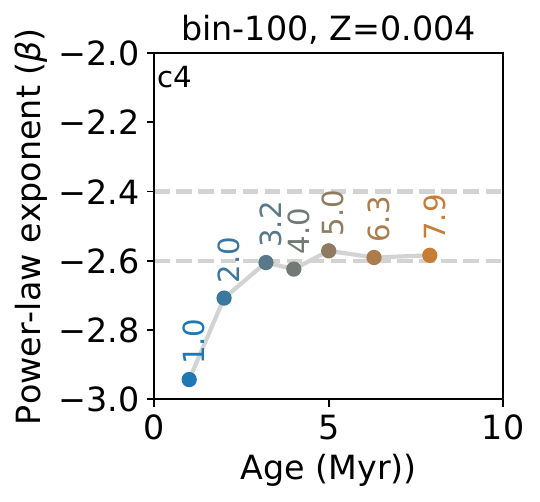}\\

    \includegraphics[width=0.24\textwidth]{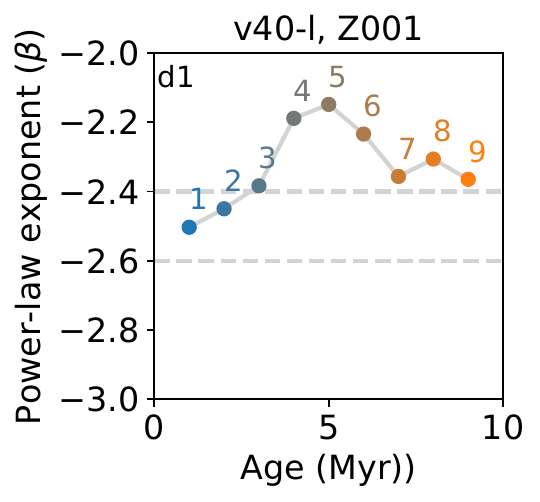}
    \includegraphics[width=0.24\textwidth]{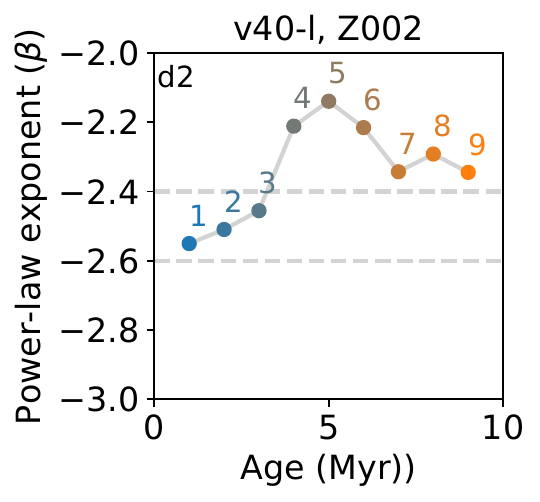}
    \includegraphics[width=0.24\textwidth]{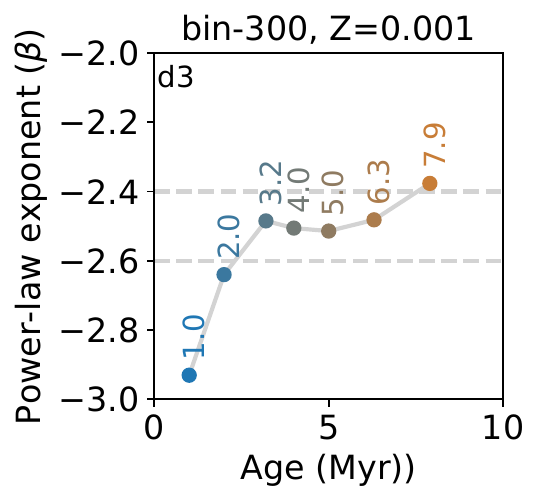}
    \includegraphics[width=0.24\textwidth]{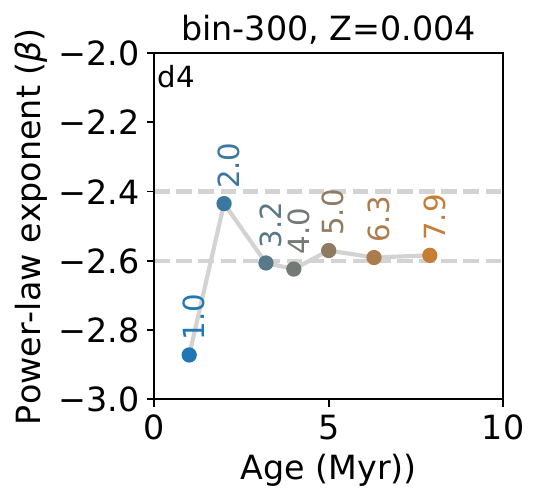}

    \caption{Evolution of SSP stellar continuum slope, $\beta$, from 1-9 Myr, obtained from the $\approx$1190-1670 \AA~range, for the models listed in Tab.~\ref{tab:model_summary}. Horizontal lines at -2.4 and -2.6 are shown for reference, as in some cases slopes are all more negative (e.g. b1) or more positive (e.g. c1) than these values.
    } 
    \label{fig:slopes}
\end{figure*}

We computed the evolution of $\beta$ for the models listed in Tab.~\ref{tab:model_summary}. The result is shown in Fig.~\ref{fig:slopes}. We find that over the wavelength range, $\approx1190-1670$ \AA, a power-law gives a reasonable fit for ages 1-9 Myr. The Fig. confirms that for ages from 1-9 Myr, the intrinsic value of $\beta$ fluctuates between -3 and -2. In addition the S99 models (panels c1-c2 and d1-d2) all have $\beta\gtrsim-2.6$. Furthermore, most BPASS models (except those displayed in panels a4 and b4) and the CB19 models at Z=0.001 (panels a1 and b1) all have steeper slopes of $\beta\lesssim-2.4$. Finally, the CB19 models of Z=0.004 (a2 and b2) and BPASS-sin models of Z=0.001 (a3 and b3): span  almost the entire range from -3 to -2 and show a similar steepening of $\beta$ from 3-5 Myr. In summary, at a fixed  metallicity and age, models with similar astrophysics, e.g., single non-rotating models with similar IMF mass limits, may have different values of the $\beta$ slope and result in different reddening corrections based on this slope. 

\section{Properties of OB-star populations}\label{sec:stars}

Stellar lines in integrated UV galaxy spectra can be used to characterize the properties of the massive-star populations that dominate these features. As shown in Fig.~\ref{fig:g140l-sp}, galaxies J1016, J1230, and J2230 have P-Cygni like N\5~1240 and C\4~1550 profiles, which we attribute to massive-star winds. To estimate light-weighted ages of the populations, we follow an approach similar to that of \cite{Pellerin2007MNRAS.381..228P}, but use spectral synthesis models that are described in Appendix~\ref{app:model_details}, account for different astrophysical processes, and were generated with independent codes. The input parameters of the models are summarized in Tab.~\ref{tab:model_summary}. Another difference with \cite{Pellerin2007MNRAS.381..228P} is that in addition to P\5~1128 and C\3~1175 we use these additional constraints: N\5~1240, Si\4~1400, and C\4~1550. Our data are not of sufficient quality to distinguish between nebular and stellar-wind components to the C\4~1550 emission, which is why in our fits to the observations we give more weight to the N\5~1240 profile. We determine the masses of the stellar populations using the approach described in \citep{Wofford2013ApJ...765..118W}, which relies on the fact that for a given age, metallicity, IMF, and SFH, the stellar continuum luminosity at 1500 \AA, L1500, only depends on the mass of the stellar population. Prior to determining the mass, it is important to know the age of the stellar population, so that the best possible reddening correction can be done. This is because as shown in Fig.~\ref{fig:slopes}, there is a dependency of the $\beta$-slope used for the correction, on age.  

\subsection{Light-weighted ages}\label{subsec:ages}

We determine the light-weighted ages of the dominant OB-star populations in our three galaxies with P-Cygni like profiles following the steps listed below.

\begin{enumerate} 
    \item We rectify the spectrum by dividing the observed flux array by a fit to the continuum. We obtain this fit by interpolating between nodes that we manually position in order to avoid strong stellar and ISM features. To guide the normalization and location of the nodes, we use a rectified S99 model of LMC metallicity. We adjust the tension of the cubic spline between the nodes as needed to closely match the rectified luminosities of the model. 
    \item We divide the flux errors by the above fit to the observed fluxes.
    \item We correct the observation for the redshift of the galaxy.
    \item We trim the noisy edges of the spectrum and extract the rest-frame $1100-1950$\AA~wavelength range.
    \item Although the standard output of S99 includes rectified models, this is not the case for the CB19 and BPASS models. We thus rectify the latter two models. For this purpose, we first resample their wavelength arrays to that of the S99 array. We then rectify each model by dividing its luminosity array by a  polynomial fit to the continuum, obtained from nodes located at: 1137, 1188, 1349, 1433, 1468.5, 1678, 1785, 1805, 1878, and 1945 \AA, which are free of strong spectral lines in the S99 models.  
    \item We do not smooth the COS G140L spectra to avoid the introduction of correlated flux errors. We resample all models to the rest-frame wavelength array of the observations. 
    \item We smooth the models to match the effective spectral resolution of the observations.
    \item We determine the reduced $\chi^2$ value, $\chi_r^2=\frac{\chi^2}{N-1}$ resulting from the comparison of the theoretical and observed spectral regions of interest, avoiding  contamination with strong foreground MW absorptions. The latter mostly affect the Si\4~and C\4~profiles. Here $N$ is the number of degrees of freedom, i.e., the number of points fitted minus one. The value of N is between 50 and 77, depending on the galaxy. We subtract one because we are constraining the age at a fixed metallicity. The fitted regions around each spectral line of interest have different sizes because we try to avoid contamination with unwanted lines. After experimenting with slightly different wavelength ranges to fit, we find that in general, the best-fitting ages differ by $\pm2$ Myr or less, depending on the choice of the fitting ranges.
\end{enumerate}

\begin{figure*}
    \centering
    \includegraphics[width=1\textwidth]{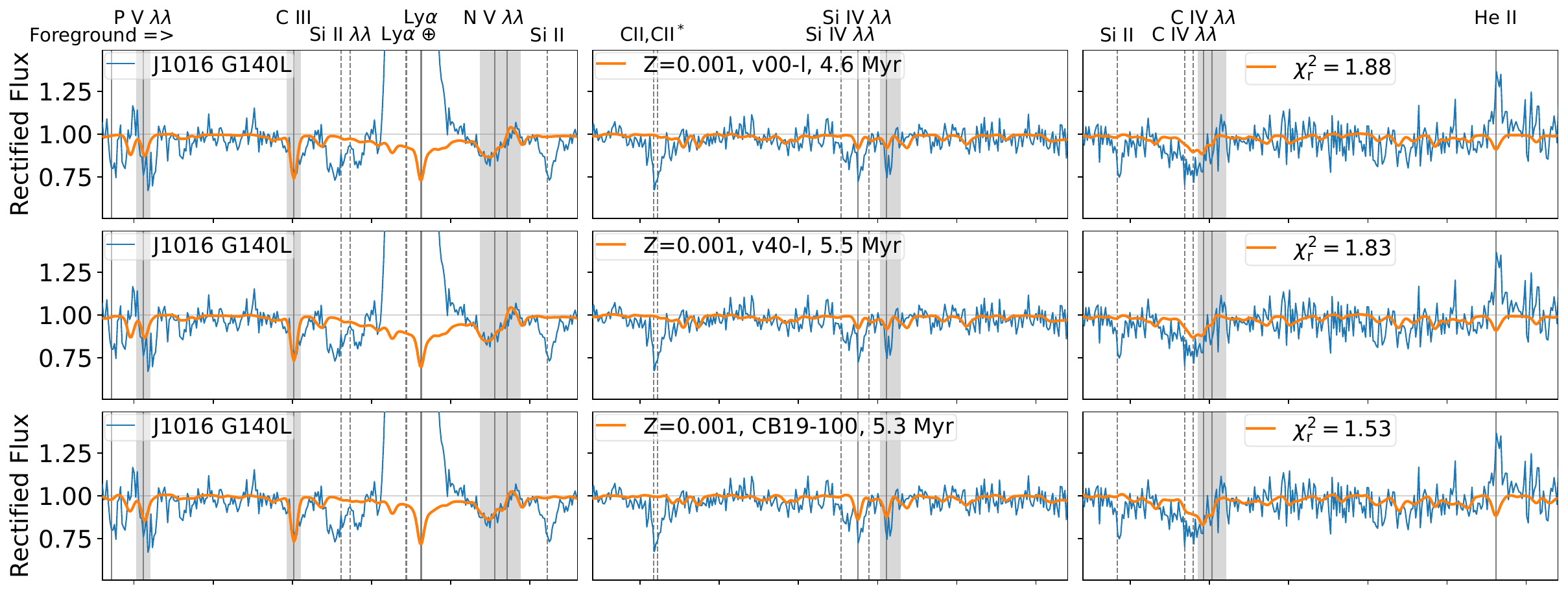}\\
    \includegraphics[width=1\textwidth]{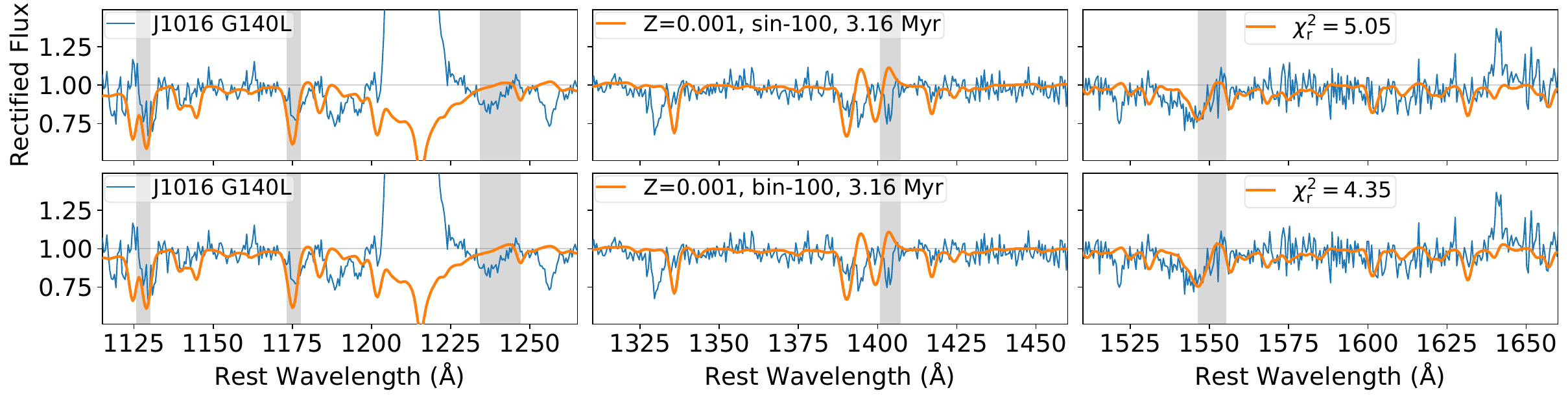}
    \caption{For galaxy J1016, comparisons between portions of the G140L spectrum (blue curves) and predictions from the models given by the legends (orange curves). The legends in the right-most column give the value of $\chi_r^2$. We mark the positions of important foreground (geocoronal and MW ISM) features with dashed vertical lines; and intrinsic features with solid vertical lines. We use the regions highlighted with gray vertical bands to compute the $\chi_r^2$ values.}
    \label{fig:j1016}
\end{figure*}

\begin{figure*}
    \centering
    \includegraphics[width=1\textwidth]{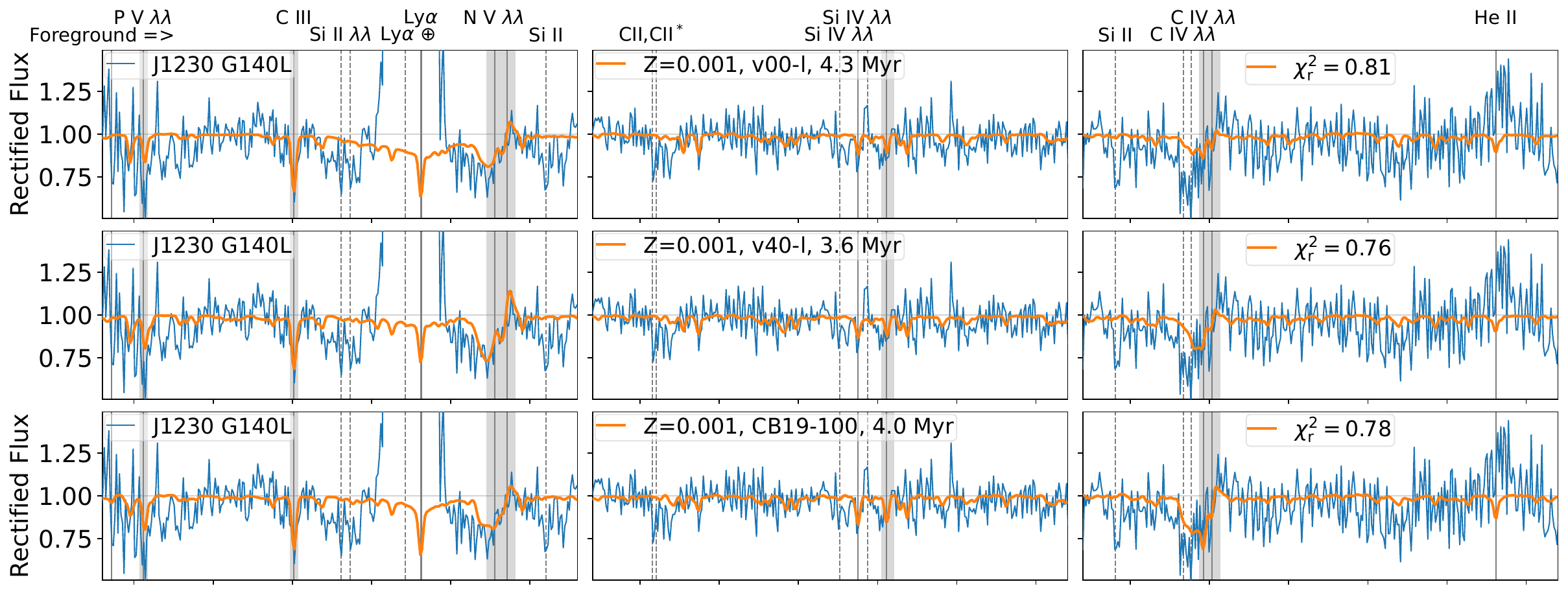}
    \includegraphics[width=1\textwidth]{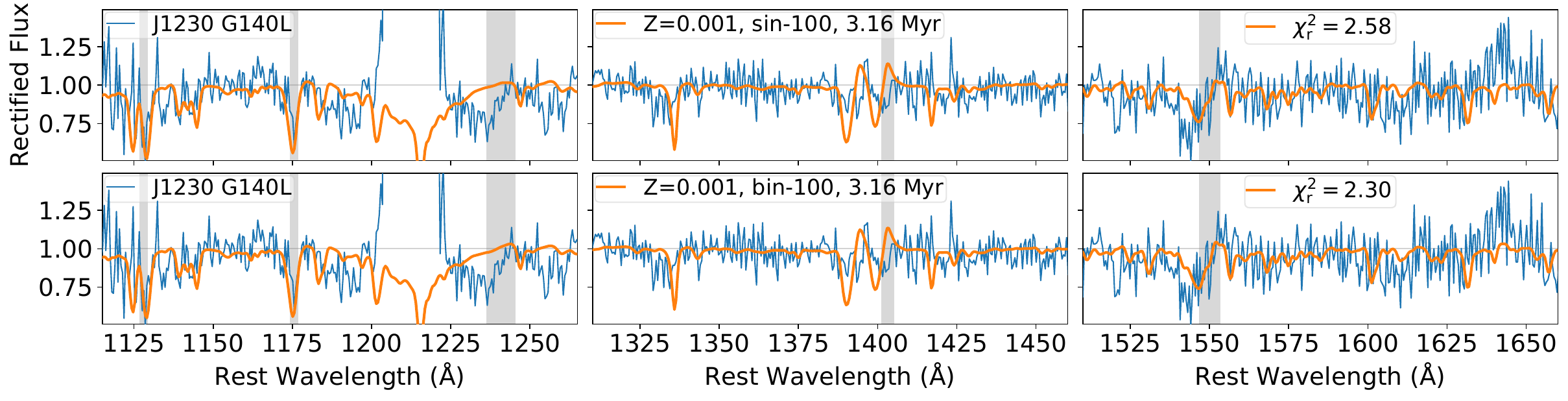}
    \caption{Similar to Fig.~\ref{fig:j1016} but for J1230.}
    \label{fig:j1230}
\end{figure*}

\begin{figure*}
    \centering
    \includegraphics[width=1\textwidth]{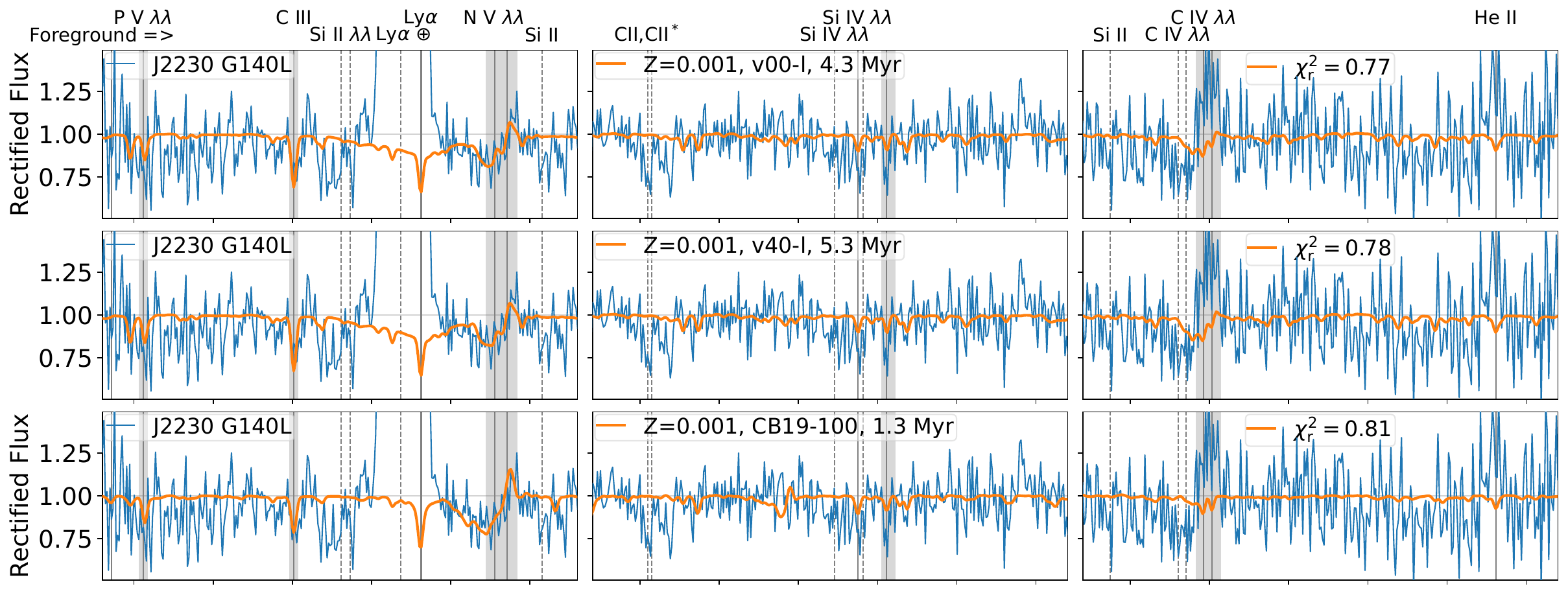}\\
    \includegraphics[width=1\textwidth]{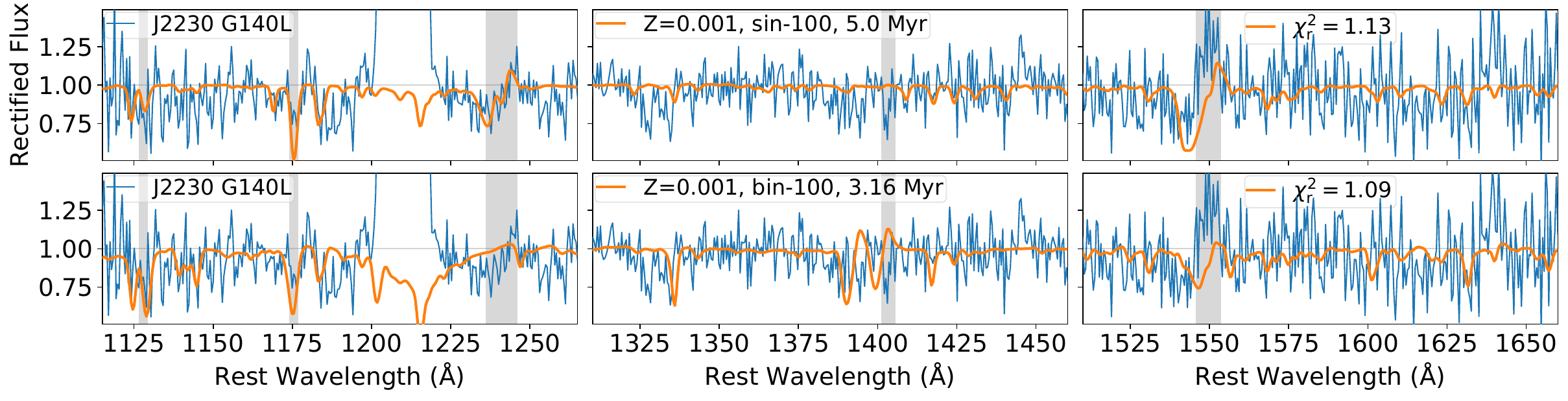}
    \caption{Similar to Fig.~\ref{fig:j1016} but for J2230.}
    \label{fig:j2230}
\end{figure*}

For galaxies J1016, J1230, and J2230, Figs.~\ref{fig:j1016}-\ref{fig:j2230} show comparisons between the observed spectra and best-fitting solutions from different fixed-metallicity models.  

In general, a model that fits the observations within the observational uncertainties will produce a $\chi_r^2\approx1$. This assumes that we can trust the uncertainties. However, in our case, these are likely to be underestimated. This is because correlated errors are introduced when resampling the observations to the nominal sampling of the COS. In addition, we rectify the observations. On the other hand, our smoothing of the model spectra to match the resolution of the observations causes blending of nearby lines in some cases, which may affect the quality of the fit. For the above reasons, only a relative comparison between the model and observation is meaningful. This is why we focus on the minimum $\chi_r^2$ value.

In Figs.~\ref{fig:j1016}-\ref{fig:j2230}, we show the models with the minimum $\chi_r^2$ values for the Z=0.001 case.  For ages of $>3$ Myr, single-star models with and without VMS yield identical solutions. All our fits with models accounting for VMS yield ages above 3 Myr, which is why we do not show them in the Figs.. We find no evidence for the presence of these stars in the above three galaxies.

Table~\ref{tab:ages} lists the ages and minimum $\chi_r^2$ values for all the models that we tried. The Tab. shows that the v40-l ages are often larger than the v00-l ages, which is consistent with rotation extending the MS lifetimes of massive stars. The fact that this behaviour is not systematic could be due to the uncertainties in the observations and/or the rectification and smoothing of the models. For galaxy J1016, at the two  metallicities, the S99 and CB19 models yield $\chi_r^2>1$, whereas $\chi_r^2<1$ for the other two galaxies. Values such that $\chi_r^2<1$ indicate larger observational uncertainties. J1016 is the galaxy with the worst effective spectral resolution and the models were smoothed accordingly. This causes the model N\5~and C\4~doublets to blend, which may be worsening the fit. For J1016, the combination CB19-100, Z=0.001 and 5.3 Myr produces the lowest $\chi_r^2$ value. For J1230, some of the Z=0.004 models seem to fit better than the Z=0.001 models. Finally, for J2230, there is a large residual when comparing the theoretical and observed emission components of the C\4~profile, which is very strong in the observation, perhaps due to contamination with nebular emission. 

The BPASS models tend to fit worse the observations in comparison to the S99 and CB19 models. In particular, with BPASS, at Z=0.001, when we succeed in fitting the C\4~1550 P-Cygni profile, we fail to simultaneously fit the N\5~1240 profile, and  in addition, the P\5~1118, C\3~1175, and Si\4~1400 profiles are too strong. Conversely, when we fit N\5~1240, then the C\4~1240 profile is too strong. The authors of BPASS suggest that this could be due details about the spectral libraries and chemical composition used in the models (private communication). Note that BPASS models also have coarser age sampling in the range of interest.  

\begin{figure*}
    \centering
    \includegraphics[width=0.75\textwidth]{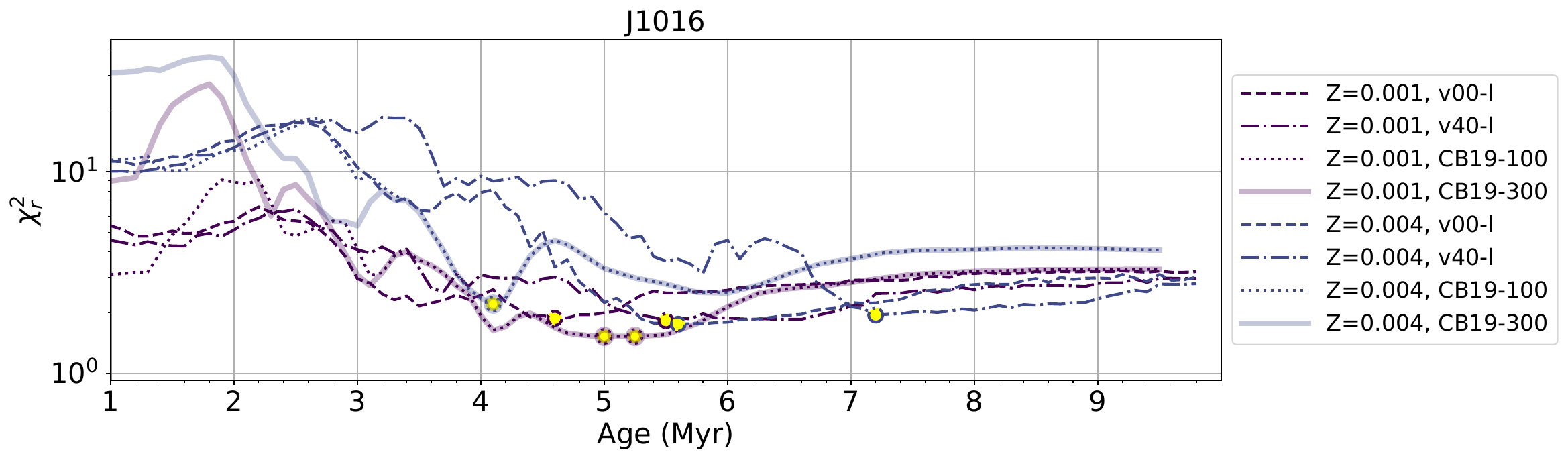}\\
    \includegraphics[width=0.75\textwidth]{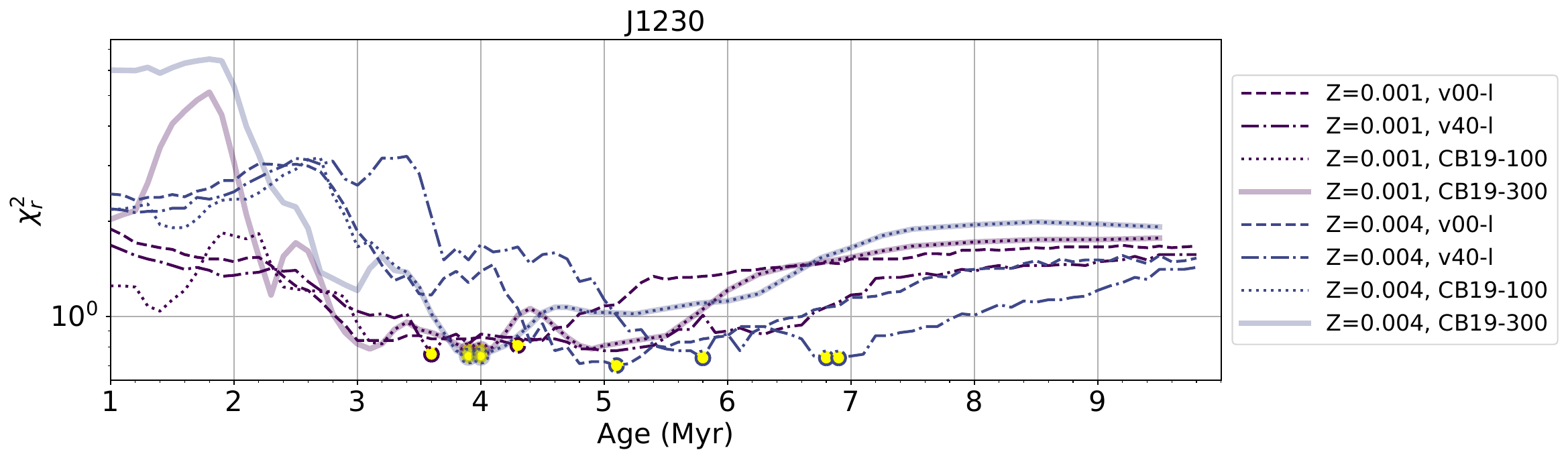}\\
    \includegraphics[width=0.75\textwidth]{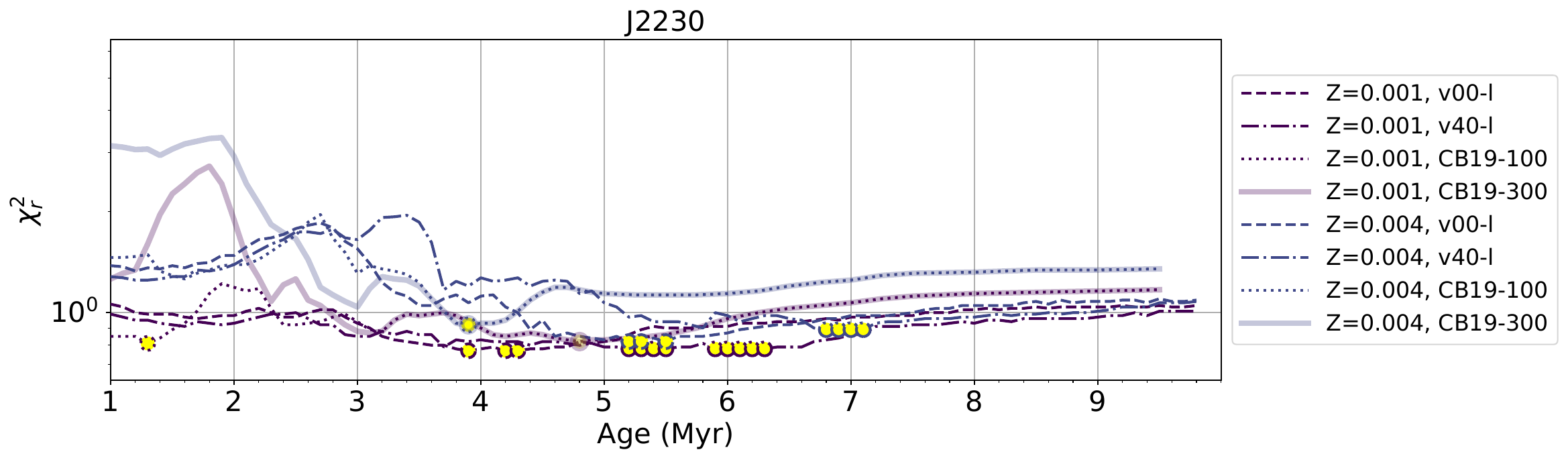}\\
    \caption{For each galaxy, evolution of $\chi_{\rm r}^2$ value for each model / metallicity combination given by the legend. The yellow circles show the ages corresponding to the minimum $\chi_{\rm r}^2$ value(s) of each curve. If the minimum $\chi_{\rm r}^2$ value is reached at multiple ages, there are multiple circles for a given curve that include but are not limited to the ages listed in Tab.~\ref{tab:ages}. This is the case for instance for the CB19-100 / Z=0.001 combination, at ages 5 and 5.25.} 
    \label{fig:rchisq_evol}
\end{figure*}

For a fixed model-metallicity combination, multiple $\chi_r^2$ values near the minimum $\chi_r^2$ may exist. This is shown in Fig.~\ref{fig:rchisq_evol} and discussed below. 

\begin{table*}
    \centering
    \setlength{\tabcolsep}{3pt}
    \begin{tabular}{cccccccccc}
    \hline
        \hfill & Z & 0.001 & 0.001 & 0.001 & 0.001 & 0.004 & 0.004 & 0.004 & 0.004\\
        \cline{3-6}
        Galaxy & Model & v00-l & v40-l & CB19-100 & CB19-300 & v00-l & v40-l & CB19-100 & CB19-300\\
    \hline                  
        J1016 & Age (Myr) & 4.6 & 5.5 & 5.3 & 5.3 & 5.6 & 7.2 & 4.1 & 4.1\\
        \hfill & $\chi_{r}^2$ & 1.88 & 1.83 & 1.53 & 1.53 & 1.76 & 1.95 & 2.21 & 2.21 \\
        \hline
        J1230 & Age (Myr) & 4.3 & 3.6 & 4.0 & 4.0 & 5.1 & 6.8 & 3.9 & 3.9 \\ 
        \hfill & $\chi_{r}^2$ & 0.81 & 0.76 & 0.78 & 0.78 & 0.70 & 0.74 & 0.75 & 0.75 \\ 
        \hline
        J2230 & Age (Myr) & 4.3 & 5.3 & 1.3 & 4.8 & 5.2 & 6.8 & 3.9 & 3.9 \\
        \hfill & $\chi_{r}^2$ & 0.77 & 0.78 & 0.81 & 0.82 & 0.82 & 0.89 & 0.92 & 0.92 \\
        \hline
        Galaxy & Model & sin-100 & sin-300 & bin-100 & bin-300 & sin-100 & sin-300 & bin-100 & bin-300 \\
        \hline
        J1016 & Age (Myr) & 3.2 & 3.2 & 3.2 & 3.2 & 2.5 & 4.0 & 2.5 & 4.0\\
        \hfill & $\chi_{r}^2$ & 5.05 & 5.05 & 4.35 & 4.35 & 4.00 & 3.68 & 4.09 & 3.89 \\
        \hline
        J1230 & Age (Myr) & 3.2 & 3.2 & 3.2 & 3.2 & 2.5 & 4.0 & 7.0 & 7.0 \\ 
        \hfill & $\chi_{r}^2$ & 2.58 & 2.58 & 2.30 & 2.30 & 2.26 & 2.20 & 2.26 & 2.26\\ 
        \hline
        J2230 & Age (Myr) & 5.0 & 5.0 & 3.2 & 3.2 & 1.3 & 1.6 & 1.3 & 4.0 \\
        \hfill & $\chi_{r}^2$ & 1.13 & 1.13 & 1.09 & 1.09 & 1.18 & 1.07 & 1.19 & 1.11 \\
        \hline
    \end{tabular}
    \caption{Ages and reduced $\chi_{r}^2$ values obtained from the best-fitting models listed in Tab.~\ref{tab:model_summary}, and shown in Figs.~\ref{fig:j1016}-\ref{fig:j2230}.}
    \label{tab:ages}
\end{table*}

\begin{table*}
    \centering
    \setlength{\tabcolsep}{4pt}
    \begin{tabular}{lccccc}
    \hline
        Comparison & Max $|\Delta|$ & Galaxy, Z & Min $|\Delta|$ & Galaxy, Z & Mean $\Delta$ \\
    \hline
        v00-l $-$ CB19-100     & 3.0 & J2230, 0.001 & 0.3 & J1230, 0.001 & $+1.10$ \\
        v00-l $-$ sin-100      & 3.9 & J2230, 0.004 & 0.7 & J2230, 0.001 & $+1.90$ \\
        sin-100 $-$ CB19-100   & 3.7 & J2230, 0.001 & 0.8 & J1230, 0.001 & $-0.80$ \\
        v00-l $-$ v40-l        & 1.7 & J1230, 0.004 & 0.7 & J1230, 0.001 & $-1.02$ \\
        CB19-100 $-$ CB19-300  & 3.5 & J2230, 0.001 & 0.0 & several      & $-0.58$ \\
        CB19-300 $-$ sin-300   & 2.3 & J2230, 0.004 & 0.1 & J1016, J1230, 0.004 & $+0.83$ \\
        sin-100 $-$ bin-100    & 4.5 & J1230, 0.004 & 0.0 & several      & $-0.45$ \\
        sin-300 $-$ bin-300    & 3.0 & J1230, 0.004 & 0.0 & several      & $-0.60$ \\
    \hline
    \end{tabular}
    \caption{Age differences (Myr) between pairs of models, computed for the same galaxy at the two metallicities ($Z = 0.001$ and $0.004$; six pairs per comparison). $\Delta$ is the age from the first model minus that from the second.}
    \label{tab:agediffs}
\end{table*}

In Fig.~\ref{fig:rchisq_evol}, we show the evolution of $\chi_r^2$ for the three galaxies and some models considered in this work. For a given model, multiple ages may have similarly low values of $\chi_{\rm r}$ and may provide fits within the observational uncertainties. These are shown in Fig.~\ref{fig:rchisq_evol} with open circles, which include some of the solutions given in Table~\ref{tab:ages}.

In summary, we tried models with and without rotation, with and without massive mass-exchanging binaries, and with and without VMS, but none is currently able to simultaneously reproduce all selected stellar age-sensitive features. As shown in Tab.~\ref{tab:agediffs}, which gives the age differences between pairs of models for the three galaxies and two metallicities, age differences can be as large as 4.5 Myr and as low as 0 Myr. 

\subsection{Reddening due to dust and mass in stars}

Using the ionized-gas O/H ratio of the galaxies as an independent gauge of metallicity, we set Z=0.001. At this metallicity, the light-weighted ages of J1016, J1230, and J2230 when averaging over the best-fitting v00-l, v40-l and CB19-100 ages shown in Figs.~\ref{fig:j1016}-\ref{fig:j2230} are: 5.1, 4.0, and 3.6 Myr, respectively. Using the v00-l, v40-l, and CB19-100 models at the latter metallicity and the corresponding ages for each galaxy, we determine the reddenings due to dust intrinsic to the galaxies and the masses in stars. For the former, we use the $\beta$ slope predicted by the model at the relevant age (see Fig.~\ref{fig:slopes}), which are given in Tab.~\ref{tab:population_properties}. We compute the mass in stars after correcting for reddening due to dust in the MW (using the MW extinction law of \cite{Fitzpatrick1999PASP..111...63F}), redshift, and reddening due to dust intrinsic to the galaxy (using the SMC extinction law of \citep{Gordon2003ApJ...594..279G}. For the MW E(B-V) value and the redsfhit we use the values of Tab.~\ref{tab:known_galaxy_properties}. The intrinsic E(B-V) values are given in Tab.~\ref{tab:population_properties}. For the latter purpose, we compare the observed luminosity at 1500 $\AA$ computed by using the distance of Tab.~\ref{tab:known_galaxy_properties}, with that predicted for an SSP model of 1e6 M$_\odot$ of the same metallicity and age. We do not determine average ages, intrinsic reddenings and masses in stars using the BPASS models, since these provide worst fits compared to the previous models, at Z=0.001. Similarly, we exclude the CB19-300 models as for our galaxies, these yield ages of $>3$ Myr, when the CB19-300 and CB19-100 models yield a similar spectral evolution. The results are shown in Tab.~\ref{tab:population_properties}. There is consistency between the three models regarding which is the galaxy with the lowest and largest intrinsic E(B-V) value. In addition, for J1016, J1230, and J2230, the masses in stars given by the three models in Tab.~\ref{tab:population_properties} are within factors of 1.3, 2, and 2.5 from each other, respectively.

\begin{table*}
    \centering
    \begin{tabular}{lllllll}
    \hline
        ID & Model & Age & $\beta$ & E(B-V) & L$_{1500}$ & log\\
        \hfill & \hfill & Myr & intr & intr. [mag] & 10$^{38}$ erg/s & (M$_*$/M$_\odot$)\\
        \hline
         J1016 & v00-l  & 4.6 & -2.29 & 0.020 & 5.95 & 5.68 \\
         J1016 & v40-l  & 5.5 & -2.19 & 0.014 & 5.57 & 5.62 \\
         J1016 & CB19-100  & 5.0 & -2.58 & 0.035 & 7.17 & 5.74 \\
         \hline
         J1230 & v00-l  & 4.3 & -2.13 & 0.038 & 1.01 & 4.85\\
         J1230 & v40-l  & 3.6 & -2.21 & 0.042 & 1.07 & 4.66\\
         J1230 & CB19-100  & 4.0 & -2.60 & 0.063 & 1.40 & 5.01\\
         \hline
         J2230 & v00-l  & 4.3 & -2.15 & 0.078 & 3.54 & 5.39\\
         J2230 & v40-l  & 5.3 & -2.15 & 0.078 & 3.56 & 5.39\\
         J2230 & CB19-100  & 1.3 & -2.95 & 0.114 & 5.50 & 5.78\\
         \hline
    \end{tabular}
    \caption{Properties of stellar populations in galaxies with UV P-Cygni like profiles, assuming Z=0.001.}
    \label{tab:population_properties}
\end{table*}
\section{Properties of nebular emission lines}\label{sec:gas}

\subsection{Optical and UV emission line fluxes}

For the eight galaxies in our sample, we measure SDSS fluxes of nebular emission lines of interest. We simultaneously fit Gaussians + a linear continuum to groups of lines with a common velocity and velocity dispersion, and use independent amplitudes for the individual lines within a group. We do not correct the Balmer lines for underlying stellar absorption components, as it is generally weak. We correct line fluxes for reddening due to dust in the Milky Way (MW) using the \cite{Fitzpatrick1999PASP..111...63F} extinction law (in the observed frame) and the E(B-V) values listed in Tab.~\ref{tab:sdss_line_fluxes}; and for reddening due to dust intrinsic to the galaxies using the \cite{Gordon2003ApJ...594..279G} SMC-bar extinction law (in the rest-frame of the galaxy) and the intrinsic E(B-V) values listed in the same Tab.. We also measure the equivalent widths of He\2 4686 and H$\beta$. The fluxes and equivalent widths are given in Tab.~\ref{tab:sdss_line_fluxes}. For the same galaxies, we obtain COS G140L fluxes of UV emission lines of interest using a similar method, i.e., we tie the velocity centroid and velocity dispersion of lines that share the same kinematics, but we do not correct the observations for reddening, as the models used for comparison include dust. We provide the fluxes in Tab.~\ref{tab:uv_fluxes}, where groups of lines fitted with fixed kinematics are separated by horizontal lines. 

\begin{table*}
    \centering
    \setlength{\tabcolsep}{0.3em} 
    \begin{tabular}{lllllllll}
    \hline
\hfill  & J0944 & J1016$^*$ & J1044 & J1119 & J1132 & J1201 & J1230$^*$ & J2230$^*$\\
\hline
\multicolumn{9}{c}{Optical emission-line flux ratios, $F(\lambda) / F(H\beta)\times100$, uncorrected for reddening} \\
\hline
H$\gamma$ $\lambda$4340 & 45.9 $\pm$ 1.3 & 46.7 $\pm$ 0.8 & 46.9 $\pm$ 0.7 & 44.7 $\pm$ 1.2 & 46.9 $\pm$ 1.4 & 45.0 $\pm$ 1.1 & 44.5 $\pm$ 1.0 & 45.0 $\pm$ 1.1 \\
He\2 $\lambda$4686 & 0.4 $\pm$ 0.1 & 2.2 $\pm$ 0.2 & 1.9 $\pm$ 0.1 & 1.9 $\pm$ 0.5 & 1.6 $\pm$ 0.5 & 1.2 $\pm$ 0.4 & 3.9 $\pm$ 0.3 & 1.6 $\pm$ 0.3 \\
{[}O\3{]} $\lambda$5007 & 429.8 $\pm$ 11.8 & 453.3 $\pm$ 5.6 & 421.3 $\pm$ 4.5 & 303.9 $\pm$ 4.3 & 208.9 $\pm$ 3.9 & 354.8 $\pm$ 4.8 & 502.2 $\pm$ 6.4 & 555.4 $\pm$ 7.5 \\
H$\alpha$ $\lambda$6563 & \ldots & 302.3 $\pm$ 4.0 & 308.1 $\pm$ 3.5 & 320.8 $\pm$ 4.8 & 322.7 $\pm$ 5.6 & 326.4 $\pm$ 5.1 & 315.6 $\pm$ 4.5 & 311.6 $\pm$ 4.7 \\
{[}N\2{]} $\lambda$6584 & 1.6 $\pm$ 0.1 & 2.2 $\pm$ 0.1 & 0.9 $\pm$ 0.0 & 2.6 $\pm$ 0.2 & 4.7 $\pm$ 0.2 & 2.4 $\pm$ 0.1 & 3.0 $\pm$ 0.1 & 7.4 $\pm$ 0.2 \\
\hline
\multicolumn{9}{c}{Optical emission-line flux ratios, $I(\lambda) / I(H\beta)\times100$, corrected for reddening}\\
\hline
H$\gamma$ $\lambda$4340 & 46.8 $\pm$ 1.3 & 48.0 $\pm$ 0.8 & 48.5 $\pm$ 0.7 & 47.3 $\pm$ 1.2 & 49.7 $\pm$ 1.5 & 47.9 $\pm$ 1.2 & 46.6 $\pm$ 1.0 & 46.6 $\pm$ 1.2 \\
He\2 $\lambda$4686 & 0.4 $\pm$ 0.1 & 2.2 $\pm$ 0.2 & 1.9 $\pm$ 0.1 & 1.9 $\pm$ 0.5 & 1.6 $\pm$ 0.5 & 1.3 $\pm$ 0.4 & 3.9 $\pm$ 0.3 & 1.7 $\pm$ 0.3 \\
{[}O\3{]} $\lambda$5007 & 427.5 $\pm$ 11.7 & 450.2 $\pm$ 5.5 & 417.5 $\pm$ 4.5 & 299.7 $\pm$ 4.3 & 205.8 $\pm$ 3.8 & 349.2 $\pm$ 4.7 & 496.1 $\pm$ 6.4 & 549.8 $\pm$ 7.4 \\
H$\alpha$ $\lambda$6563 & (286.0 $\pm$ 0.6) & 286.0 $\pm$ 3.8 & 286.0 $\pm$ 3.3 & 286.5 $\pm$ 4.3 & 286.0 $\pm$ 5.0 & 287.4 $\pm$ 4.5 & 285.8 $\pm$ 4.1 & 285.8 $\pm$ 4.3 \\
{[}N\2{]} $\lambda$6584 & 1.5 $\pm$ 0.1 & 2.0 $\pm$ 0.1 & 0.8 $\pm$ 0.0 & 2.4 $\pm$ 0.2 & 4.1 $\pm$ 0.2 & 2.1 $\pm$ 0.1 & 2.7 $\pm$ 0.1 & 6.8 $\pm$ 0.2 \\
\hline
\multicolumn{9}{c}{Complementary quantities}\\
\hline
$^{\rm a}$F(H$\beta$) & 14.5 $\pm$ 0.2 & 49.8 $\pm$ 0.4 & 42.4 $\pm$ 0.3 & 12.2 $\pm$ 0.1 & 7.3 $\pm$ 0.1 & 11.8 $\pm$ 0.1 & 22.0 $\pm$ 0.2 & 16.1 $\pm$ 0.2 \\
$^{\rm b}$I(H$\beta$) & 16.5 $\pm$ 0.2 & 58.7 $\pm$ 0.5 & 52.6 $\pm$ 0.4 & 17.0 $\pm$ 0.2 & 10.4 $\pm$ 0.1 & 17.1 $\pm$ 0.2 & 29.4 $\pm$ 0.3 & 20.7 $\pm$ 0.2 \\
$^{\rm c}$EW(H$\beta$) & 250.7 $\pm$ 13.0 & 98.6 $\pm$ 5.0 & 266.2 $\pm$ 13.5 & 45.1 $\pm$ 2.3 & 70.4 $\pm$ 3.6 & 197.8 $\pm$ 10.1 & 94.0 $\pm$ 4.8 & 96.0 $\pm$ 4.9 \\
$^{\rm c}$EW(He\2 $\lambda$4686) & 0.9 $\pm$ 0.2 & 1.9 $\pm$ 0.2 & 4.8 $\pm$ 0.4 & 0.8 $\pm$ 0.2 & 1.1 $\pm$ 0.3 & 2.2 $\pm$ 0.6 & 3.4 $\pm$ 0.3 & 1.5 $\pm$ 0.3 \\
$^{\rm d}$$E(B-V)_{\rm MW}$ & 0.011 $\pm$ 0.001 & 0.013 $\pm$ 0.000 & 0.037 $\pm$ 0.001 & 0.013 $\pm$ 0.001 & 0.011 $\pm$ 0.001 & 0.021 $\pm$ 0.001 & 0.024 $\pm$ 0.000 & 0.061 $\pm$ 0.001 \\

$^{\rm e}$$E(B-V)_{\rm int}$ & 0.029 $\pm$ 0.054 & 0.040 $\pm$ 0.013 & 0.032 $\pm$ 0.011 & 0.097 $\pm$ 0.014 & 0.105 $\pm$ 0.017 & 0.101 $\pm$ 0.015 & 0.071 $\pm$ 0.013 & 0.016 $\pm$ 0.014 \\
\hline
    \end{tabular}
    \caption{Top and middle.-- Optical emission-line flux ratios derived from our fits to the SDSS spectra. Since the H$\alpha$ line of J0944 is truncated, for this galaxy we provide the H$\beta$ flux multiplied by the intrinsic ratio, H$\alpha$/H$\beta=2.86$. Bottom.-- $^{\rm a}$Reddening- and redshift-uncorrected H$\beta$ fluxes ($10^{-15}$ erg/s/cm$^2$). $^{\rm b}$Reddening- and redshift-corrected H$\beta$ fluxes ($10^{-15}$ erg/s/cm$^2$). $^{\rm c}$Equivalent widths (\AA), where positive values are for emission lines. $^{\rm d}$Colour excess due to dust in the MW from the IRSA Dust Extinction Service, based on the \citet{Schlegel1998ApJ...500..525S} dust maps with the recalibration from \citet{Schlafly2011ApJ...737..103S}. $^{\rm d}$Intrinsic colour excess obtained from the SMC-bar extinction law with Rv=2.74, after MW-dereddening. For the latter, we combined the reddening values obtained from the H$\gamma$/H$\beta$ and H$\alpha$/H$\beta$ ratios using inverse-variance weights, but only if both values were positive. Otherwise, we kept the positive value. There were no cases where both values were negative. Galaxy IDs marked with asterisks correspond to the targets with P-Cygni like C\4 UV lines, which all have EW(H$\beta$) values below 100 \AA.} 
    \label{tab:sdss_line_fluxes}
\end{table*}

\begin{table*}
\centering
\begin{tabular}{lcccccccc}
\hline
Line & J0944 & J1016$^*$ & J1044 & J1119 & J1132 & J1201 & J1230$^*$ & J2230$^*$ \\
\hline
\multicolumn{9}{c}{UV emission-line fluxes uncorrected for reddening ($10^{-15}\,\mathrm{erg\,s^{-1}\,cm^{-2}}$)}\\
\hline
C\,{\sc iv}\,$\lambda1548$ & 5.34 $\pm$ 0.75 & -9.99 $\pm$ 41.67 & 19.79 $\pm$ 1.63 & -0.43 $\pm$ 0.91 & 0.70 $\pm$ 0.56 & 1.10 $\pm$ 0.49 & -0.12 $\pm$ 279.45 & 4.06 $\pm$ 14.32 \\
C\,{\sc iv}\,$\lambda1551$ & 4.29 $\pm$ 0.67 & -4.14 $\pm$ 49.46 & 11.37 $\pm$ 1.35 & -0.90 $\pm$ 2.26 & -0.09 $\pm$ 0.31 & 0.56 $\pm$ 0.35 & 1.59 $\pm$ 21771.32 & 1.89 $\pm$ 18.50 \\
\hline
He\,{\sc ii}\,$\lambda1640$ & (0.5$\pm$0.2) & 9.75 $\pm$ 2.48 & 6.08 $\pm$ 1.29 & \ldots & \ldots & (1.6$\pm$0.1) & 9.27 $\pm$ 3.43 & 3.88 $\pm$ 1.47 \\
\hline
O\,{\sc iii}]\,$\lambda1661$ & (0.69 $\pm$ 0.33) & 1.14 $\pm$ 0.95 & 6.29 $\pm$ 0.88 & -0.94 $\pm$ 7.28 & -1.36 $\pm$ 0.97 & 0.92 $\pm$ 0.37 & 2.79 $\pm$ 1.29 & 0.52 $\pm$ 0.82 \\
O\,{\sc iii}]\,$\lambda1666$ & 4.41 $\pm$ 0.71 & 2.22 $\pm$ 1.38 & 14.66 $\pm$ 1.43 & 2.73 $\pm$ 26.68 & -0.51 $\pm$ 0.65 & (1.28 $\pm$ 0.45) & 2.98 $\pm$ 1.34 & 2.91 $\pm$ 1.42 \\
\hline
[Si\,{\sc iii}]\,$\lambda1883$ & 1.61 $\pm$ 1.65 & 3.98 $\pm$ 2.28 & 3.61 $\pm$ 2.74 & 4.43 $\pm$ 2.37 & -1.13 $\pm$ 3.89 & 4.11 $\pm$ 1.72 & 7.04 $\pm$ 4.39 & 3.27 $\pm$ 1.60 \\
Si\,{\sc iii}]\,$\lambda1892$ & 0.98 $\pm$ 1.32 & 1.75 $\pm$ 1.58 & 3.07 $\pm$ 2.53 & -1.96 $\pm$ 1.55 & 0.91 $\pm$ 1.62 & 3.93 $\pm$ 1.77 & -4.00 $\pm$ 3.28 & 0.72 $\pm$ 0.94 \\
\hline
[C\,{\sc iii}]\,$\lambda1906$ & 3.08 $\pm$ 2.67 & 16.13 $\pm$ 4.42 & 18.72 $\pm$ 2.60 & 5.03 $\pm$ 78.66 & 0.98 $\pm$ 2.30 & 5.69 $\pm$ 1.22 & 5.57 $\pm$ 190.99 & 1.65 $\pm$ 0.87 \\
C\,{\sc iii}]\,$\lambda1909$ & 8.99 $\pm$ 2.79 & 15.21 $\pm$ 4.39 & 15.41 $\pm$ 2.11 & 8.13 $\pm$ 71.26 & -0.48 $\pm$ 0.78 & 5.97 $\pm$ 1.27 & 16.63 $\pm$ 155.18 & 4.66 $\pm$ 1.39 \\
\hline
\end{tabular}
\caption{UV emission-line fluxes derived from our fits to the COS G140L data and uncorrected for reddening. Since no significant UV He\2 emission is detected for J0944 and J1201, for these galaxies, we provide He\2 $\lambda1640$ fluxes equal to seven times the optical He\2 $\lambda4686$ fluxes based on the intrinsic nebular ratio of  \citet{Huntzinger_Chen_Jones_Sanders_Senchyna_Stark_Bresolin_Charlot_Chevallard_2026}. In addition, for these two galaxies, O\,{\sc iii}]\, fluxes in parenthesis indicate lower limits due to contamination with Al\2 1670 from the MW. Galaxy IDs marked with asterisks correspond to the targets with P-Cygni like C\4 UV lines. Fluxes with negative values or large error bars are due to the weakness of the lines relative to the noisy nearby continuum.}
\label{tab:uv_fluxes}
\end{table*}

\subsection{Photoionization models}

A photoionization model predicts the ionization, thermal, and emission-line properties of gas exposed to a source of ionizing radiation, given its physical conditions and chemical composition. For the galaxies in our sample, we compare observed optical and UV emission line ratios to models of photoionization by AGN  \citep{Feltre2016MNRAS.456.3354F}, shocked regions  \citep{Alarie2019RMxAA..55..377A}, and massive OB stars. The models of photoionization by massive stars use ionizing flux predictions from either pySTARBURST99 \citep{Hawcroft2025ApJS..280....5H} or CB19. The corresponding SSP models are described in Appendix~\ref{app:model_details} and the input parameters provided in Tab.~\ref{app:model_details}. The pySTARBURST99 photoionization grid is from Aranguré et al. (in prep.) and was computed with Cloudy c25.00 \citep{Gunasekera2025RMxAA..61..120G} for ages $<10$ Myr. The pySTARBUST99 grid does not account for VMS, as these stars are not currently available at the metallicities of interest in pySTARBUST99. The models by Aranguré et al. consider variations in stellar rotation; gas geometry, volume filling factor and density; dust content; ionization parameter; and escape fraction of Lyman-continuum (LyC) photons. The CB19 grid is an updated version of the one described in \cite{Martinez2023MNRAS.525.2916M}, where the chemical abundances have been corrected (see Erratum of Martinez et al., in prep.). The grids of models of photoionization by massive stars use the input parameters summarized in Tab.~\ref{tab:photoionization_grids}. 

\begin{table*}
    \centering

    \begin{tabular}{lcc}
        \toprule
        \hfill & Arangur\'e et al. & Updated version of \\
        Parameter & (in prep.) & \citet{Martinez2023MNRAS.525.2916M} \\
        \midrule
        Stellar populations & pySTARBURST99 & CB19 \\
        Stellar rotation ($v/v_{\rm{crit}}$) & 0, 0.4 & 0 \\
        IMF shape; $M_\mathrm{up}$ ($M_\odot$) & \cite{Kroupa2002Sci...295...82K}; 120 & \cite{Kroupa2002Sci...295...82K}; 100, 300 \\
        Age range (Myr), $\Delta t$ (Myr) & $0 - 10$, 1 & $1 - 10$, 0.5 \\
        Nebular abundance scaling & \citet{Nicholls2017MNRAS.466.4403N} & \citet{Gutkin2016MNRAS.462.1757G} \\
        12+log(O/H)$_{\rm{gas+dust}}$ & 7.30, 8.00  & 6.94, 7.94 \\
        $\Delta$log(N/O) & 0 & 0.25 \\
        Depletion; parameter & \citet{Gunasekera2023RNAAS...7..246G}; $F_*=0,0.3$ & \citet{Gutkin2016MNRAS.462.1757G}; $\xi=0.3$ \\
        Grain set; size distribution & ISM; \citet{Mathis1977ApJ...217..425M}& ISM; \citet{Mathis1977ApJ...217..425M} \\
        log(C/O) & \citet{Nicholls2017MNRAS.466.4403N} & $-0.36$ \\
        log(U) & $-4.0, -3.0, -2.0$ & $-4.0, -3.5, -3.0, -2.5, -2.0, -1.5$  \\
        log(n$_{\rm{H}}$ [cm$^{-3}$]) & 1, 3, 5 & 1, 2, 3, 4 \\
        HbFrac & $0.1-1.0$, every 0.1 & 0.17, 0.33, 0.5, 0.67, 0.83, 1.0 \\
        $\epsilon$ & 0.01, 1.0 & 1.0 \\
        $f_S$ & 0.03, 3.0 & 1.0 \\
        \bottomrule
    \end{tabular}
    \caption{Parameters of the Cloudy models of photoionization by OB stars. Row definitions--. Stars: the stellar population synthesis code. Stellar rotation: the value assumed by the stellar isochrones. IMF: the initial stellar mass function (IMF) and its upper mass limit M$_{\rm{up}}$. Age: the range of ages and time step $\Delta t$ of the photoionization models. ISM abundance scaling: reference of the scaling relations for the interstellar abundances. 12+log(O/H$)_{\rm{total}}$: the total oxygen abundance (gas+dust) of the photoionization models. Depletion: the assumed depletion method and the respective input parameter values. Grains: the Cloudy ISM dust composition (graphites+silicates) with abundances set by the metallicity, the depletion parameter, and the grain size distribution. $\Delta$log(N/O): shift from the canonical log(N/O) vs log(O/H) value. It is set to 0.25 for the updated version of the models described in \citet{Martinez2023MNRAS.525.2916M}, to be consistent with the relation of \citet{Nicholls2017MNRAS.466.4403N}. log(C/O): the value assumed by the models; from the \citet{Nicholls2017MNRAS.466.4403N} relations in Arangur\'e et al., and takes a fixed value of -0.36 for all metallicities in \citet{Martinez2023MNRAS.525.2916M}. log(U): the input mean ionization parameter as defined in \citet{Stasinska2015A&A...576A..83S}. log(n$_{\rm{H}}$): the hydrogen density. HbFrac: the fraction of ionization-bounded H$\beta$ flux; models with HbFrac $<1$ are density-bounded (i.e., have LyC escape). $\epsilon$: the volume-filling factor. $f_{\rm{S}}$: form factor defined in \citet{Stasinska2015A&A...576A..83S}.}
    \label{tab:photoionization_grids}
\end{table*}

\subsection{Diagnostic diagrams}

Figures~\ref{fig:diagnostics} and~\ref{fig:diagnostics_vms} show comparisons of models to our observations in diagnostic diagrams that use optical or UV nebular emission line ratios computed from the fluxes reported in Tabs.~\ref{tab:sdss_line_fluxes} and~\ref{tab:uv_fluxes}. To the observations, we add XMPG SBS0335-052E, which we analyse with older models in \cite{Wofford2021MNRAS.500.2908W}. This XMPG is an extreme He\2~emitter and serves as reference for comparison with the rest of the galaxies in our sample. A main difference between Figures~\ref{fig:diagnostics} and~\ref{fig:diagnostics_vms} is whether or not VMS are included in the SSPs. In the first two columns of Fig.~\ref{fig:diagnostics} and first column of Fig.~\ref{fig:diagnostics_vms}, models of photoionization by massive stars have no escape of LyC photons, while they have some fraction of escape given by the HbFrac parameter in the rest of columns (see Tab.~\ref{tab:photoionization_grids}). In Fig.~\ref{fig:diagnostics_vms} (models with VMS), we exclude the three galaxies in our sample with P-Cygni profiles, for which we find no clear evidence of the presence of VMS (J1016, J1230, and J2230; see \S~\ref{sec:stars}). In addition, we exclude models at ages > 3 Myr, as no VMS are expected to be present beyond 3 Myr. 

\subsubsection{Optical diagrams}

The top rows of Figures~\ref{fig:diagnostics} and~\ref{fig:diagnostics_vms} show the location of the galaxies in the BPT diagnostic diagram of \citet{Baldwin1981}. In this diagram, the empirical line of \cite{Kauffmann2003} and theoretical lines of \cite{Kewley2001} and \cite{Stasinska2015MNRAS.449..559S} show the limits above which photoionization by OB stars is excluded. Above the dashed curve of \cite{Kewley2001}, one mostly finds AGN or retired galaxies; while between the dashed and dotted-dashed curves one finds a mixture of star formation and Seyfert nuclei, or star formation and LINERs \citep{Kewley2006}. The dotted line was proposed by \citet{Schawinski2007MNRAS.382.1415S} to separate Seyferts from LINERs. Objects below the lines of \cite{Kewley2001} and \cite{Stasinska2015MNRAS.449..559S} can be dominated by photoionization by stars, AGN \cite{Feltre2016MNRAS.456.3354F}, or shocked regions \citep{Alarie2019RMxAA..55..377A}, particularly at low metallicity. In Fig.~\ref{fig:diagnostics}, we overlay low-Z models of photoionization by OB stars, AGN, and shocks. At low-Z, these models partially overlap. The different branches of the cyan models correspond to different values of log U (each branch) and different ages. SBS0335-052E, J0944, and J1044, which are the most metal-poor galaxies considered, are closest to the overlapping region in the BPT. Models of photoionization by stars with no LyC escape are unable to reach the positions of SBS0335-052E and J1044, even when VMS are included (see Fig.~\ref{fig:diagnostics_vms}). We computed models (not displayed in the Figs.) that show that lowering the value of N/O helps, but only when the value is below what is observed for XMPGs by \cite{Arellano2025MNRAS.544.1588A}. Only models with LyC escape reach the location of the above two and other galaxies in our sample, in the BPT.

In Figs.~\ref{fig:diagnostics} and~\ref{fig:diagnostics_vms}, the second row of panels shows the location of the galaxies in the He\2/H$\beta$ vs. age plot. For the galaxies with derived age determinations reported in Tab.~\ref{tab:population_properties}, there are three symbols per galaxy corresponding to these ages. For the rest of galaxies, we use horizontal lines to indicate the values of their H\2/H$\beta$ ratios. If one considers models without LyC escape, PyStarburst99 is unable to reach the galaxies while CB19 can do it in some cases for both metallicities. However, this is not the case for SBS0335-052E, in agreement with what we found for this galaxy \citep[][see fig. 10]{Wofford2021MNRAS.500.2908W}. The increase in the He\2/H$\beta$ ratio predicted by the CB19 Z=0.002 models (with and without LyC escape) at 3.5 Myr is attributed to cWR stars. At the lower Z=0.0002 metallicity, this effect is absent, as expected due to the weaker winds of the massive stars, which are driven by metal lines. Only CB19 models with LyC escape or VMS can reach the He\2/H$\beta$ ratio of SBS0335-052E. Interestingly, for galaxies with P-Cygni profiles, in most cases, even the CB19 models with LyC escape fail to reproduce the observed He\2/H$\beta$ ratios at the ages that we determined for these galaxies, except for J1230. 

\subsubsection{UV diagrams}

As previously mentioned, UV diagnostic diagrams break some degeneracies that the BPT suffers from, especially in the XMP regime, where models of photoionization by AGN, shocks and hard stellar fields overlap,. \cite{Plat2019MNRAS.490..978P} combined the AGN models of \cite{Feltre2016MNRAS.456.3354F}, shock models of \citep{Alarie2019RMxAA..55..377A}, and models of density-bounded nebulae ionized by massive stars into UV + optical diagrams for metal-poor galaxies. Regarding photoionization by AGN versus massive stars, \cite{Feltre2016MNRAS.456.3354F} explicitly compares UV and optical diagnostics and identifies UV diagrams that separate AGN from stars robustly, such as C\4/He\2 vs C\3]/He\2. \cite{Gutkin2016MNRAS.462.1757G} presents the companion set of massive-star photoionization models. Some of the above  models have been explicitly confronted to XMPGs in \cite{Senchyna2019b} and \cite{Wofford2021MNRAS.500.2908W}, who both found that observations fall at the edge of model grids that only account for one type of ionizing source. \cite{Nakajima2018} proposed EW(C\3]) vs C\3]/He \2 and EW(C\4) vs C\4/He\2 as AGN vs. massive-star classifiers. \cite{Nakajima2022MNRAS.513.5134N} extended UV diagnostics to Pop III and direct-collapse black hole scenarios. Finally, using the CLASSY survey \cite{Berg2022ApJS..261...31B, Mingozzi2022ApJ...939..110M, Mingozzi2024ApJ...962...95M} tested UV diagnostics on local high-z analogues with HST/COS. They found that some proposed UV diagrams work well but others misclassify extreme star-forming galaxies. 

Note that UV diagnostics have their own degeneracies. C\4 is affected by resonant scattering and stellar wind absorption, which is why \cite{Mingozzi2022ApJ...939..110M} recommends avoiding this line. As previously mentioned, He\2 can come from Wolf-Rayet stars, X-ray binaries or very metal-poor stellar populations. C\3] depends on C/O abundance and density. Shocks with precursors can mimic AGN in several UV ratios. This is why combining UV and optical diagnostics might be useful.

\begin{figure*}
    \centering
    \includegraphics[height=0.7\textheight]{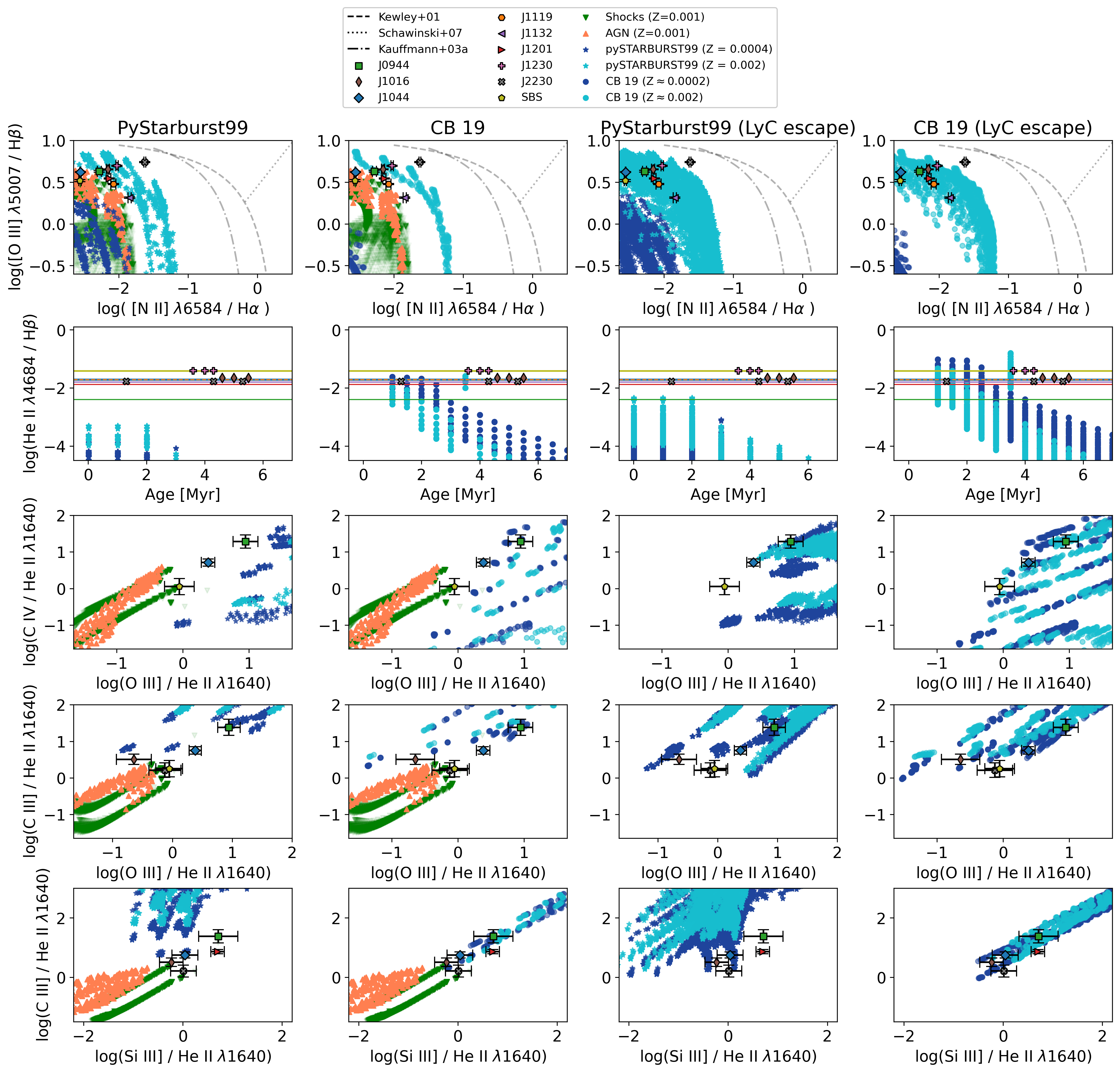}
    \caption{Comparison of models and observations in optical and UV diagnostic diagrams. In all panels, we add the extreme XMPG SBS 0335-052E \citep[olive pentagon;][]{Wofford2021MNRAS.500.2908W} for reference. From top to bottom, the diagrams are: (i) BPT \citet{Baldwin1981} with lines of \citet{Kewley2001}, \citet{Kauffmann2003} and \citet{Schawinski2007MNRAS.382.1415S} overlaid; 
    (ii) He\,{\sc ii} $\lambda4686/$H$\beta$ at the ages listed in Table~\ref{tab:ages}. If no age is available from the comparison of spectral synthesis models with direcet stellar lines, we use a horizontal line to indicate the value of He\,{\sc ii} $\lambda4686/$H$\beta$; 
    (iii) C\,{\sc iv}\,$\lambda\lambda1548,51$/He\,{\sc ii}\,$\lambda1640$ versus O\,{\sc iii}]\,$\lambda1666$/He\,{\sc ii}\,$\lambda1640$; 
    (iv) [C\,{\sc iii}],C\,{\sc iii}]\,$\lambda\lambda1906,9$/He\,{\sc ii}\,$\lambda1640$ versus O\,{\sc iii}]\,$\lambda1666$/He\,{\sc ii}\,$\lambda1640$; 
    and (v) [C\,{\sc iii}],C\,{\sc iii}]\,$\lambda\lambda1906,9$/He\,{\sc ii}\,$\lambda1640$ versus [Si\,{\sc iii}],Si\,{\sc iii}]\,$\lambda\lambda1883,92$/He\,{\sc ii}\,$\lambda1640$. 
    Columns 1 and 2 (3 and 4) show models of photoionization by SSPs without VMS or binaries, without (with) LyC escape. In columns 1 and 3, we overlay the $Z=0.002$ (cyan stars) and $Z=0.0004$ (dark blue stars) models with SSPs from PyStarburst99. The metallicities of our galaxies are intermediate between these two values. In columns 2 and 4, we show models using SSPs from CB19 with $Z\approx0.002$ (cyan circles) and $Z\approx0.0002$ (dark blue circles). In cases (i) and (iii) - (v), we also show the $Z=0.001$ models of photoionization by AGN of \citep[orange triangles;][]{Feltre2016MNRAS.456.3354F} and by shocks of \citep[green triangles;][]{Alarie2019RMxAA..55..377A}. 
    } 
    \label{fig:diagnostics}
\end{figure*}

\begin{figure}
    \centering
    \includegraphics[width=0.95\linewidth]{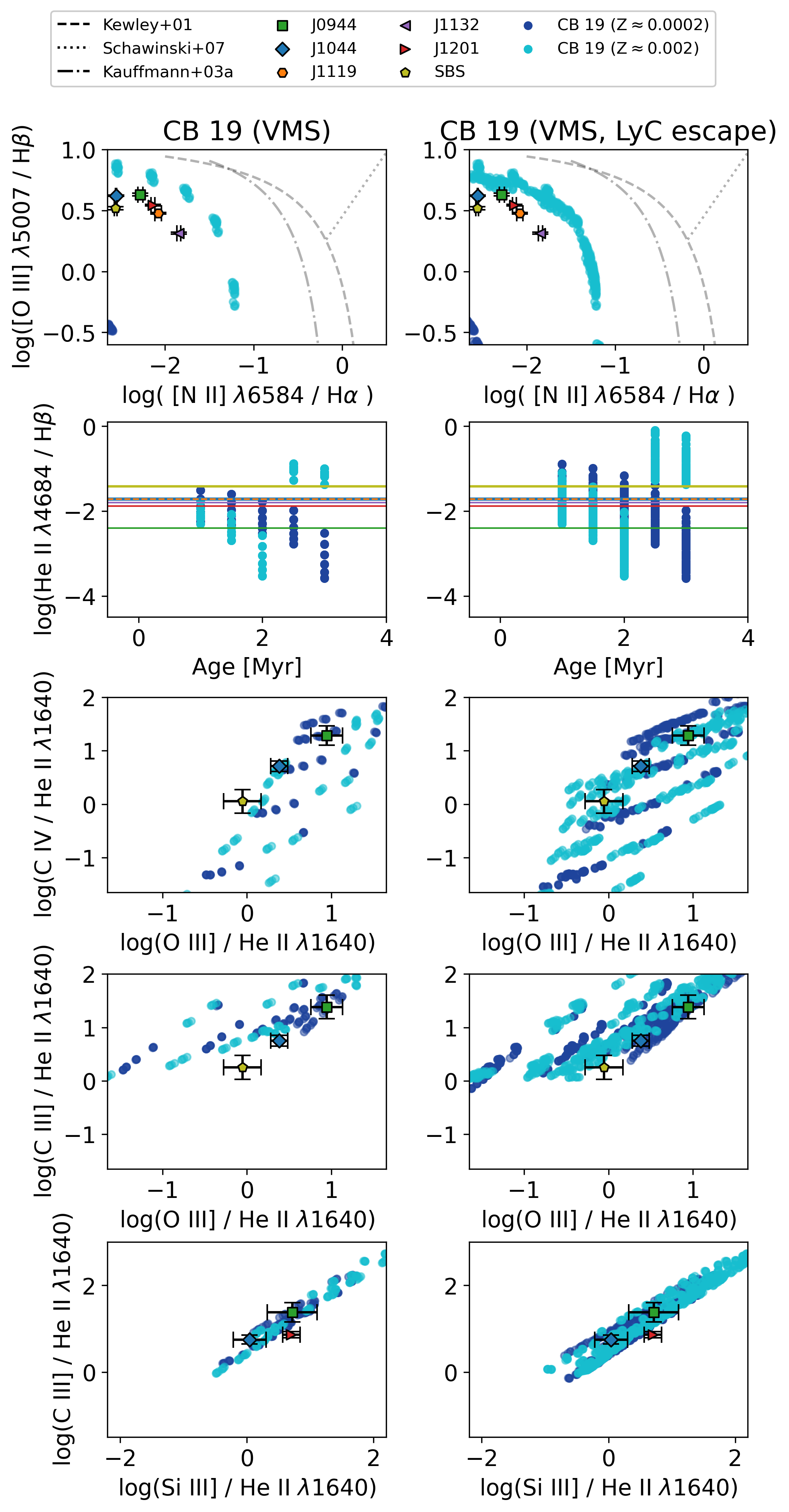}
    \caption{Comparison of six XMPGs against models of photoionization by SSPs that include VMS of up to 300 M$_\odot$. The models are an updated version of the models presented in \citet{Martinez2023MNRAS.525.2916M} and use CB19 spectra as input (see text for details). We plot the $Z=0.002$ (cyan circles) and $Z=0.0002$ (dark blue circles) models. Models without (with) LyC escape are shown in columns 1 (2).}
    \label{fig:diagnostics_vms}
\end{figure}

The last three rows of Figs.~\ref{fig:diagnostics} and~\ref{fig:diagnostics_vms} show UV diagnostic diagrams based on different combinations of line ratios. Only galaxies with measurements of He\2 1640 are shown. The first two columns show that in these UV diagrams, the XMP models of photoionization by stars are better separated from predictions of photoionization by XMP AGN and shocks. In addition, the bottom row of Fig.~\ref{fig:diagnostics} shows that according to the CB19 models, photoionization by massive stars might be dominant. Let us now focus on the two most metal-poor galaxies in our sample, J0940 and J1044. Figure~\ref{fig:diagnostics} shows that without LyC escape, the CB19 models approach them better at the lowest metallicity, and that with and without LyC escape, these galaxies are the outer edge of the grid of models in the top two UV diagrams. This is also the case when VMS are included with no LyC escape.

\section{Summary and conclusions}\label{sec:conclusion}

We analyse new COS G140L UV spectra and optical SDSS observations of eight nearby metal-poor galaxies. Seven are extremely metal-poor galaxies (XMPGs), while the remaining galaxy has sub-SMC metallicity. All galaxies exhibit narrow He\2 $\lambda4686$ emission and lack classical Wolf-Rayet (cWR) blue and red bumps. The spectra cover the wavelength range from P\5 $\lambda1118$ to C\3] $\lambda1909$. Two galaxies have UV spectra with insufficient S/N for our purposes; therefore, we only use their optical data in our analysis of the ionized gas.

The six galaxies with useful UV spectra can be divided into two groups (Fig.~\ref{fig:g140l-sp}). The first group comprises galaxies with weak N\5 $\lambda1240$ and C\4 $\lambda1550$ P-Cygni-like doublets. Their C\4 profiles may include a contribution from nebular emission. These galaxies have smaller EW(H$\beta$) values than those in the second group and exhibit visible stellar Balmer absorption components. The second group comprises the three most metal-poor galaxies, whose C\4 $\lambda1550$ profiles are dominated by narrow emission, with no clear absorption components. These galaxies have EW(H$\beta)\gtrsim200$ \AA~and undetected stellar Balmer absorption lines.

We test the spectral consistency of STARBURST99, GALAXEV, and BPASS predictions at $Z=0.001-0.004\,Z_\odot$ and ages of 2 and 4 Myr. In models that account for VMS, these stars are still present at 2 Myr. To this end, we compare stellar-wind lines sensitive to metallicity, the presence of O-type stars, and their evolved descendants (Fig.~\ref{fig:model_comparison}). The BPASS predictions differ from those of STARBURST99 and GALAXEV because of differences in the chemical composition adopted in the BPASS models (private communication). At 2 Myr, in the CB19-300 (GALAXEV) and sin-300 and bin-300 (BPASS) models, the classical He\2 $\lambda1640$ emission line characteristic of VMS is either absent or too weak to serve as a useful VMS signpost. The most notable difference in the CB19-300 model is the strength of C\4 $\lambda1550$, although this line is also sensitive to age.

Although the models considered here are not suitable for directly identifying candidate VMS through their stellar He\2 and C\4 lines at $Z=0.001-0.004\,Z_\odot$, their predicted ionizing fluxes are useful inputs to photoionization models. These models enable indirect searches for VMS through the high-ionization nebular emission lines they produce. 

To search indirectly for candidate VMS through nebular emission lines and investigate the nebular gas properties of our XMPGs, we compare their optical and UV emission-line ratios with photoionization models of OB-star populations from pySTARBURST99 and GALAXEV, including models with VMS (GALAXEV). Since the ages inferred for galaxies with weak P-Cygni-like profiles are too old for them to potentially host VMS, we exclude these galaxies from part of the analysis (see \S~\ref{sec:gas}). We can only reproduce the extreme optical and UV nebular emission-line ratios of J1044 and SBS 0335-52E by adopting unrealistic values of one or more parameters: a very low N/O ratio, a very high ionization parameter, and/or a very high Lyman continuum escape fraction. We find no indirect evidence for candidate VMS in the galaxies belonging to the second group.

We attribute our failure to find direct or indirect evidence of VMS at $Z\lesssim0.1\,Z_\odot$ primarily to their rarity, which results from the shape of the IMF, which disfavours the formation of the most massive stars, and their very short lifetimes, which make them difficult to detect. Our relatively small galaxy sample may also contribute to this result. Alternatively, different types of ionizing sources may be acting simultaneously in some or all of our XMPGs, contrary to our assumption of a single dominant ionizing source. Finally, the VMS models currently available in spectral synthesis codes may be inadequate. In particular, \cite{Grafener2015A&A...578L...2G} present a model of narrow-line XMP VMS that has not yet been implemented in spectral synthesis models. If some of the narrow He\2 emission observed in our galaxies is stellar in origin, less extreme photoionization conditions may be sufficient to reproduce the nebular He\2 emission. In this context, \cite{Martins2025A&A...698A.262M} present a spectral synthesis model that predicts stellar He II emission from VMS at $Z=0.1\,Z_\odot$. Their model assumes that the stellar mass-loss rate does not scale with metallicity. The width of the predicted He\2 emission is not clearly established in that work. It would have to be narrow to match our observations.

Improvements to spectral synthesis models that include VMS at $Z\lesssim0.1\,Z_\odot$ are therefore needed to enable more reliable searches for these stars with future observatories such as the Habitable Worlds Observatory \citep{Martins2026ASPC..542...37M}.

\section{Acknowledgements}
We thank the referee for the careful revision of our work and comments that helped us greatly improve the quality of our paper. We also thank J. Eldridge and E. Stanway for providing useful insights regarding the BPASS v2.2 models. A. W. and O. A. acknowledge the support of UNAM's DGAPA through project PAPIIT IN106922 and IN107725. O. A. acknowledges support from CONAHCyT Beca Nacional para estudios de Posgrado, CVU 1142736. C.M acknowledges the support of UNAM's DGAPA through projects PAPIIT IG101223 and IN113226.  


\section*{Data Availability}
 
The HST/COS data underlying this article are available in the Mikulski Archive for Space Telescopes (MAST) at \url{https://www.stsci.edu/hst-program-info/program/?program=15220}, under program ID 15220.



\bibliographystyle{mnras}
\bibliography{bib} 


\appendix

\section{Spectral Synthesis Models}\label{app:model_details}

In our work, we use SSP spectral synthesis models generated with STARBUST99, pySTARBURST99, GALAXEV, and BPASS, which are described below. Main inputs of all models are the stellar evolution tracks, stellar spectral libraries, stellar initial mass function (IMF), and metallicity (Z). Although no VMS have been reported in nearby XMPGs, this could be beacause they are rare due to the shape of the IMF and because they are very short-lived (<3 Myr). In this work we consider some models with VMS. Some important similarities and differences in the model parameters are summarized below. We focus on massive stars which dominate the UV emission of young stellar populations. 

\subsection{STARBURST99 and pySTARBURST99}

We use S99 models for comparison with direct stellar lines and pySTARBURST99 models as input to models of photoionization by massive stars.

For OB stars, our S99 models use the Geneva 2012/2013 evolutionary tracks for single stars. We try the tracks with zero rotation as well as those with a rotation velocity equal to 40\% of the break-up velocity on the zero-age main-sequence \citep{Ekstrom2012A&A...537A.146E, Georgy2013A&A...558A.103G}. Note that the only subsolar metallicity that is available for these tracks, is Z=0.002. Following the notation of \citet{Leitherer2014ApJS..212...14L}, we refer to the rotating and non-rotating S99 subsolar-metallicity models as v00-l and v40-l, respectively.

For MS OB stars, S99  uses a theoretical stellar library that includes subsolar metallicities: Z=0.001, 0.004 and 0.008. The library was generated with the radiation hydrodynamics stellar atmospheres code WM-Basic \citep{Pauldrach2001A&A...375..161P} and is described in \cite{Leitherer2010ApJS..189..309L}. We compute v00-l and v40-l models at the lowest-two metallicities. Although the metallicities of the evolutionary tracks and library do not exactly match, as explained in  \citet{Leitherer2014ApJS..212...14L}, this is not a concern because the tracks are only used to select a reasonable grid of luminosities and effective temperatures for the stars. On the other hand, for Wolf-Rayet (WR) stars, S99 uses PoWR atmosphere models \citep{Grafener2002A&A...387..244G, Hamann2003A&A...410..993H, Sander2015A&A...577A..13S, Todt2015A&A...579A..75T}, but is limited to the solar-metallicity set described in \S3 and tab. 2 of \cite{Leitherer2014ApJS..212...14L}, which was the only available at the time. 

Although there is an updated version of S99 called pySTARBURST99 that is python based and is presented in \citet{Hawcroft2025ApJS..280....5H}, for MS OB-type stars, pySTARBURST99 uses the FASTWIND v10 stellar atmosphere code \citep{Santolaya-Rey1997A&A...323..488S, Puls2005A&A...435..669P} and a low spectral resolution stellar library that does not account for stellar wind lines. Thus, for the comparison with observed stellar wind profiles, we use S99, which is the older FORTRAN version. However, for the photoionization models, we use the pySTARBURST99 version, since it has more consistent metallicities between the stellar evolution and stellar atmosphere models.

Models with VMS are unavailable with both S99 and PySTARBURST99 at the metallicities of interest in this work.

The S99 models follow the \cite{Kroupa2002Sci...295...82K} IMF with exponents of 1.3 and 2.3, and mass boundaries of 0.1, 0.5, and 100 M$_\odot$. 

\subsection{GALAXEV}\label{sec:models}

We use the so-called Charlot \& Bruzual 2019 (CB19) models. For stars through the WR phase, CB19 uses single non-rotating PARSEC \citep{Bressan2012MNRAS.427..127B} tracks computed by \cite{Chen2015MNRAS.452.1068C}. The tracks are available for several metallicities, which are listed in appendix A/tab. 8  of \citealt{Sanchez2022ApJS..262...36S}. In our case, the relevant tracks have Z=0.001 and 0.004. 

The CB19 models use the same theoretical spectral library, as S99, but for the atmospheres of WR stars, it uses the PoWR models listed in tab.~A1 of \citet{Plat2019MNRAS.490..978P}. The relevant WR models associated with the above tracks in this work have log$_{10}$(Z/Z$_\odot$) = -1 and -0.7, where Z$_\odot$=0.017 \citep{Sanchez2022ApJS..262...36S}.

For VMS, the CB19 models use the \cite{Chen2015MNRAS.452.1068C} PARSEC tracks and dedicated stellar spectral models. In practice, since the Chen et al. models are parametrized by (Teff, log g and Z) and not by stellar mass, Chen et al. models are used in GALAXEV whenever a MS star is hotter than 50,000 K, irrespective of its mass. In practice this high Teff in the MS happens mostly for VMS. The Chen et al. models cover from 50 A to the far infrared with relatively low resolution and must be interpolated to the same resolution as the PoWR models, for compatibility.

The CB19 models follow a \cite{Chabrier2003PASP..115..763C} IMF. We try models with a lower mass limit of $0.1$ M$_\odot$ and and upper mass limit of either 100 or 300 M$_\odot$, where the latter account for VMS. We refer to these models as CB19-100 and CB19-300, respectively.  

\subsection{BPASS}

In this work, we use Binary Population and Spectral Synthesis (BPASS) data v2.2 models, which builds on the detailed modelling described in \cite{Eldridge2017PASA...34...58E}, and includes improved or modified prescriptions for older stellar populations, as described in \cite{Stanway2018MNRAS.479...75S}. This release includes stellar evolution models for isolated single stars (hereafter, ``sin'' models), and treatment of binary star evolution with a range of periods and mass ratios (hereafter, ``bin'' models), in particular, massive mass-exchanging stars. We use models with Solar-scaled metal abundances and stellar metallicities of Z=0.001 and 0.004. 

BPASS models use their own O-star WM-Basic spectral library, which is complemented with the PoWR WR models that are described in \S2.3 of \cite{Eldridge2017PASA...34...58E}. For reference,  \cite{Eldridge2017PASA...34...58E} compare predictions from the solar-metallicity S99 v00-h model of \cite{Leitherer2014ApJS..212...14L} to UV spectra from single non-rotating BPASS models with the same metallicity and IMF, and find that they yield  similar results. 

Although VMS evolutionary tracks are considered in BPASS v2.2, the spectra used for VMS remain the same as those for normal OB and WR stars. For stars of 100-300 $M_\odot$, both the evolutionary tracks and the stellar atmospheres were computed with the mass-loss rates of \cite{Vink2001A&A...369..574V}. BPASS does not currently incorporate enhanced mass-loss for VMSs resulting from their proximity to the Eddington limit.

Finally the BPASS models follow a  \cite{Chabrier2003PASP..115..763C} IMF with a lower mass limit of 0.1 M$_\odot$. We try BPASS models with and without VMS, i.e., with IMF upper mass limits of 100 and 300 M$_\odot$, respectively. Hereafter, we refer to the sin/bin models with these IMF limits as sin-100/bin-100 and sin-300/bin-300, respectively.

The BPASS models are only available at 51 ages after the initial
starburst, ranging from 1 Myr to 100 Gyr in increments of
log(age/years)=0.1. Thus, their age sampling is coarser than for the CB19 and S99 models for first 10 Myr. Note that according to \cite{Beryl2025arXiv251116752H}, the contribution of stars with atmospheres removed via mass exchange with a binary companion (stripped stars) to the non-ionizing continuum is negligible. In addition, according to \cite{Gotberg2019A&A...629A.134G}, the contribution of stripped stars to the ionizing spectrum is significant only after 10 Myr.

\section{Galaxies excluded from UV diagnostic diagrams}\label{app:bad_galaxies}

\begin{figure*}
    \centering
    \includegraphics[width=1\linewidth]{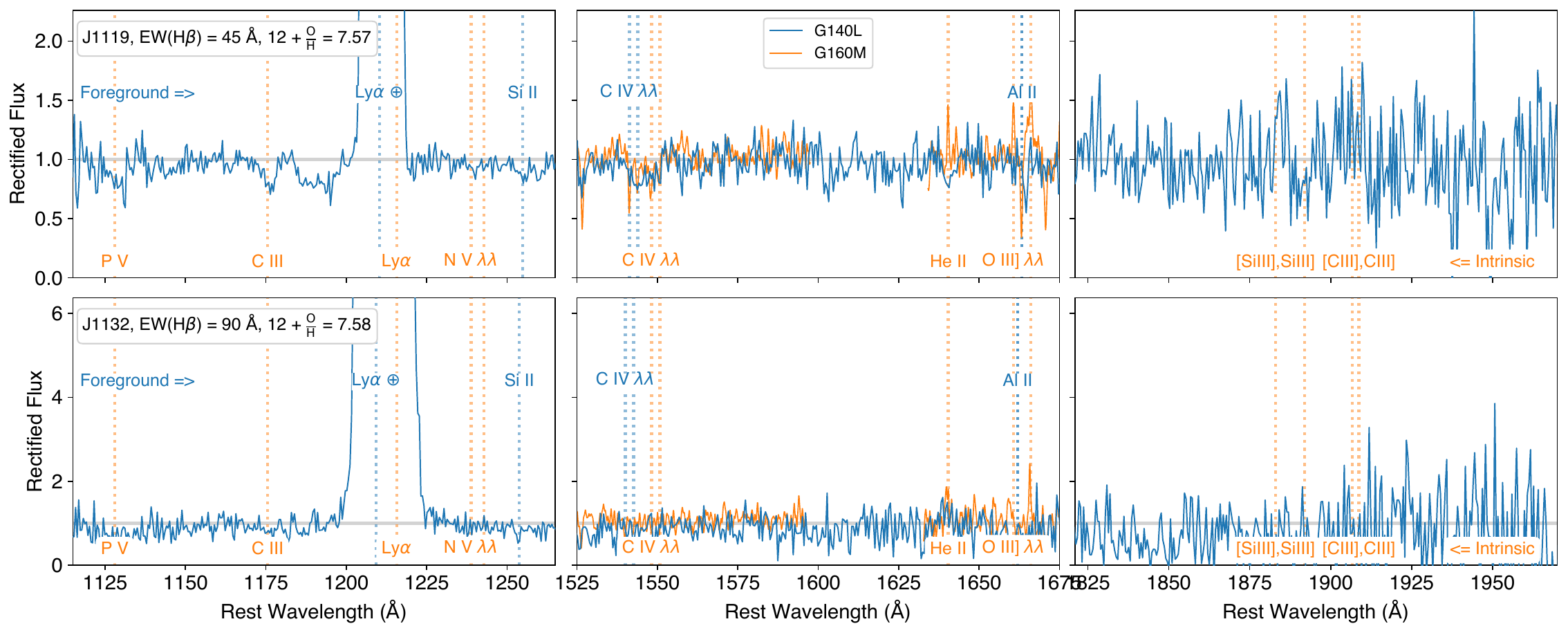}
    \caption{Similar to Fig.~\ref{fig:g140l-sp}, but for the two galaxies excluded from UV diagnostic plots.}
    \label{fig:bad_uv_sp}
\end{figure*}

The UV spectra of extremely-metal-poor galaxies J1119 and J1132 pose some challenges. As shown in Fig.~\ref{fig:bad_uv_sp}, which is similar to  Fig.~\ref{fig:g140l-sp}, the UV spectrum of J1132 is mostly noise and a reliable He\2 1640 emission line flux cannot be obtained from these data. This is expected given the lack of a bright source in the COS TA image (see Fig.~\ref{fig:ta_images}). Only the fluxes of some nebular UV lines were measured for this galaxy (see Tab.~\ref{tab:uv_fluxes}). On the other hand, the UV spectrum of J1119 is also noisy. It has no detectable stellar N\5 1240 doublet and has a C\4 1550 absorption that seems to be mostly from the ISM. In addition, although  He\2 1640 is detected in absorption in this galaxy, the higher-resolution G160M spectrum of his galaxy shows it in emission. This could be an effect of the different PA of the G140L and G160M observations, but we are interested in He\2 emitters. For this galaxy, some nebular line UV fluxes are reported in Tab.~\ref{tab:uv_fluxes}. Given the lack of reliable He\2 1640 emission line measurements for these two galaxies (from the G140L data), we excluded the galaxies from our UV diagnostic diagrams (Figs.~\ref{fig:diagnostics} and~\ref{fig:diagnostics_vms}). However, we used their optical emission lines in the latter two figures.

\bsp	
\label{lastpage}
\end{document}